%% file: manuscript.tex
\documentclass{arxivpreprint}

\usepackage{amsmath,amssymb,amsfonts,bm}
\usepackage{graphicx}
\usepackage[utf8]{inputenc}
\usepackage[T1]{fontenc}
\usepackage[protrusion=true,expansion=false]{microtype} % reduce overfull lines
\usepackage{booktabs}
\usepackage{enumitem}
\usepackage{xcolor}
\usepackage{float}
\usepackage{ragged2e}
\usepackage{rotating}
\usepackage{longtable}
\usepackage[round,sort&compress,numbers]{natbib} % numeric citations to match IOP house style

\graphicspath{{figures/}{supplementary_figures/}}

\providecommand{\br}{\toprule}
\providecommand{\mr}{\midrule}

\newcommand{\bx}{\mathbf{x}}
\newcommand{\bu}{\mathbf{u}}
\newcommand{\beps}{\boldsymbol{\varepsilon}}
\newcommand{\dA}{\Delta A}
\newcommand{\Thalf}{T_{1/2}}
\newcommand{\kBT}{k_B T}
\newcommand{\Vm}{V_m}

\begin{document}
\justifying
\fancyhead[R]{G F Pinton}

\articletype{Paper}

\title{An open-source framework for predicting ultrasound neuromodulation: bridging tissue elastomechanics and neuron firing dynamics}

\author{Gianmarco F Pinton$^{1,*}$\orcid{0000-0002-4896-1439}}

\affil{$^1$Lampe Joint Department of Biomedical Engineering, University of North Carolina at Chapel Hill and North Carolina State University, Chapel Hill, NC, USA}

\affil{$^*$Author to whom any correspondence should be addressed.}

\email{gia@email.unc.edu}


\begin{abstract}
Transcranial focused ultrasound is a non-invasive neuromodulation modality with millimetre-scale spatial resolution, but its biophysical mechanism of action remains unresolved. Exposure is conventionally specified by transducer surface pressure or derated focal pressure, quantities that are only indirectly related to the variables that ultimately matter for therapy, namely which neurons fire and through which biophysical pathway they are recruited. A meaningful definition of acoustic dose for neuromodulation would thus ideally reach the cellular level, identifying the neuron types being driven and the mechanism by which they are driven. Current ultrasound simulation tools alone can neither distinguish between competing mechanistic hypotheses nor estimate the parameters that govern them. We address this gap with an end-to-end computational framework that maps a transcranial acoustic field to per-voxel neural firing maps registered to anatomy. The pipeline couples heterogeneous nonlinear full-wave acoustic propagation, viscoelastic shear-wave propagation, Pennes bioheat thermal diffusion, a bilayer-mechanics conversion from tissue strain to membrane tension, and a multi-compartment Hodgkin--Huxley neuron carrying mechanosensitive, cavitation-coupled, calcium-coupled, thermosensitive, astrocytic-gliotransmitter, and mechanosensitive-synaptic channel pathways. Together these stages constitute a single volumetric spatio-temporal representation that spans acoustic, elastic, and thermal physics in rigid (skull) and soft (brain) media at sub-millimetre voxel resolution ($dx = 0.308$\,mm, $\sim 28$ million voxels over a $130 \times 80 \times 80$\,mm head volume), bridging macroscopic field propagation, mesoscopic per-voxel strain and tension fields, and microscopic ion-channel gating and single-neuron firing. Six candidate mechanisms are implemented as interchangeable components on a shared neuron model so that their firing predictions can be compared directly on the same field, and every numerical parameter is classified by source and bracketed by sensitivity analysis. We apply the framework to a theta-burst sonication delivered through a micro-CT human-skull specimen targeting the left dorsal anterior cingulate cortex (one of the two deep cortical targets sonicated by Yaakub et al.), predicting a focal firing zone of approximately 8\,500\,mm$^3$ at a focal thermal rise well within the ITRUSST consensus safety envelopes. The principal output is a per-voxel firing-volume map registered to anatomy and resolved jointly with the underlying acoustic, elastic, and thermal field histories that drive it, providing spatially resolved, falsifiable predictions that can be tested against high-density extracellular recordings. By linking acoustic exposure to cellular firing, the framework supports parameter estimation, cell-type-resolved mechanism identification, and quantitative safety assessment for ultrasound neuromodulation.
\end{abstract}

% =============================================================================
\section{Introduction}
\label{sec:intro}

Transcranial focused ultrasound (tFUS) has emerged as a non-invasive neuromodulation modality with millimetre spatial resolution and the ability to reach deep brain structures~\citep{deffieux2013lowintensity,legon2014transcranial,tufail2010transcranial,verhagen2019offline,folloni2019manipulation,lee2016transcranial}. Despite a decade of clinical and animal studies and a growing body of behavioural evidence~\citep{naor2016ultrasonic,blackmore2019ultrasound,fomenko2018lowintensity}, the biophysical mechanisms by which acoustic energy is translated into changes in neural firing remain unresolved. Several candidate pathways have been argued for in the literature, including direct mechanosensitive ion-channel gating by tissue strain through Piezo1 and the K2P family (TRAAK, TREK-1, TREK-2)~\citep{kubanek2018ultrasound,qiu2019piezo1,yoo2022focused,zhu2023piezo1}, intramembrane cavitation in the form of the Plaksin--Krasovitski sonophore whose carrier-frequency leaflet motion produces a cycle-averaged perturbation of membrane capacitance~\citep{plaksin2014intramembrane,plaksin2016celltype,lemaire2019understanding}, intracellular calcium-coupled inhibitory pathways acting independently of the K2P channels, and thermosensor (TRP) activation under high-intensity protocols where focal heating exceeds a few Kelvin~\citep{lee2016transcranial,verhagen2019offline}.

The difficulty in arriving at a mechanistic interpretation of tFUS results is less a shortage of plausible candidates than an absence of tools that bridge the disciplines involved. Ultrasound exposure is naturally described in the language of acoustics and continuum mechanics, whereas neural responses are measured and modelled in the language of cellular electrophysiology, and no quantitative framework currently carries an exposure prescription from the transducer through to a per-voxel statement about which neurons fire and how they are driven. This raises the barrier to interdisciplinary interpretation of experimental results and makes mechanism comparisons difficult, because each candidate pathway has so far been studied with a different upstream model. Acoustic propagation has been well-modelled by validated FDTD solvers that handle heterogeneous tissue and skull aberration~\citep{pinton2009fullwave,treeby2010kwave,aubry2022itrusst}, the radiation-force-to-shear-wave coupling has been captured by recent shear-FDTD and continuum-mechanics work~\citep{salahshoor2020transcranial}, intramembrane cavitation has been reduced to tractable cycle-averaged form by SONIC~\citep{lemaire2019understanding} on a single-compartment Hodgkin--Huxley neuron without strain-driven channels, and mechanosensitive-channel models are typically demonstrated in isolation on a single cell type~\citep{qiu2019piezo1,yoo2022focused,zhu2023piezo1}. Each segment of the chain from pressure to firing exists in the literature, but no published framework chains them at the per-voxel level needed for spatially resolved mechanism comparison.

Recent advances in acoustic and elastomechanical simulation have begun to make such a bridge feasible. Heterogeneous angular-spectrum solvers and full-wave finite-difference models now propagate ultrasound through skull-aberrated head anatomy at clinically relevant frequencies in minutes rather than hours~\citep{pinton2009fullwave,treeby2010kwave,aubry2022itrusst}, and a complementary class of viscoelastic and shear-wave propagators, including finite-element models built on LS-DYNA~\citep{palmeri2005finite} and recent shear-wave finite-difference solvers~\citep{salahshoor2020transcranial,pinton2025shearfdtd}, converts the resulting radiation-force distribution into voxel-resolved tissue displacement and strain. Transcranial propagation is itself difficult, because skull bone introduces mode conversion, attenuation, and aberration that vary on the scale of millimetres, and the elastic response of brain tissue is viscoelastic and anisotropic at the resolutions where membrane-level effects emerge. These tools resolve the transcranial acoustic, viscoelastic, and shear-strain fields with sufficient spatial fidelity that the bilayer-mechanics conversion from local strain to membrane tension, on which the cellular pathways depend, becomes tractable at every voxel of a whole-brain volume rather than at only a handful of points.

The present work develops an acousto-mechanical-to-cellular-electrophysiology bridge. The framework takes as input a transcranial pressure or displacement field from any compatible upstream solver and produces per-voxel neural firing maps registered to anatomy, with consistent units, a single time-stepping discipline, and a per-parameter source classification that distinguishes literature-anchored measurements, calibrated values, modelling assumptions, and parameters derived from others. Six candidate mechanisms are implemented as interchangeable modules on a shared multi-compartment Hodgkin--Huxley neuron model. The first is direct strain-driven mechanosensitive-channel gating involving Piezo1 and the K2P family~\citep{kubanek2018ultrasound,qiu2019piezo1,yoo2022focused,zhu2023piezo1}. The second is intramembrane cavitation in its Plaksin--Krasovitski sonophore form, in which the lipid bilayer's carrier-frequency leaflet motion produces a cycle-averaged perturbation of membrane capacitance~\citep{plaksin2014intramembrane,plaksin2016celltype,lemaire2019understanding}. The third is a Piezo1-inactivation, intracellular-calcium, and small-conductance-potassium counter-current pathway~\citep{coste2010piezo,kohler1996sk} that captures the high-pressure inhibition observed by Legon and colleagues~\citep{legon2014transcranial} without invoking the K2P channels. The fourth is thermosensitive TRP-channel activation coupled to a Pennes bioheat field, providing a parallel thermal pathway for high-intensity protocols~\citep{lee2016transcranial,verhagen2019offline}, and the radiation-force and bioheat pathways feed the same neuron so that the spatial relationship between thermal and mechanical foci is a quantitative output of the pipeline rather than an input assumption. The fifth is the astrocytic relay of Oh et al.~\citep{oh2019ultrasonic}, in which ultrasound gates mechanosensitive TRPA1 channels in astrocytes, the resulting Ca\textsuperscript{2+} influx triggers glutamate release, and glutamate drives neuronal NMDA receptors, a slow, non-Piezo pathway whose drive accumulates over the sonication rather than tracking the carrier, making its cumulative (duty $\times$ on-time) dependence a distinguishing prediction. The sixth is mechanosensitive synaptic and voltage-gated transmission, in which membrane tension raises presynaptic vesicle-release probability~\citep{tyler2008remote,tyler2012mechanobiology}, enhances post-synaptic NMDA-receptor gating at fixed glutamate and voltage~\citep{maneshi2017mechanical}, and hyperpolarises the activation of voltage-gated Na\textsuperscript{+}/Ca\textsuperscript{2+} channels so as to increase their current~\citep{kubanek2016ultrasound}, acting on the fast synaptic and spike-generating machinery rather than through astrocytes, and bearing on stretch-enhanced synaptic plasticity. Vectorised integration over voxels combined with a tension-threshold pre-filter makes it tractable to evaluate roughly $10^5$ neurons on a transcranial grid, the throughput required for whole-brain firing-map predictions.

The framework is intentionally agnostic with respect to the mechanism debate. Rather than committing to a single hypothesis, it allows the same input field to be re-run under each pathway combination so that the resulting firing-map predictions can be compared directly to one another and, in principle, to experimental neural recordings. This is a framework paper rather than a mechanism paper. It does not adjudicate between the Piezo1, intramembrane-cavitation, and radiation-force-via-cytoskeleton accounts, and it leaves the multi-scale strain-to-tension coupling factor as a fitted nuisance parameter rather than calibrating it from first principles. Reported neuromodulatory protocols span a broad parameter space, with carrier frequencies from $\sim$250\,kHz to $\sim$2\,MHz, focal pressures from $\sim$50\,kPa to several MPa, duty cycles from below 1\% to continuous wave, interstimulus intervals from milliseconds to tens of seconds, ON-window durations from microseconds to minutes, and target sites distributed across cortical, thalamic, basal-ganglia, and brainstem structures~\citep{tufail2010transcranial,deffieux2013lowintensity,legon2014transcranial,verhagen2019offline,folloni2019manipulation,lee2016transcranial}. This protocol diversity motivates a framework that admits arbitrary acoustic exposures rather than one tuned to a single protocol. The demonstration application is a theta-burst sonication~\citep{yaakub2023transcranial} delivered through the Halle micro-CT human-skull specimen with the bowl aimed at the left dorsal anterior cingulate cortex (dACC), one of the two deep cortical regions targeted by Yaakub et al., with placement auto-selected on the outer-skull mesh (nearest 10--20 cap landmark AF3 at 28~mm). Quantitative reproduction of in-vivo behavioural datasets~\citep{tufail2010transcranial,deffieux2013lowintensity,legon2014transcranial} is part of a larger, ongoing validation effort.

\begin{figure}[htbp]
  \centering
  \includegraphics[width=\textwidth]{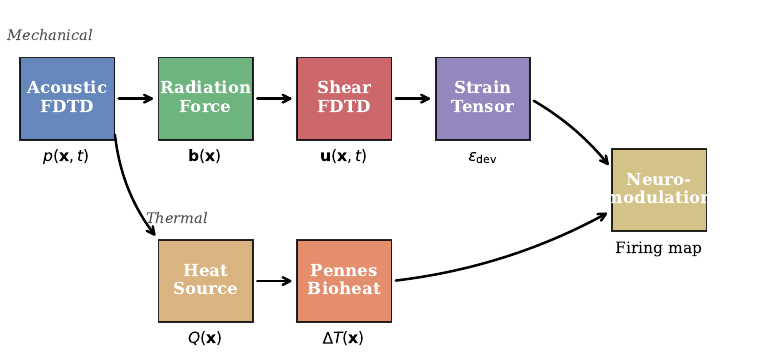}
  \caption{End-to-end simulation pipeline. An upstream Fullwave~2 solver produces the pressure field $p(\bx, t)$, which branches into two pathways. \textbf{Mechanical.} Radiation force $\mathbf{b}(\bx)$ drives a shear FDTD solver to produce tissue displacement $\bu(\bx, t)$, from which the strain tensor and von~Mises equivalent strain $\varepsilon_{\mathrm{eq}}(\bx)$ are extracted. \textbf{Thermal.} The volumetric heat source $Q(\bx)$ is computed from acoustic absorption, and the Pennes bioheat equation yields the temperature rise $\Delta T(\bx)$. Both pathways converge in the multi-pathway Hodgkin--Huxley neuromodulation block that is the contribution of this paper, which predicts spatial firing maps via mechanosensitive ion channels (strain), Q\textsubscript{10} thermal coupling (temperature), and the optional NICE, astrocytic-relay, and TRP pathways described in section~\ref{sec:methods}.}
  \label{fig:pipeline}
\end{figure}

The end-to-end pipeline is summarised in Figure~\ref{fig:pipeline}. The Fullwave~2 solver provides a pressure field that branches into a radiation-force and shear-FDTD path producing tissue displacement and strain, and an acoustic-absorption path producing volumetric heat that drives the Pennes bioheat equation. Both branches feed the multi-pathway neuromodulation block presented in this paper. The paper is organised as follows. Section~\ref{sec:methods} presents the core strain-to-tension-to-Hodgkin--Huxley pipeline (sections~\ref{sec:methods_strain}--\ref{sec:methods_multicomp}), three additional mechanism modules covering intramembrane cavitation, Piezo1 inactivation with Ca\textsuperscript{2+} / SK coupling, and TRP thermosensors (sections~\ref{sec:methods_nice}--\ref{sec:methods_trp}), the parameter-source classification (section~\ref{sec:methods_provenance}), and integrator choices (section~\ref{sec:methods_numerics}). Section~\ref{sec:results} reproduces five published validations (section~\ref{sec:results_validation}), applies the framework to the Yaakub canonical demonstration on the Halle micro-CT skull (auto-pick placement aimed at the left dACC) plus a real-field mechanism-comparison sweep that scales the same field across the threshold region (section~\ref{sec:results_tips}), and reports a source-classified sensitivity analysis (section~\ref{sec:results_sensitivity}). Section~\ref{sec:discussion} discusses the mechanism-discrimination outlook, limitations, and follow-up directions. An open-source implementation, with unit tests covering each pathway and a per-parameter source classification, accompanies the paper.

% =============================================================================
\section{Methods}
\label{sec:methods}

The framework begins with an upstream acoustic solve. The resulting pressure field $p(\bx, t)$ branches into two parallel paths, namely a radiation-force-driven viscoelastic shear-wave solve that yields the tissue-displacement field $\bu(\bx, t)$ and its strain tensor, and an acoustic-absorption-driven Pennes bioheat solve that yields the temperature field $T(\bx, t)$. The strain and temperature fields then feed the cellular-electrophysiology block presented in this paper, which returns per-voxel spike-count and firing maps under a chosen pathway combination. The pipeline is implemented to accept displacement, pressure-amplitude, and temperature inputs on any compatible Cartesian grid, so that alternative upstream solvers can be substituted without modifying the downstream cellular stages.

\subsection{Parameter source classification}
\label{sec:methods_provenance}

Every numerical parameter exposed by the framework is classified by its source in one of five categories. \emph{Literature-anchored} parameters are fit directly to a published experimental measurement and carry a citation. \emph{Calibrated} parameters are chosen so the model reproduces a qualitative literature observation, with the calibration target named explicitly. \emph{Assumed} parameters represent modelling choices that are not pinned by data. \emph{Estimate} parameters are best guesses from sparse or indirect data, a weaker class than an assumption, used for the TRPA1 mechanogating constants where no direct tension-clamp fit exists. \emph{Derived} parameters are computed from others. An automated check enforces that every literature-anchored entry carries a non-empty reference and every calibrated entry carries non-empty calibration-target notes. This classification makes mechanism comparisons traceable on a per-parameter basis. Differences in predicted firing between mechanism candidates are only meaningful if the underlying parameters are themselves grounded in the same way. The full per-parameter table, with worked-example assignments for each pathway, is given in supplementary~\S\ref{sec:supp_provenance}.

\subsection{Upstream acousto-elasto-thermal solvers}
\label{sec:methods_upstream}

The framework's design intent is to accept any FDTD or finite-element output that produces a displacement field, a pressure-amplitude field, and (optionally) a temperature field on a Cartesian grid. In practice the demonstration application uses three custom solvers from our group, namely Fullwave (acoustic), the reduced-viscoelastic shear-FDTD solver, and a Pennes-bioheat solver, that we summarise here for completeness.

\subsubsection{Fullwave~2 (acoustic solver)}

The Fullwave solver~\citep{pinton2009fullwave,pinton2020fullwave,pinton2021fullwave} integrates the heterogeneous nonlinear wave equation
\begin{equation}
  \frac{1}{c^2(\bx)}\,\frac{\partial^2 p}{\partial t^2}
  - \nabla^2 p
  + \mathcal{A}\bigl[p; \alpha(\bx, \omega)\bigr]
  = \frac{\beta(\bx)}{\rho_0(\bx)\,c^4(\bx)}\,\frac{\partial^2 p^2}{\partial t^2},
  \label{eq:fullwave}
\end{equation}
on a Cartesian grid with 10 points per wavelength. Heterogeneous sound speed $c(\bx)$, density $\rho_0(\bx)$, and the nonlinearity coefficient $\beta(\bx)$ are assigned per voxel from the segmented micro-CT geometry. Power-law absorption $\mathcal{A}[p; \alpha(\bx, \omega)]$ is implemented through a multiple-relaxation expansion that produces frequency-dependent attenuation $\alpha(\bx, \omega) \propto \omega^y$ with tissue-specific exponent $y$~\citep{pinton2021fullwave}. The right-hand side is the Westervelt-form quadratic nonlinearity that captures harmonic generation on transmit. Perfectly-matched-layer boundaries absorb outgoing waves. The solver passes to the downstream framework the acoustic pressure field $p(\bx, t)$ on the simulation grid, from which the radiation force $\mathbf{b}(\bx) = 2\alpha(\bx)\,\langle\mathbf{I}\rangle(\bx)/c(\bx)$ (with $\langle\mathbf{I}\rangle$ the time-averaged acoustic intensity vector) and the volumetric heat source $Q(\bx) = 2\alpha(\bx)\,I_{\mathrm{SPPA}}(\bx)$ are computed.

\subsubsection{Elastic shear-FDTD (reduced viscoelastic)}

The shear-wave solver~\citep{pinton2025shearfdtd} exploits the strong modulus hierarchy in soft tissue ($K \gg \mu$, with $K/\mu \sim 10^5$--$10^6$ for typical ultrasound conditions) to project the full Kelvin--Voigt momentum equation onto its solenoidal subspace, eliminating the bulk-pressure degrees of freedom and yielding a reduced shear-only equation
\begin{equation}
  \rho\,\frac{\partial^2 \bu}{\partial t^2}
  = \mu(\bx)\,\nabla^2 \bu
  + \eta(\bx)\,\nabla^2 \dot{\bu}
  + \mathbf{f}_S(\bx, t),
  \label{eq:shearwave}
\end{equation}
with shear modulus $\mu(\bx)$, viscosity $\eta(\bx)$, and the solenoidal projection $\mathbf{f}_S$ of the body force from the acoustic step. The reduced form runs on a 4$\times$-downsampled grid relative to the acoustic step (acoustic and shear timescales are separated by three orders of magnitude, so this is a clean operator-split coupling), retains Kelvin--Voigt frequency-dependent attenuation, and matches the analytical shear-wave dispersion to grid-error tolerance. The solver passes to the downstream framework the displacement field $\bu(\bx, t)$ on the focal slice, from which the strain tensor of section~\ref{sec:methods_strain} is computed. This reduced solver is a separate contribution of our group and underpins the present pipeline.

\subsubsection{Thermal Pennes bioheat}

Volumetric heat $Q(\bx)$ from acoustic absorption drives the Pennes bioheat equation~\citep{pennes1948analysis}, the de-facto standard for thermal modelling of therapeutic and diagnostic ultrasound~\citep{aubry2022itrusst},
\begin{equation}
  \rho\,c_p\,\frac{\partial \Delta T}{\partial t}
  = \nabla\cdot[k(\bx)\,\nabla \Delta T]
  - w_b(\bx)\,c_b\,\Delta T
  + Q_{\mathrm{avg}}(\bx),
  \label{eq:pennes}
\end{equation}
where $\Delta T(\bx, t) = T(\bx, t) - T_a$ is the temperature rise above the arterial baseline $T_a$, $k(\bx)$ is the heterogeneous tissue thermal conductivity, $w_b(\bx)$ the perfusion rate, $c_b$ the blood specific heat, and $Q_{\mathrm{avg}}(\bx) = D_c \cdot f_{\mathrm{abs}}(\bx)\, 2\alpha(\bx)\,I_{\mathrm{SPPA}}(\bx)$ the duty-cycle-averaged volumetric heat source over the PRF period (with $D_c$ the protocol duty cycle, e.g.\ $0.10$ for the Yaakub schedule). Here $f_{\mathrm{abs}}(\bx)$ is the fraction of attenuated longitudinal energy deposited locally as heat. It equals $f_{\mathrm{abs}} = 1$ in soft tissue, which is essentially purely absorptive, and $f_{\mathrm{abs}} = 0.30$ in cortical bone, where a large share of the attenuation is scattering and mode conversion to shear rather than local heating~\citep{pinton2012attenuation}. This bone fraction is referenced near 1\,MHz. Its $0.5$\,MHz frequency dependence is not separately scaled, and a $\pm$ sweep of $f_{\mathrm{abs}}^{\mathrm{bone}}\in[0.05,0.30]$ leaves the brain-tissue $\Delta T$ essentially unchanged while scaling the skull-bone peak. The PDE is integrated by an explicit finite-difference scheme with thermal-CFL-stable timestep over the full protocol session (80\,s for the Yaakub canonical run). The solver passes to the downstream framework the absolute temperature field $T(\bx) = T_a + \Delta T(\bx)$, used to scale the HH gating kinetics via the Q\textsubscript{10} factor and, when enabled, the TRP thermosensor activations.

\subsection{Strain tensor and membrane tension}
\label{sec:methods_strain}

The infinitesimal strain tensor is computed from the displacement field by central differences,
\begin{equation}
  \varepsilon_{ij} = \tfrac{1}{2}(\partial_j u_i + \partial_i u_j),
  \label{eq:strain_tensor}
\end{equation}
and stored in Voigt notation $\beps = [\varepsilon_{xx}, \varepsilon_{yy}, \varepsilon_{zz}, \gamma_{yz}, \gamma_{xz}, \gamma_{xy}]$ with engineering shear strains $\gamma_{ij} = 2\varepsilon_{ij}$. Of the three principal invariants of $\beps$, the membrane-tension input is driven by the von~Mises equivalent strain
\begin{equation}
  \varepsilon_{\mathrm{eq}} = \sqrt{\tfrac{2}{3}\bigl[e_{xx}^2 + e_{yy}^2 + e_{zz}^2 + \tfrac{1}{2}(\gamma_{yz}^2 + \gamma_{xz}^2 + \gamma_{xy}^2)\bigr]},
  \label{eq:strain_vm}
\end{equation}
with $e_{ii} = \varepsilon_{ii} - \tfrac{1}{3}(\varepsilon_{xx}+\varepsilon_{yy}+\varepsilon_{zz})$ the deviatoric diagonal components.

The lipid bilayer is modelled as a sheet embedded in the tissue continuum. Effective tension arises from the bilayer area dilation driven by the local equivalent strain~\citep{rawicz2000effect} and follows
\begin{equation}
  T = K_A \cdot \alpha \cdot \varepsilon_{\mathrm{eq}}\;[\mathrm{mN/m}],
  \label{eq:tension}
\end{equation}
with $K_A$ the bilayer area-expansion modulus (0.1--0.5~N/m for typical lipid membranes) and $\alpha$ a dimensionless multi-scale coupling factor relating far-field tissue strain to local membrane area change~\citep{sukharev2012mechanosensitive}. Both $K_A$ and $\alpha$ are recorded as modelling assumptions in the parameter source classification (\S\ref{sec:methods_provenance}). Calibration to a specific tissue, cell type, or exposure is the user's responsibility.

\subsection{Mechanosensitive ion channels}
\label{sec:methods_channels}

Each channel is a two-state Boltzmann-gated population~\citep{howard1988compliance} with steady-state open probability
\begin{equation}
  P_o(T) = \bigl(1 + \exp[-\dA\,(T - \Thalf)/\kBT]\bigr)^{-1},
  \label{eq:boltzmann}
\end{equation}
where $\Thalf$ is the half-activation tension, $\dA$ is the effective gate area, and $\kBT = 4.114$~pN$\cdot$nm. $k_BT$ is held at its 25\,$^\circ$C value throughout because the published $\Delta A$ and $T_{1/2}$ values from patch-clamp recordings are themselves calibrated at room temperature. Rescaling $k_BT$ to 37\,$^\circ$C without re-fitting those parameters would shift the Boltzmann curve by an unjustified $\sim$4\%. The macroscopic mechanosensitive current is the sum over all co-expressed mechano channels,
\begin{equation}
  I_{\mathrm{mech}} = \sum_k \bar{g}_k\,w_k\,i_k\,(\Vm - E_{\mathrm{rev},k}),
\end{equation}
where each dynamic activation gate $w_k$ obeys $dw_k/dt = (P_{o,k} - w_k)/\tau_{\mathrm{act},k}$ with a fixed per-channel activation time constant. An optional second Boltzmann gate $i_k(t)$ models inactivation under sustained tension and is held at $i_k \equiv 1$ for non-inactivating channels. For Piezo1 this gate closes within $\tau_{\mathrm{inact}} \approx 12$~ms~\citep{coste2010piezo}, so a sustained tension produces a brief depolarising transient rather than a maintained inward current (full equation in supplementary~\S~\ref{sec:supp_channel_extensions}). The library covers four channels. The cation channel Piezo1~\citep{coste2010piezo,lewis2015mechanical} is the primary depolarising mechanosensor, and the K\textsuperscript{+}-selective two-pore channels TRAAK~\citep{brohawn2014mechanosensitivity}, TREK-1, and TREK-2~\citep{honore2007neuronal} are the principal hyperpolarisers. Six neuron-type-specific channel profiles cover CNS targets relevant to tFUS. Peripheral sensory neurons rely on Piezo2 (not in the library) and are out of scope. Full per-channel parameter values, neuron profile compositions, and the Piezo1 causal-role evidence summary are tabulated in supplementary~\S~\ref{sec:supp_channel_library}.

\subsection{Hodgkin--Huxley neuron model}
\label{sec:methods_hh}

The membrane potential follows the standard HH equation~\citep{hodgkin1952quantitative} augmented with the mechanosensitive current,
\begin{equation}
  C_m\,\frac{d\Vm}{dt} = -I_{\mathrm{Na}} - I_K - I_L - I_{\mathrm{mech}} + I_{\mathrm{stim}}.
  \label{eq:hh}
\end{equation}
$I_{\mathrm{Na}} = \bar{g}_{\mathrm{Na}}\,m^3\,h\,(\Vm - E_{\mathrm{Na}})$, $I_K = \bar{g}_K\,n^4\,(\Vm - E_K)$, and $I_L = \bar{g}_L\,(\Vm - E_L)$. The gating variables follow the standard rate equations $dx/dt = \phi[\alpha_x(1-x) - \beta_x x]$ scaled by the Q\textsubscript{10} factor $\phi = Q_{10}^{(T - T_{\mathrm{ref}})/10}$. Two parameterisations are available. The first is the original squid model~\citep{hodgkin1952quantitative} ($Q_{10} = 3$, $T_{\mathrm{ref}} = 6.3\,^\circ$C), and the second is the regular-spiking cortical pyramidal cell of Pospischil et~al.~\citep{pospischil2008minimal} ($V_T = -56.2$~mV, $Q_{10} = 2.3$, $T_{\mathrm{ref}} = 36\,^\circ$C). The Pospischil parameterisation is the widely adopted minimal-HH description of mammalian cortical pyramidal cells and is also used by the SONIC reduction~\citep{lemaire2019understanding} of intramembrane cavitation, making it a reference for body-temperature single-compartment cortical-cell modelling. All transcranial simulations in this paper therefore use the Pospischil cortical parameterisation, for which the body-temperature scaling factor remains physiological ($\phi = 1.087$). The squid model is retained for validation only. It serves as a positive control in the Q\textsubscript{10}-temperature-scaling check (supplementary~\S\ref{sec:supp_verif_q10}), where the framework reproduces the textbook result that, scaled to body temperature ($\phi \approx 29$), the original squid kinetics enter depolarisation block, confirming the fidelity of the Q\textsubscript{10} implementation.

\subsection{Three-compartment dendrite--soma--AIS extension}
\label{sec:methods_multicomp}

A single iso-potential compartment cannot resolve two features that matter for tFUS predictions. One is the dendrite-heavy distribution of Piezo1~\citep{lewis2015mechanical}, and the other is AIS-first action-potential initiation, which depends on a 10--40$\times$ higher Na density at the AIS than the soma~\citep{kole2008action,hu2009distinct,stuart1995initiation}. We therefore couple a dendrite, soma, and AIS compartment by axial coupling conductances area-normalised to the soma in the Pinsky--Rinzel manner~\citep{pinsky1994intrinsic} (full equations in supplementary~\S~\ref{sec:supp_multicomp}). The literature-parameterised cortical-pyramidal preset uses $\bar{g}_{\mathrm{Na}} = 30, 120, 2000$~mS/cm$^2$ for dendrite, soma, AIS~\citep{stuart1995initiation,mainen1996influence,kole2008action}, soma-normalised coupling $g_{ds} = 1$~mS/cm$^2$ and $g_{sa} = 50$~mS/cm$^2$, area ratio $A_d : A_s : A_a = 5 : 1 : 0.05$, and dendrite/soma/AIS Piezo1 density scaling $5\times / 1\times$ / no Piezo1 on the AIS. The AIS Na density sits at the lower end of the cited 10--40$\times$ ratio (modelled ratio $\approx 17\times$) because explicit RK4 integration becomes the stability bottleneck above $\bar{g}_{\mathrm{Na}} = 2000$~mS/cm$^2$ at the AIS membrane time constant (supplementary~\S~\ref{sec:supp_multicomp}). Mechanosensitive channel kinetics are shared across compartments but conductance densities scale per-compartment. Setting all three compartments identical with zero coupling recovers the single-compartment trace bit-for-bit (regression check).

\subsection{Intramembrane cavitation coupling}
\label{sec:methods_nice}

The NICE model of Plaksin \& Krasovitski~\citep{plaksin2014intramembrane} represents the bilayer as two leaflets that can separate under the carrier pressure $P_{\mathrm{ac}}(t)$, producing a cycle-averaged change in membrane capacitance $\langle \Delta C_m(P_{\mathrm{env}}) \rangle$. This is mechanistically distinct from the strain-driven Boltzmann-channel pathway and gives rise to the duty-cycle excitation/inhibition flip predicted in~\citep[Fig.~6]{plaksin2014intramembrane} and attributed to differences between excitatory and inhibitory cortical-cell channel complements~\citep{plaksin2016celltype}. Following the SONIC reduction of Lemaire et~al.~\citep{lemaire2019understanding}, we resolve the carrier-frequency leaflet dynamics with a damped oscillator augmented with Lennard-Jones-like and surface-tension non-linear walls (full ODE in supplementary~\S~\ref{sec:supp_nice}). The cycle-averaged capacitance is integrated at 256 sub-steps per cycle over 8 cycles and averaged, then enters the per-compartment HH equation by replacing $C_m$ with the slow envelope-modulated effective capacitance $C_m + \langle \Delta C_m \rangle(P_{\mathrm{env}}(t))$, together with a displacement-current term $V \cdot d\langle \Delta C_m \rangle/dt$. At 100~kPa the default parameters give $\langle \Delta C_m \rangle / C_{m,0} \approx 0.05$, qualitatively matching SONIC. At 1~MPa it rises to $\approx 0.55$, qualitatively matching Fig.~2 of~\citep{plaksin2014intramembrane}. The displacement-current term is non-zero only at envelope ramps and carries the duty-cycle excitation/inhibition signal.

\subsection{Calcium pool and SK counter-current}
\label{sec:methods_calcium}

Piezo1 inactivates within $\sim$10--15~ms in real recordings~\citep{coste2010piezo} and admits a Ca\textsuperscript{2+} flux ($P_{\mathrm{Ca}}/P_{\mathrm{Na}} \approx 5$) that charges an intracellular Ca\textsuperscript{2+} pool gating small-conductance Ca\textsuperscript{2+}-activated K\textsuperscript{+} (SK) channels, providing an inhibitory pathway distinct from the K2P channels and a candidate route for high-pressure inhibition outcomes. Each compartment maintains an intracellular Ca\textsuperscript{2+} concentration with single-pool decay ($\tau_{\mathrm{Ca}} = 50$~ms~\citep{mainen1996influence,borggraham1999interpretations}, $[\mathrm{Ca}^{2+}]_{\mathrm{rest}} = 0.05~\mu$M) charged by the Ca\textsuperscript{2+} fraction of the mechanosensitive current. An SK-type Ca\textsuperscript{2+}-activated K\textsuperscript{+} channel population on the soma membrane delivers a Hill-gated counter-current ($[\mathrm{Ca}^{2+}]_{1/2} = 0.5~\mu$M~\citep{kohler1996sk}, Hill coefficient $n_H = 4$, $\bar{g}_{\mathrm{KCa}} = 0.3$~mS/cm$^2$). Full equations and per-parameter source classification entries appear in supplementary~\S~\ref{sec:supp_calcium}.

\subsection{TRPV1 / TRPV4 thermosensors}
\label{sec:methods_trp}

For protocols with focal heating exceeding a few Kelvin~\citep{lee2016transcranial,verhagen2019offline}, an optional TRP pathway adds two thermosensitive channels~\citep{caterina1997capsaicin} with Boltzmann temperature-gating
\begin{equation}
  P_o(T) = \bigl(1 + \exp[(T_{1/2} - T)/k_T]\bigr)^{-1},
  \label{eq:trp_boltzmann}
\end{equation}
and macroscopic current $I_{\mathrm{TRP},j} = \bar{g}_j\,P_{o,j}(T)\,(V_m - E_{\mathrm{rev},j})$ entering the HH equation as $-I_{\mathrm{TRP},j}$. Default parameter values are TRPV1 ($T_{1/2} = 43\,^\circ$C, $k_T = 1.5\,^\circ$C, $\bar{g} = 0.05$~mS/cm$^2$, $E_{\mathrm{rev}} = 10$~mV) and TRPV4 ($T_{1/2} = 30\,^\circ$C, $k_T = 3\,^\circ$C, $\bar{g} = 0.02$~mS/cm$^2$, $E_{\mathrm{rev}} = 0$~mV). At 37\,$^\circ$C TRPV1 is essentially closed ($P_o = 0.018$) and TRPV4 acts as a tonic depolariser ($P_o = 0.91$). At 45\,$^\circ$C TRPV1 reaches $P_o = 0.79$, contributing a substantial inward current. The Boltzmann steepness $k_T$ for TRPV1 is calibrated to the Caterina Q\textsubscript{10} $\geq 20$ floor~\citep{caterina1997capsaicin}. TRPV4 is polymodal (temperature, osmotic, and reported mechanosensitive activation), but only the temperature-gated branch is modelled here. The mechanical drive into the neuron is already routed through Piezo1 and TREK-1 via the calibrated membrane-tension input ($T = K_A\,\alpha\,\varepsilon_{\mathrm{eq}}$), and adding a parallel TRPV4 mechanosensitive branch with its own gating constants would make the mechanical-pathway parameters jointly unidentifiable from the experimental observations the framework targets.

\subsection{Astrocytic TRPA1 $\to$ glutamate $\to$ NMDA relay}
\label{sec:methods_astrocyte}

The leading non-Piezo account of ultrasonic neuromodulation routes the mechanical stimulus through astrocytes rather than directly through neuronal mechanosensors~\citep{oh2019ultrasonic}. Ultrasound gates mechanosensitive TRPA1 in astrocytes, the Ca\textsuperscript{2+}-permeable TRPA1 current charges an astrocytic Ca\textsuperscript{2+} pool, the rising Ca\textsuperscript{2+} triggers glutamate release, and glutamate activates neuronal NMDA receptors whose depolarising current excites the neuron. TRPA1 is the necessary sensor in this account, since knockout or antagonists (HC-030031, A-967079) abolish the ultrasound response, and the relay is slow, operating on the hundreds-of-ms-to-seconds timescale of astrocytic Ca\textsuperscript{2+} signalling rather than per carrier cycle. We model it as an optional single-pool relay on the same shared neuron. TRPA1 reuses the Boltzmann-in-tension gate of Eq.~\ref{eq:tension} with a Ca\textsuperscript{2+}-permeability fraction $f_{\mathrm{Ca}} = 0.18$ ($P_{\mathrm{Ca}}/P_{\mathrm{Na}} \approx 5$--$8$). Its astrocytic Ca\textsuperscript{2+} pool follows
\begin{equation}
  \frac{d[\mathrm{Ca}^{2+}]}{dt} = k_{\mathrm{flux}}\,\bar{g}_{\mathrm{TRPA1}}\,P_o(T)\,f_{\mathrm{Ca}}\,(E_{\mathrm{rev}} - V_{\mathrm{astro}}) - \frac{[\mathrm{Ca}^{2+}] - [\mathrm{Ca}^{2+}]_{\mathrm{rest}}}{\tau_{\mathrm{Ca}}},
  \label{eq:astro_ca}
\end{equation}
with a slow astrocytic decay $\tau_{\mathrm{Ca}} = 300$~ms and a non-spiking driving potential $V_{\mathrm{astro}} = -80$~mV. Here $k_{\mathrm{flux}}$ is calibrated so the pool peaks at $\sim 1~\mu$M at full ultrasound drive. Glutamate release is a Hill function of astrocytic Ca\textsuperscript{2+} ($[\mathrm{Ca}^{2+}]_{1/2} = 0.3~\mu$M, $n = 3$), and the NMDA conductance combines glutamate saturation with the Jahr--Stevens voltage-dependent Mg\textsuperscript{2+} block~\citep{jahr1990voltage},
\begin{equation}
  g_{\mathrm{NMDA}}(V, [\mathrm{glu}]) = \bar{g}_{\mathrm{NMDA}}\,\frac{[\mathrm{glu}]}{[\mathrm{glu}] + K_{\mathrm{glu}}}\,\frac{1}{1 + [\mathrm{Mg}^{2+}]\,e^{-V/16.13}/3.57}.
  \label{eq:nmda}
\end{equation}
Blocking the receptor, the AP5 condition of Oh et al., is recovered by setting $\bar{g}_{\mathrm{NMDA}} = 0$, so that the astrocytic Ca\textsuperscript{2+} and glutamate transients persist but no neuronal excitation results. Because the relay integrates the duty-gated drive over the slow $\tau_{\mathrm{Ca}}$, its excitation grows with cumulative on-time, in contrast to the per-pulse channel mechanisms. For inexpensive mechanism scoring the relay also exposes a closed-form scalar drive as a function of peak tension, duty cycle, and on-time. The TRPA1 mechanogating half-tension and gate area are flagged \emph{Estimate}, since no direct tension-clamp fit exists (\S\ref{sec:methods_provenance}). The Ca\textsuperscript{2+}, glutamate, and NMDA constants are literature-anchored or calibrated, as classified in the parameter source table (supplementary~\S\ref{sec:supp_provenance}) and detailed in supplementary~\S\ref{sec:supp_astrocyte}.

\subsection{Mechanosensitive synaptic and voltage-gated transmission}
\label{sec:methods_synaptic}

The sixth candidate acts on synaptic transmission and the spike-generating conductances themselves rather than on a single post-synaptic membrane, bundling three literature-grounded effects on the shared neuron. (i)~Presynaptic vesicle release is modulated by presynaptic membrane tension~\citep{tyler2008remote,tyler2012mechanobiology}. The release probability follows a saturating sigmoid normalised so $P_r(0) = P_{r,0}$ and $P_r(\infty) = P_{r,\max}$, scaling the cleft glutamate transient and hence synaptic gain. (ii)~Post-synaptic NMDA-receptor gating is stretch-enhanced~\citep{maneshi2017mechanical}, so the NMDA conductance of Eq.~\ref{eq:nmda} is multiplied by a tension factor
\begin{equation}
  f_{\mathrm{stretch}}(T) = 1 + \alpha_{\mathrm{NMDA}}\,\frac{\sigma\!\left(\tfrac{T - T_{1/2}^{\mathrm{s}}}{k_{\mathrm{s}}}\right) - \sigma\!\left(\tfrac{-T_{1/2}^{\mathrm{s}}}{k_{\mathrm{s}}}\right)}{1 - \sigma\!\left(\tfrac{-T_{1/2}^{\mathrm{s}}}{k_{\mathrm{s}}}\right)},
  \label{eq:nmda_stretch}
\end{equation}
with $\sigma$ the logistic function, normalised so $f_{\mathrm{stretch}}(0) = 1$ exactly and $f_{\mathrm{stretch}}(\infty) = 1 + \alpha_{\mathrm{NMDA}}$. Setting $\alpha_{\mathrm{NMDA}} = 1$ reproduces the roughly doubled NMDA current Maneshi et al.\ report at saturating stretch. (iii)~The activation of the voltage-gated Na\textsuperscript{+}/Ca\textsuperscript{2+} channels that carry the action potential is shifted by membrane tension. Kubanek et al.~\citep{kubanek2016ultrasound} patch-clamped these currents directly under ultrasound, and the activation half-voltage moves as
\begin{equation}
  V_{1/2}(T) = V_{1/2}^{0} - k_{\mathrm{mechano}}\,T,
  \label{eq:nav_shift}
\end{equation}
a hyperpolarising shift that increases the Na\textsuperscript{+}/Ca\textsuperscript{2+} current at fixed voltage. The constant $k_{\mathrm{mechano}}$ is \emph{Calibrated} so a few-mN/m tension produces a $\sim$mV-scale shift and a tens-of-percent current change, and $k_{\mathrm{mechano}} = 0$ recovers the unmodified neuron exactly. The released glutamate drives an AMPA conductance and the stretch-enhanced NMDA receptor on a post-synaptic Hodgkin--Huxley neuron, so the pathway produces fast, per-pulse excitation in contrast to the slow astrocytic relay. As with the astrocytic pathway, each of the three effects exposes a closed-form per-case scalar gain, namely $P_r(T)/P_r(0)$, $f_{\mathrm{stretch}}(T)$, and the Na\textsuperscript{+} activation-shift current ratio, for inexpensive mechanism scoring. The presynaptic-release, stretch, and shift constants are flagged \emph{Estimate} or \emph{Calibrated}, since no direct in-tissue tension-clamp fits exist (\S\ref{sec:methods_provenance}), and are tabulated in supplementary~\S\ref{sec:supp_synaptic}. Stretch-enhanced long-term potentiation, the synaptic-plasticity consequence that motivates NMDA mechanosensitivity, is a documented downstream effect that we do not model. The framework reports the acute per-pulse excitation only.

\subsection{Default mechanism scenario and when to enable the full pathway}
\label{sec:methods_scenario_default}

The framework enumerates five mechanism scenarios (S1--S5) that include progressively more pathway components, with S1 the single-compartment baseline and S2--S5 layered on a shared dendrite--soma--AIS neuron model. S2 (3-compartment Hodgkin--Huxley with Piezo1 and TREK-1, but neither intramembrane cavitation nor Piezo1 inactivation / Ca\textsuperscript{2+} pool / SK / TRP) is the default for the canonical Yaakub-style transcranial demonstration. At the body-temperature, sub-MPa carrier, 20~ms tone-burst regime studied here the additional mechanism modules contribute quantitative corrections of order 5--35\% to the focal firing zone, but do not change the qualitative outcome (the focal voxel sits in the upper Boltzmann knee at $T \approx 2.42\times \Thalf$ in either S2 or the full S5, so the firing-zone topology is set by the strain field rather than by which mechanism modules are enabled). The source-classified sensitivity analysis in \S~\ref{sec:results_sensitivity} confirms that the dominant uncertainties (the Piezo1 half-activation tension $\Thalf$ and the strain-to-tension product $K_A \cdot \alpha$) are present in S2 and unaffected by S3--S5's added parameters. We therefore retain S2 as the default and treat S3--S5 as optional extensions, selectable through the framework's scenario configuration, for protocols where the omitted pathways become quantitatively dominant. S3 (NICE intramembrane cavitation) applies above ${\sim}1$~MPa focal carrier, where the cycle-averaged $\langle \Delta C_m \rangle / C_{m,0}$ becomes substantial ($\gtrsim 50\%$). S4 (Piezo1 inactivation with Ca\textsuperscript{2+} pool and SK counter-current) applies to protocols with ON windows ${\gtrsim} 100$~ms or as the modelled mechanism for Legon-style high-pressure inhibition. S5 (full pathway with TRPV1 and TRPV4) applies to protocols with focal heating of several Kelvin, where TRPV1 begins to open noticeably ($P_o$ rises from 0.018 at 37\,$^\circ$C to 0.12 at 40\,$^\circ$C and 0.79 at 45\,$^\circ$C). The mechanism-comparison sweep in Figure~\ref{fig:tips_mechanisms} exercises all five scenarios on the same field to make these regime-dependent contributions quantitatively legible.

\subsection{Numerical integration}
\label{sec:methods_numerics}

All neural-pipeline ODE systems are integrated with explicit fourth-order Runge--Kutta (RK4). The upstream PDE solvers (acoustic, shear-FDTD, Pennes) use their own time-stepping schemes described in \S\ref{sec:methods_upstream}. Gating variables are clamped to $[0, 1]$ after each step. The single-compartment HH default is $\Delta t = 10$\,$\mu$s. The three-compartment system is significantly stiffer because the AIS Na conductance ($\bar{g}_{\mathrm{Na,AIS}} = 2000$~mS/cm$^2$) and the area-amplified soma--AIS coupling ($g_{sa}\,A_s/A_a = 1000$~mS/cm$^2$) drive the local membrane time constant during a spike to $\tau_V = C_m/g_{\mathrm{total}} \approx 0.3~\mu$s, while the slow gating, Ca\textsuperscript{2+}-pool, and NICE-envelope dynamics evolve on millisecond timescales, giving a stiffness ratio of $\sim 10^4$. The explicit-RK4 stability bound $|\lambda \Delta t| \lesssim 2.78$ then requires $\Delta t \le 1$~$\mu$s, and the multi-compartment integrator default is $\Delta t = 1$\,$\mu$s. Implicit or operator-split integrators would relax this constraint at the cost of solver complexity. The explicit RK4 path is adequate for the per-voxel firing-map sweeps the framework targets.

\subsection{Computational performance and implementation notes}
\label{sec:methods_performance}

The framework was designed for whole-brain transcranial cases at clinically relevant frequencies on a single workstation, with end-to-end run times that make per-voxel firing-map sweeps tractable as part of an iterative protocol-design loop rather than as offline batch jobs. All timings below are for the canonical Yaakub demonstration on the Halle micro-CT skull (Fullwave~2 volume $422 \times 260 \times 260 \approx 28.5$~M voxels at $dx = 0.308$~mm, 4\,221 acoustic time-steps, 6\,018 shear-FDTD time-steps, 5\,468 voxels firing over the 20~ms ON window, Table~\ref{tab:yaakub_summary}) on a workstation with 72 CPU cores and one NVIDIA RTX A6000 48~GB GPU.

The Fullwave~2 solver runs on the GPU, integrating the heterogeneous nonlinear wave equation with an explicit eighth-order finite-difference scheme. Saved pressure snapshots are decimated on the fly to bring the post-run analysis grid to $\sim 100^3$ voxels, which is what feeds the downstream stages. The Pennes bioheat solver runs against the same NumPy API on either CPU or GPU and is normally executed on the GPU alongside the acoustic solver. The Kelvin--Voigt shear-FDTD solver~\citep{pinton2025shearfdtd} runs on the CPU and applies a one-time FFT-based Helmholtz projection of the radiation-force input, reducing the per-step cost from $O(N \log N)$ to $O(N)$. The per-voxel Hodgkin--Huxley pipeline is a multi-threaded JIT-compiled kernel parallelised over voxels, with all per-channel state held in registers and pre-allocated spike-time buffers so the kernel is allocation-free. Recording spike times inside the same kernel eliminates the second pass that earlier implementations needed to extract them. Compared with a pure-Python reference, the JIT kernel is $\approx 50\times$ faster at bit-for-bit equivalence, verified across all manuscript-relevant feature combinations (supplementary~\S\ref{sec:supp_verification}).

Per-stage wall-clock and memory for the canonical run are summarised in Table~\ref{tab:perf}, totalling approximately 17~min end-to-end for a single transcranial scenario. Iterating mechanism scenarios on a cached upstream field re-uses the pressure, displacement, and temperature outputs and runs the neural pipeline alone in $\sim 3$~min per scenario.

\begin{table}
\caption{Per-stage wall-clock and memory for the canonical Yaakub run (Halle micro-CT skull, auto-pick aim at left dACC, 72 CPU cores plus one NVIDIA RTX A6000 GPU). The Memory column is the persistent working set per stage. The total peak RAM is dominated by the full-resolution slab pressure history (decimated on the fly during the Fullwave~2 loop) plus transient buffers and Python runtime overhead.}
\label{tab:perf}
\begin{tabular}{lllr}
\br
Stage & Hardware & Memory & Wall-time \\
\mr
Fullwave~2 (acoustic) & GPU & 2 GB & 8 min \\
Radiation force and Pennes bioheat & GPU (CuPy-mirrored, CPU fallback) & 0.1 GB & 1 min \\
Shear-FDTD (Kelvin--Voigt) & CPU (multi-threaded) & 0.5 GB & 5 min \\
Hodgkin--Huxley neurons & CPU (JIT, 72 cores) & 0.1 GB & 3 min \\
\mr
Total (peak RAM $\sim 12$~GB) & & & $\sim 17$ min \\
\br
\end{tabular}
\end{table}

The framework is implemented in Python with NumPy, SciPy, and a JIT-compiled inner loop for the neural pipeline. Upstream FDTD solvers use CUDA where available. All code, data, and the parameter source classification are openly available (see Data Availability).

% =============================================================================
\section{Results}
\label{sec:results}

\subsection{Framework validations}
\label{sec:results_validation}

Each pathway component is anchored to a published experimental or theoretical observation. Quantitative details and the underlying numerical experiments live in the supplementary verification log (\S~\ref{sec:supp_verification}). In aggregate, the single-compartment HH baseline (Pospischil cortical regular-spiking at 37\,$^\circ$C) supports repetitive firing with the standard $m,h,n$ gating waveform and converges as $O(\Delta t^4)$. The multi-compartment integrator reproduces AIS-first initiation~\citep{stuart1997action} with a $\sim$9~$\mu$s lead at the default densities and recovers the single-compartment trace bit-for-bit when compartments are made identical with zero coupling. The intramembrane-cavitation pathway reproduces the duty-cycle excitation/inhibition direction across four (pressure, stimulus) regimes~\citep{plaksin2014intramembrane}. The Piezo1 inactivation gate matches the patch-clamp timecourse~\citep{coste2010piezo} with the $1-1/e$ crossing in [6, 18]~ms (nominal 12~ms). The intracellular Ca\textsuperscript{2+} pool follows its analytical exponential decay to relative tolerance $10^{-4}$. The SK channel reduces the spike count under sustained near-threshold drive, the inhibitory direction predicted for the Piezo1-Ca\textsuperscript{2+}-SK route without invoking the K2P channels. And TRPV1 is essentially closed at body temperature ($P_o = 0.018$) while reaching $P_o = 0.79$ at 45\,$^\circ$C with $P_o(47\,^\circ\mathrm{C})/P_o(37\,^\circ\mathrm{C}) \approx 50$, comfortably above the Q\textsubscript{10} $\geq 20$ floor of Caterina~\citep{caterina1997capsaicin}. The astrocytic TRPA1 relay is anchored to the three mechanism controls of Oh et al.~\citep{oh2019ultrasonic} (supplementary~\S\ref{sec:supp_verif_astrocyte}). Abolishing TRPA1 collapses the relay output (necessary-sensor / knockout control), the AP5 condition abolishes neuronal excitation while the astrocytic Ca\textsuperscript{2+} and glutamate transients persist, and the astrocytic Ca\textsuperscript{2+} rises with a $\sim$200~ms half-time, five orders of magnitude slower than the carrier cycle, confirming the slow integrating character of the pathway. Finally, the three mechanosensitive synaptic / voltage-gated gains (supplementary~\S\ref{sec:supp_verif_synaptic}) each reduce to the unmodified neuron at zero tension and rise monotonically thereafter. Presynaptic release climbs from its resting probability toward saturation~\citep{tyler2008remote}, the NMDA stretch factor saturates to a current doubling~\citep{maneshi2017mechanical}, and the voltage-gated Na\textsuperscript{+} current gain increases with the Kubanek activation shift~\citep{kubanek2016ultrasound}, with the no-ultrasound no-op (every gain exactly unity or baseline at zero tension) enforced as a regression control.

\subsection{Yaakub canonical demonstration on the Halle micro-CT skull}
\label{sec:results_tips}

We applied the framework to a transcranial focused-ultrasound configuration in two complementary modes. The first is a theta-burst protocol~\citep{yaakub2023transcranial} sonicated through the Halle micro-CT human-skull specimen with the bowl aimed at the left dorsal anterior cingulate cortex (dACC), one of the two deep cortical targets sonicated by Yaakub et al. (the canonical pipeline output, exercising the upstream Fullwave, shear-FDTD, and Pennes solvers and the downstream multi-pathway Hodgkin--Huxley neuron at every above-threshold brain voxel). The second is a real-field mechanism-comparison sweep that scales the same Halle\,/\allowbreak\,dACC $\varepsilon_{\mathrm{eq}}$ field across the threshold region to isolate the mechanism-comparison axis. The two together establish that the framework operates on upstream-solver output and discriminates between mechanism candidates on the same neuron model.

\begin{figure}[htbp]
  \centering
  \includegraphics[width=0.95\textwidth]{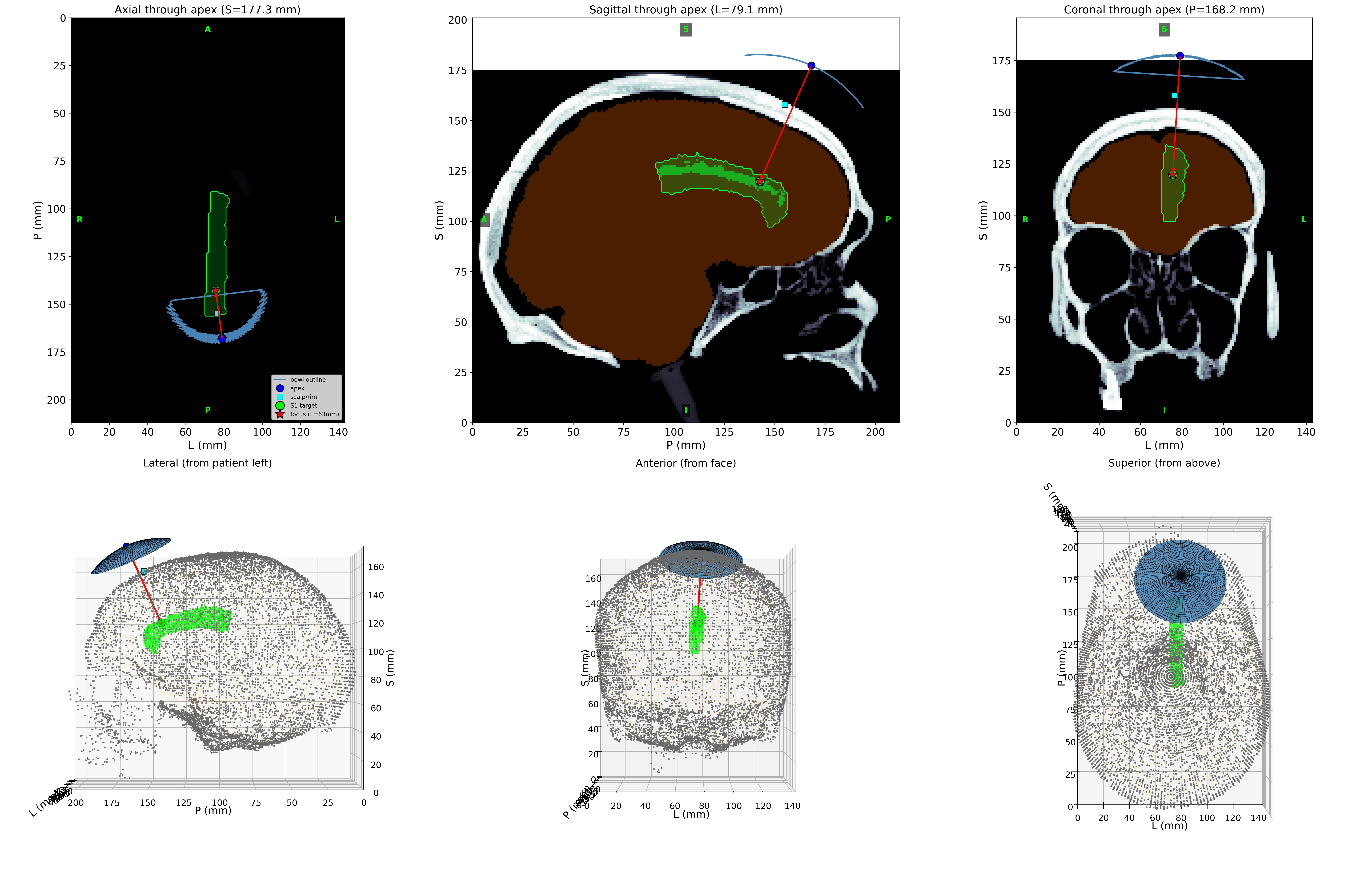}
  \caption{Auto-pick placement aimed at the left dACC on the Halle micro-CT skull. \textbf{Top row.} Orthogonal slice views with the bowl outline drawn in plane. \textbf{Bottom row.} Three 3-D camera angles (lateral, anterior, superior) with the bowl wireframe (steelblue), outer-bone surface (grey point cloud), pial brain mesh (translucent tan), target ROI (lime, Harvard--Oxford anterior-cingulate label warped to subject space via the SyN MNI-to-Halle map), beam axis (red), and apex / scalp / target / focus markers.}
  \label{fig:placement_3d}
\end{figure}

The Halle micro-CT temporal-bone NRRD (Hounsfield-thresholded at 700--1973 to separate cortical bone from soft tissue, with sound speed 1540--2900~m/s, density 1000--2200~kg/m$^3$, power-law attenuation 0.5--20~dB/(MHz$\cdot$cm), and nonlinearity coefficient $\beta = 4.8$--5.5 mapped voxelwise) is resampled onto the Fullwave~2 grid at $dx = 0.308$~mm (PPW $= 10$ at 1540~m/s, 500~kHz), giving a domain of $130 \times 80 \times 80$~mm = $422 \times 260 \times 260$ voxels. CFL is held at 0.20 and the time step is $dt = 40$~ns over $4{,}221$ steps (168.8~$\mu$s simulated, round-trip at 1540~m/s), satisfying the stability bound at the highest skull speed. The CTX-500 bowl ($R_{\mathrm{outer}} = 32$~mm, $R_{\mathrm{inner}} = 0$, $\mathrm{ROC} = F = 63$~mm, four annular elements distributed across three voxel-thickness layers, $73{,}760$ source voxels in total) is voxelised on the slab grid by sampling the spherical-shell parametric surface. The bowl is positioned and oriented from the anatomical placement (Fig.~\ref{fig:placement_3d}, perpendicular skin-contact residual 0.65~mm, scalp-to-target distance 40.2~mm, apex-to-target distance 63.0~mm, sagitta 8.7~mm). The per-voxel excitation pulse drives every source voxel with the same waveform, a 2-cycle 500~kHz burst, drop-off 2 (192-tic envelope, 7.68~$\mu$s active duration), continuous-wave correction factor 2.609 to lift the Gaussian-burst peak to a sinusoidal-RMS reference, calibrated to a safety-envelope-compliant source amplitude $p_0 = 0.10$~MPa.

\begin{figure}[htbp]
  \centering
  \includegraphics[width=\textwidth]{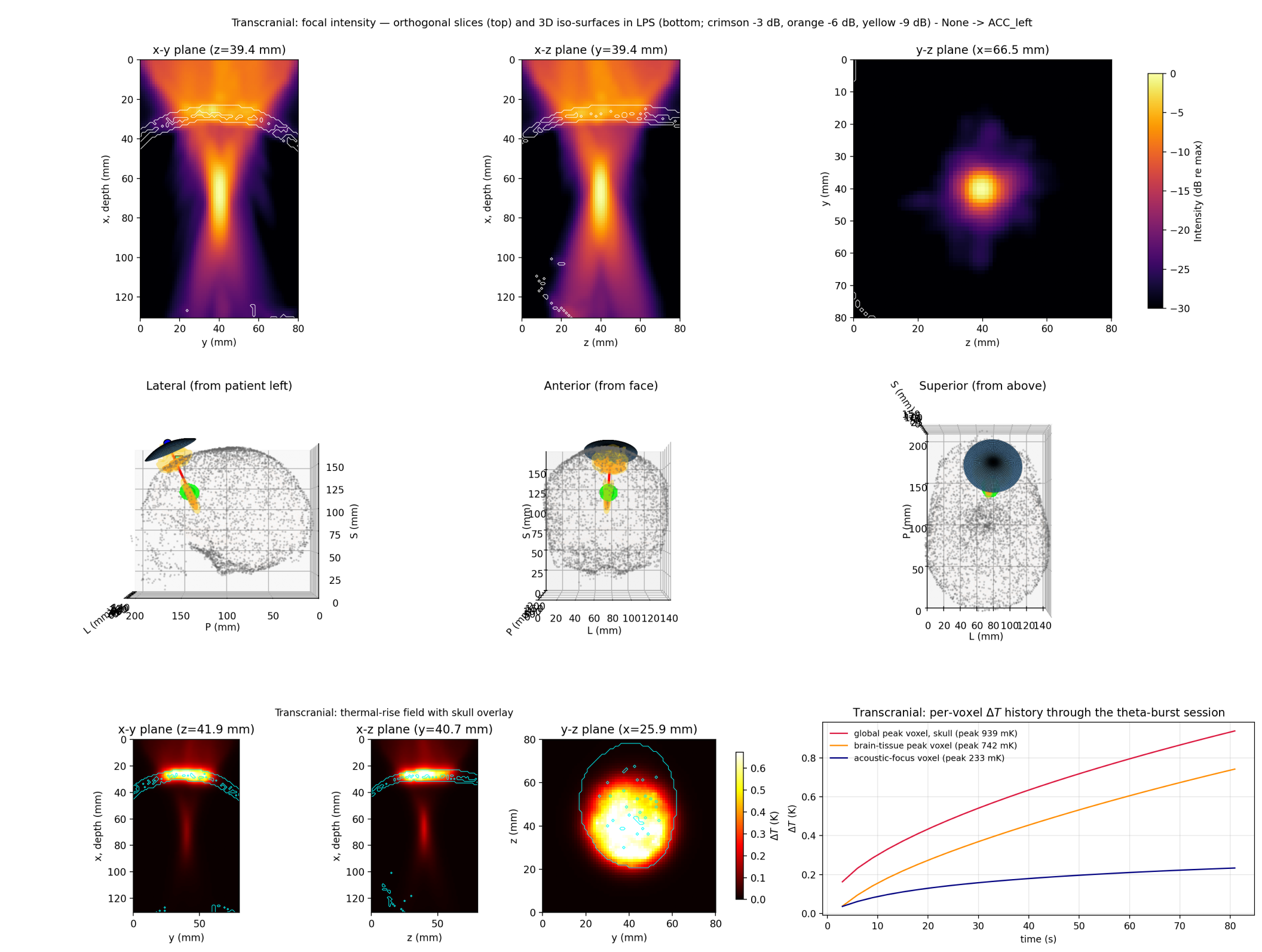}
  \caption{Fullwave~2 (acoustic) output and the parallel Pennes thermal branch driven by it. \textbf{Top.} Three orthogonal slices of focal intensity in dB re max (skull mask contoured in white). \textbf{Bottom.} Three orthogonal $\Delta T$ slices through the global peak-temperature voxel at the end of the 80\,s session (left, skull contoured in cyan), and per-voxel $\Delta T$ traces over the session at the global peak voxel (crimson, in skull, 0.94\,K), the brain-tissue peak voxel (orange, 0.74\,K, a $\sim$1\,mm-thick conduction layer abutting the hot skull, where the direct acoustic intensity is $\sim$5$\times$ lower than at the focus, rather than focal heating), and the acoustic-focus voxel (navy, 233\,mK). The skull-bone peak (0.94\,K) sits below the ITRUSST 2\,K thermal bound. The target focal-tissue rise (233\,mK), which is the safety-relevant brain temperature, stays within Yaakub's reported $\leq 0.5$\,K, and the cumulative thermal dose (CEM43 $= 8.1{\times}10^{-4}$\,min) remains far below any damage threshold.}
  \label{fig:yaakub_intensity}
\end{figure}

\begin{figure}[htbp]
  \centering
  \includegraphics[width=\textwidth]{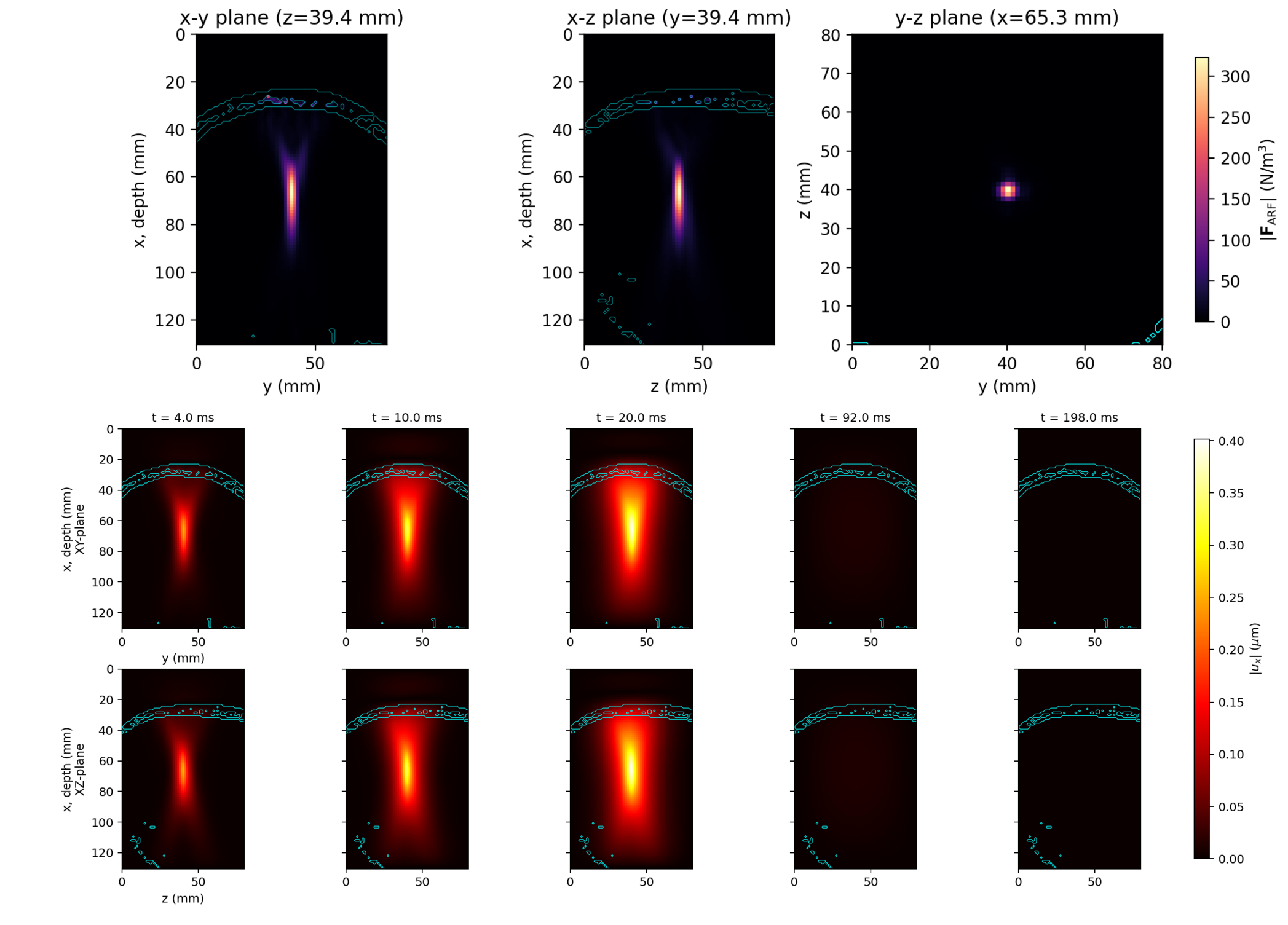}
  \caption{Mechanical pipeline outputs. \textbf{Top.} Acoustic radiation-force body-force density $|\mathbf{F}_{\mathrm{ARF}}|$ on three orthogonal slices through the brain-restricted peak voxel, the volumetric body force $\mathbf{F} = 2\alpha\mathbf{I}/c$ that drives the shear-FDTD solver. \textbf{Bottom.} Resulting transverse displacement $|u_x|$ on the focal $x$--$y$ plane (upper sub-row) and $x$--$z$ plane (lower sub-row) at five times spanning one Yaakub PRF period ($t = 4, 10, 20, 92, 198$\,ms, with the first three columns tracing the 20\,ms ON ramp-up and the last two tracing the 180\,ms OFF visco-elastic relaxation). Field-wide peak $|u_x| \approx 0.40~\mu$m at $t = 20$\,ms. Cyan contour, skull mask.}
  \label{fig:yaakub_arf_disp}
\end{figure}

\begin{figure}[htbp]
  \centering
  \includegraphics[width=\textwidth]{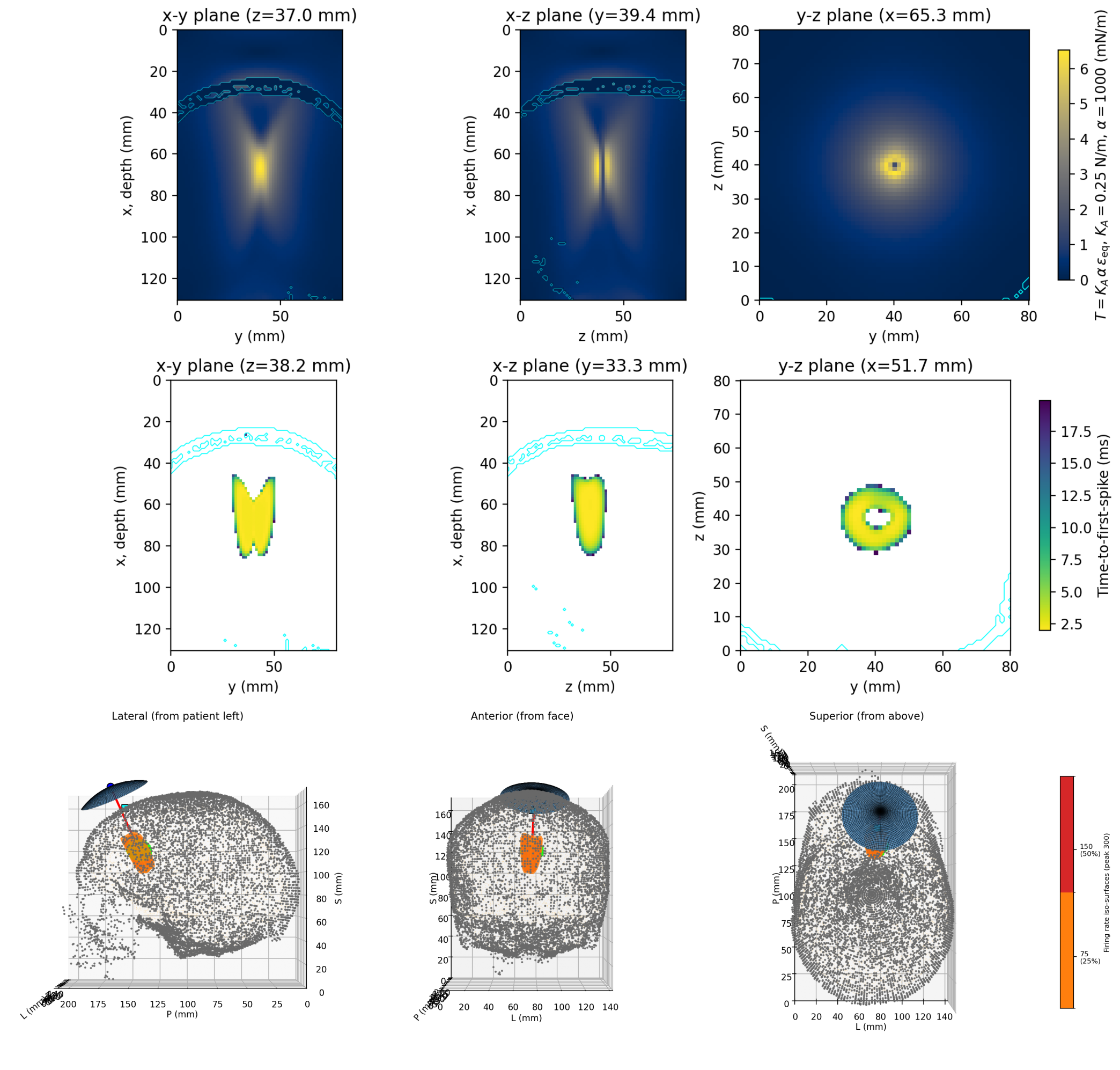}
  \caption{From mechanical drive to per-voxel firing. \textbf{Top.} Membrane tension $T = K_A\,\alpha\,\varepsilon_{\mathrm{eq}}$ on three orthogonal mid-planes through the peak-tension voxel at end-of-pulse, the proximal input to the mechanosensitive Boltzmann gates. \textbf{Middle.} Time-to-first-spike on three orthogonal slices through the peak firing-rate voxel (the slice coordinates differ slightly from the top row because the peak-rate voxel sits at the upper margin of the focal lobe, not at its tension maximum). Reversed colour map so faster latencies are brighter, transparent where no spike is recorded. \textbf{Bottom.} Three-camera 3-D rendering of iso-surfaces of the per-voxel firing-rate field at 50\,\% (red) and 25\,\% (orange) of peak, with the skull silhouette as a grey point cloud, locating the firing zone in anatomy.}
  \label{fig:yaakub_neural_latency_iso}
\end{figure}

\begin{figure}[htbp]
  \centering
  \includegraphics[width=\textwidth]{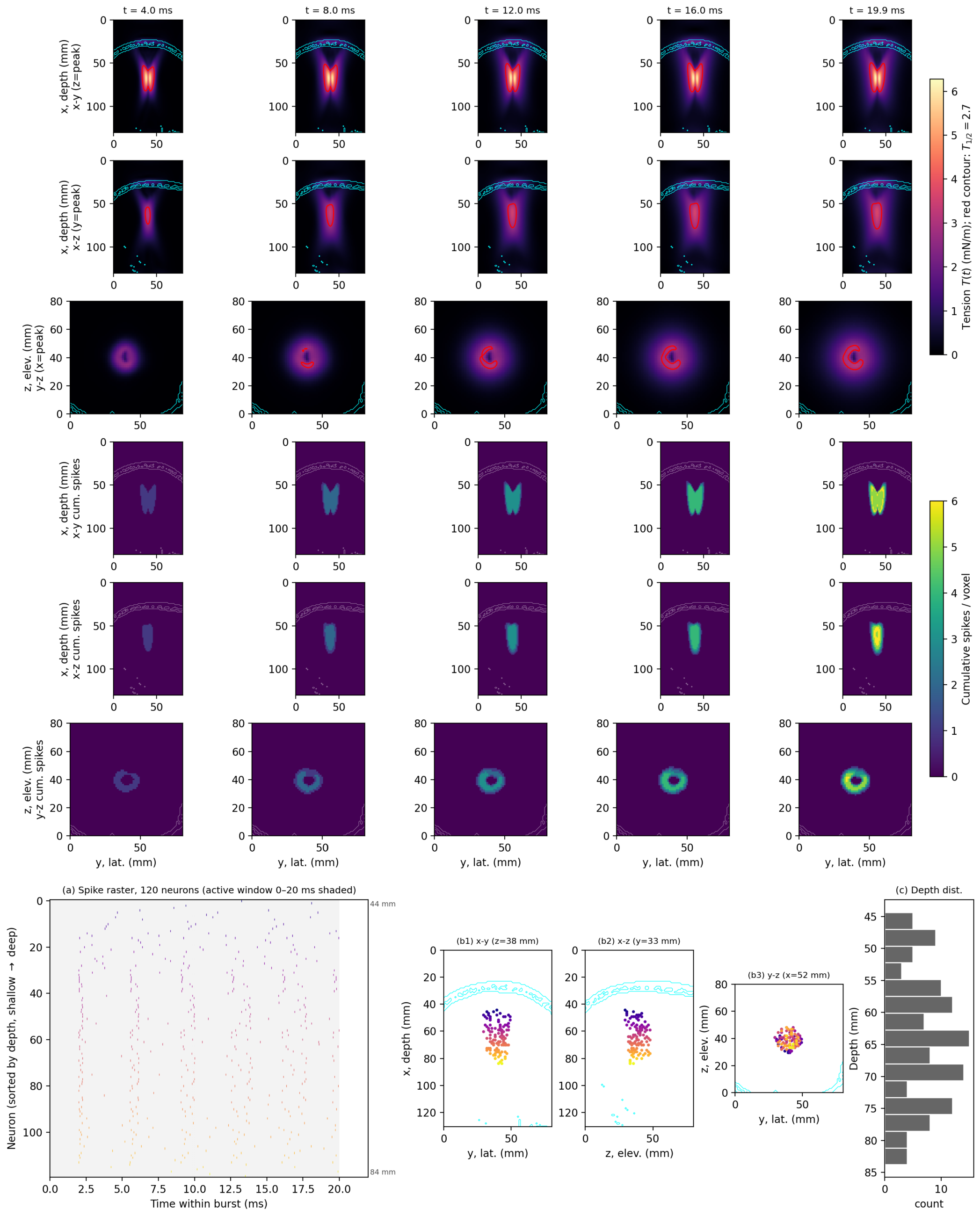}
  \caption{Within-burst recruitment dynamics. \textbf{Top six rows.} Instantaneous membrane tension $T(t)$ and cumulative spike count per voxel at five within-burst times ($t = 4, 8, 12, 16, 19.9$~ms), each on three orthogonal slices ($x$--$y$, $x$--$z$, $y$--$z$) through the brain-restricted peak voxel. Tension panels (rows 1--3) carry a red contour at the Piezo1 half-activation $\Thalf = 2.7$~mN/m. Spike panels (rows 4--6) show the running cumulative count, making the mechano-to-spike lag visible (voxels recruit a few~ms after their tension crests). \textbf{Bottom.} Single-burst spike raster across 120 firing voxels sampled by depth (tick colour encodes depth, shallow to deep). The three ROI panels (b1--b3) on three orthogonal slices and the depth histogram (c) tie each raster row back to anatomy.}
  \label{fig:yaakub_neural_snap_rast}
\end{figure}

\begin{figure}[htbp]
  \centering
  \includegraphics[width=0.85\textwidth]{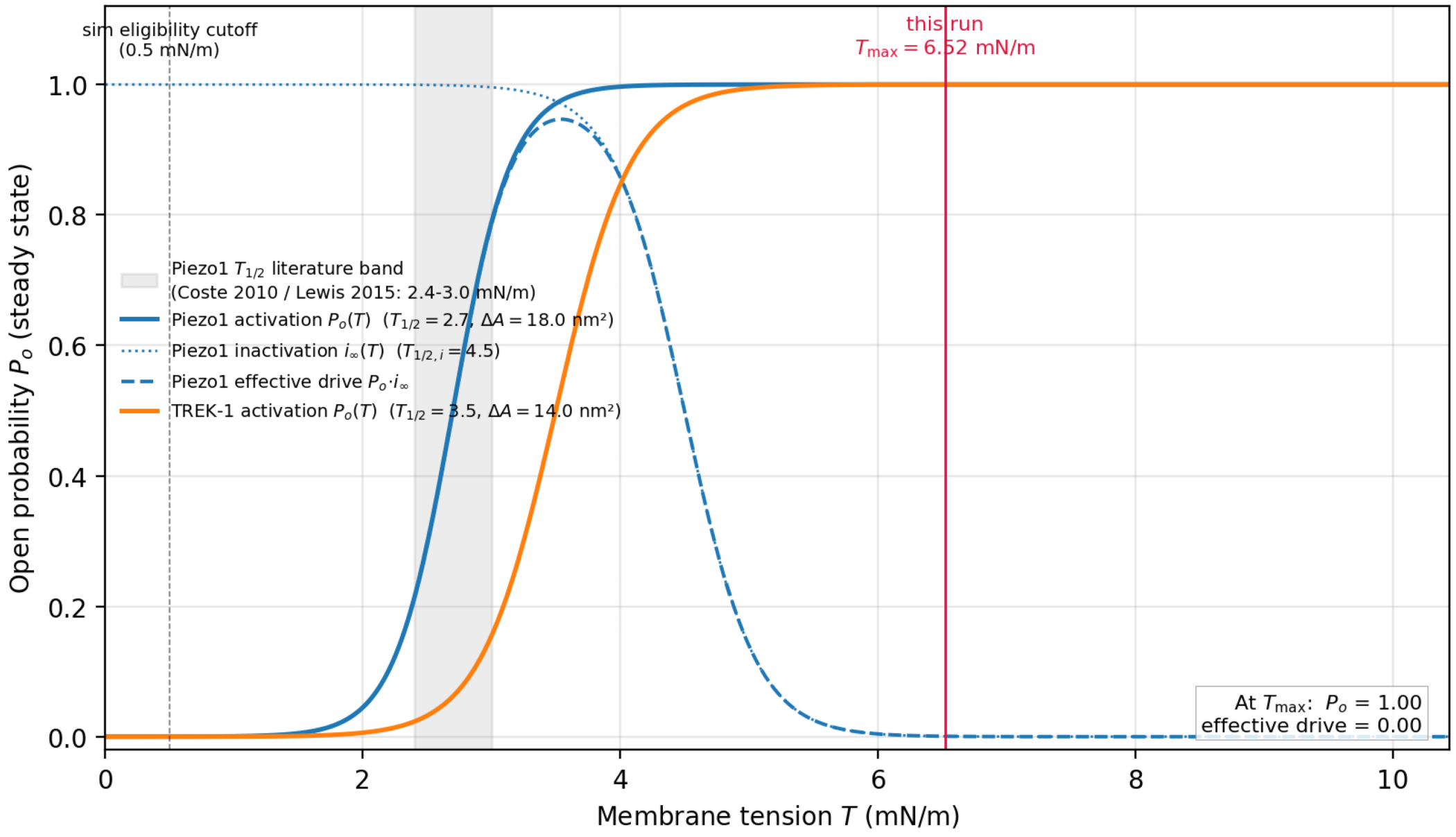}
  \caption{Mechanosensitive gating curves used in this simulation. Solid blue, Piezo1 activation $P_o(T)$ ($T_{1/2} = 2.7$~mN/m, $\Delta A = 18$~nm$^2$). Dotted blue, Piezo1 inactivation $i_\infty(T)$ ($T_{1/2,i} = 4.5$~mN/m). Dashed blue, Piezo1 effective drive $P_o \cdot i_\infty$, which peaks at $T \approx 3$~mN/m and falls off above $\sim 5$~mN/m. Solid orange, TREK-1 activation $P_o(T)$ ($T_{1/2} = 3.5$~mN/m, $\Delta A = 14$~nm$^2$). Grey band, literature confidence interval on the Piezo1 $T_{1/2}$~\citep{coste2010piezo,lewis2015mechanical}. Crimson vertical, per-simulation working point ($T_{\max} = 6.52$~mN/m, the focal-voxel peak tension on this run). Black dashed, 0.5~mN/m eligibility cut-off. The focal voxel sits about $2.42\times \Thalf$ on the Piezo1 curve, where Piezo1 activation saturates ($P_o = 1.00$), while the inactivation gate drops the effective drive to $\approx 0.01$. This is the mechanism reason the S4\,/\allowbreak\,S5 scenarios in Figure~\ref{fig:tips_mechanisms} produce $\sim$30--33\% fewer spikes than S2\,/\allowbreak\,S3 at the same focal drive.}
  \label{fig:yaakub_neural_gating}
\end{figure}

The theta-burst protocol~\citep{yaakub2023transcranial} is the canonical human translation of the NHP theta-burst paradigm of~\citep{verhagen2019offline}. We use it because the 500\,kHz NeuroFUS CTX-500 four-element annular array on which it is based is widely used in offline-effect human tFUS studies. The reported parameters are carrier 500~kHz, aperture 64~mm, geometric focus 63~mm, 20~ms rectangular tone-bursts repeated at 5~Hz (yielding an overall duty cycle of 10\%) over an 80~s session, free-field peak pressure 1.0~MPa, derated to a transcranial peak negative pressure of approximately 0.5~MPa. Yaakub et al. report two deep cortical cohorts (dACC at 60~mm focal depth and PCC at 69~mm focal depth, both with the CTX-500 protocol above). We adopt the dACC cohort for the canonical demonstration because its 60~mm depth aligns most closely with the CTX-500's 63~mm geometric focus. The bowl is aimed at the left anterior cingulate cortex (Harvard--Oxford ACC ROI, MNI RAS = $(-7, 35, 20)$, anatomically validated via the SyN MNI-to-Halle warp), with the apex pose auto-selected on the outer-skull mesh by minimising the perpendicular-aim residual to the target (nearest 10--20 cap landmark AF3 at 27.9~mm offset, with placement shown in Figure~\ref{fig:placement_3d}). Heterogeneous acoustic propagation through the Halle micro-CT skull specimen is computed by the Fullwave solver (Eq.~\ref{eq:fullwave}). Figure~\ref{fig:yaakub_intensity} shows both the resulting focal-intensity lobe (top) in dB and in 3-D LPS coordinates relative to the skull, brain mesh, and target ROI, and the parallel Pennes-bioheat thermal rise driven by the same field (bottom). The acoustic radiation force then drives the Kelvin--Voigt shear-FDTD solver (Eq.~\ref{eq:shearwave}) over a single 20\,ms tone-burst integrated to 200\,ms (one PRF period). The resulting body-force and displacement fields are shown in Figure~\ref{fig:yaakub_arf_disp}. The brain tissue properties are density $\rho_0 = 1000$~kg/m$^3$, shear modulus $\mu = 4580$~Pa, shear-wave speed $c_s = 2.14$~m/s, and Kelvin--Voigt viscosity $\eta = 0.5$~Pa$\cdot$s. The resulting shear-FDTD time step is $\Delta t = 33.2~\mu$s. The peak equivalent strain $\varepsilon_{\mathrm{eq}}$ at the focal voxel, taken from the full transient displacement trajectory at end-of-pulse ($t = 20$\,ms, just after the ON window closes), is $\varepsilon_{\mathrm{eq}}^{\max} \approx 2.61 \times 10^{-5}$, with the focal voxel reaching peak displacement $|u_x| \approx 0.40~\mu$m at a safety-envelope-compliant surface drive that reproduces the Yaakub transcranial focal target ($p_0 = 0.10$~MPa free-field calibration, with a 1.087$\times$ post-hoc linear rescale to land the in-brain peak pressure (PPP) at the paper's reported 0.50~MPa, with supplementary~\S~\ref{sec:supp_alpha} discussing the rescaling step). With $K_A = 0.25$~N/m (Rawicz et al. lipid-bilayer area-expansion modulus~\citep{rawicz2000effect}) and the multi-scale coupling factor $\alpha$ treated as a free calibrated parameter on a literature-bracketed sweep (Salahshoor / Rawicz / Sukharev iso-potential preparations report $\alpha \approx 200$~\citep{rawicz2000effect,sukharev2012mechanosensitive,salahshoor2020transcranial}, and supplementary~\S~\ref{sec:supp_alpha} shows the firing-zone size as a function of $\alpha$ across the [200, 2000] range, with $\alpha \lesssim 460$ collapsing firing to the focal voxel only and $\alpha \gtrsim 2000$ saturating), the headline run reports numbers at the cortical-pyramidal calibration point $\alpha = 1000$ (motivated in supplementary~\S~\ref{sec:supp_alpha} by dendrite-heavy Piezo1 expression and viscoelastic stress concentration above the iso-potential reference). At this working point the focal membrane tension is $T_{\max} \approx 6.52$~mN/m, about $2.42\times$ the Piezo1 half-activation $\Thalf = 2.7$~mN/m. The single-push 20~ms ON window produces firing at 5\,468 brain voxels (the firing zone summarised in Table~\ref{tab:yaakub_summary}), with the field-wide tension above the 0.5~mN/m Piezo1 eligibility threshold contracting back below threshold within $\sim 100$~ms of relaxation as the visco-elastic response decays with $\tau \approx 33$~ms. Spatially-resolved detail is shown in Figs.~\ref{fig:yaakub_intensity}--\ref{fig:yaakub_neural_gating}.

The focal acoustic metrics in Table~\ref{tab:yaakub_summary} (PNP = 0.48\,MPa, PPP = 0.50\,MPa, MI = 0.69, $I_{\mathrm{SPPA}} = 8.1$\,W/cm$^2$) are the Fullwave~2 focal-peak values through the skull at the calibrated surface drive (free-field $p_0 = 0.10$~MPa with a 1.087$\times$ linear rescale that sets the in-brain peak pressure (PPP) to 0.50~MPa, matching Yaakub's reported ${\approx}0.5$~MPa transcranial focal pressure, and is applied directly to the pressure field downstream of the FDTD since acoustic propagation is linear). $I_{\mathrm{SPPA}}$ is computed from the focal peak pressure via the continuous-wave intensity relation $I_{\mathrm{SPPA}} = \mathrm{PPP}^2/(2\rho c)$ since the 20\,ms ON window carries a continuous 500\,kHz sinusoid at the reported PPP. The duty-cycle-averaged $I_{\mathrm{SPTA}}$ over the PRF period is $0.10 \times I_{\mathrm{SPPA}} \approx 0.81$\,W/cm$^2$. Both quantities sit well within the ITRUSST consensus safety envelopes for transcranial ultrasound neuromodulation (MI $\leq 1.9$, $I_{\mathrm{SPPA}} \leq 14$\,W/cm$^2$)~\citep{aubry2025safety}, and they are consistent with the derated transcranial PNP of approximately 0.5\,MPa that Yaakub~\citep{yaakub2023transcranial} reports.

\begin{table}[H]
\centering
\footnotesize
\setlength{\tabcolsep}{6pt}
\renewcommand{\arraystretch}{1.05}
\caption{Yaakub canonical-run quantitative summary on the Halle micro-CT skull, auto-pick placement aimed at the left dACC (ACC\_left, MNI RAS = $(-7, 35, 20)$). Acoustic block, spatial-peak metrics from the Fullwave FDTD output, with $I_{\mathrm{SPPA}}$ computed as $\mathrm{PPP}^2/(2\rho c)$ at the focal voxel, the continuous-wave-equivalent in-pulse intensity for the continuous 500\,kHz sinusoid during the 20\,ms ON window (time-averaging over the 10\% duty-cycle PRF period gives $I_{\mathrm{SPTA}} \approx 0.81$\,W/cm$^2$). Thermal block, Pennes bioheat at session end (10\% duty-cycle Yaakub schedule, 80\,s session). Shear-wave block, end-of-pulse displacement at the peak ARF voxel and visco-elastic relaxation fit. Mechano-to-neural block, peak-over-transient tension and the resulting per-voxel firing population over the 20\,ms ON window. All numbers from the Halle\,/\allowbreak\,dACC validation report shown in Figs.~\ref{fig:yaakub_intensity}--\ref{fig:yaakub_neural_gating}.}
\label{tab:yaakub_summary}
\begin{tabular}{@{}lr|lr@{}}
\toprule
\multicolumn{2}{c|}{\textbf{Fullwave~2 (acoustic)}} & \multicolumn{2}{c}{\textbf{Pennes bioheat (80\,s session)}} \\
\midrule
PNP               & 0.48\,MPa      & $\Delta T_{\mathrm{focus}}$ (final)  & 233\,mK \\
PPP               & 0.50\,MPa      & $\Delta T_{\mathrm{global}}$ (final) & 0.94\,K \\
MI                & 0.69         & $\Delta T$ per single burst      & 22\,mK \\
$I_{\mathrm{SPPA}}$       & 8.1\,W/cm$^2$    & CEM43 (peak voxel)          & $8.1{\times}10^{-4}$\,min \\
$-6$\,dB focal volume      & 161\,mm$^3$     & Skull absorption fraction       & 0.30 \\
Axial FWHM            & 22.2\,mm       & Session                & 80\,s, 400 bursts \\
Lat.\,/\allowbreak\,elev.\ FWHM       & 3.7\,/\allowbreak\,3.7\,mm   & Skull-bone peak $\Delta T$ (final)  & 0.94\,K \\
Peak ARF density         & 323\,N/m$^3$     & Near-skull brain peak $\Delta T$ (final) & 0.74\,K \\
\midrule
\multicolumn{2}{c|}{\textbf{Shear FDTD}} & \multicolumn{2}{c}{\textbf{Mechano-to-neural (20\,ms ON window)}} \\
\midrule
Peak $|u_x|$ (trace voxel)            & 0.40\,$\mu$m        & Firing voxels            & 5\,468 \\
Time to peak                   & 20.0\,ms          & Total spikes per burst       & 20\,416 \\
Field-wide max $|u_x|$              & 0.40\,$\mu$m        & Mean spikes per firing voxel    & 3.73 \\
$\tau_{\mathrm{relax}}$ (OFF window)       & 32.6\,ms          & Peak firing rate          & 300\,Hz \\
Peak $\varepsilon_{\mathrm{eq}}$ (end-of-pulse)  & $2.61{\times}10^{-5}$    & Iso-50\,\% firing volume      & 7\,025\,mm$^3$ \\
Peak $T = K_A \alpha\,\varepsilon_{\mathrm{eq}}$ & 6.52\,mN/m         & Iso-25\,\% firing volume      & 8\,523\,mm$^3$ \\
$\Delta t_{\mathrm{shear}}$ / $n_{\mathrm{steps}}$ & 33.2\,$\mu$s / 6018    & $T_{\max} / T_{1/2}$         & 2.42$\times$ \\
\bottomrule
\end{tabular}
\end{table}

Each above-threshold brain voxel runs the dendrite--soma--AIS Pospischil cortical-pyramidal neuron at body temperature, with Piezo1 ($\bar{g} = 0.1$~mS/cm$^2$) and TREK-1 K\textsuperscript{+} ($\bar{g} = 0.04$~mS/cm$^2$) in the dendrite-heavy cortical-pyramidal preset (mechanosensitive densities elevated above the library defaults of Table~\ref{tab:channels} to place the focal voxel at the Boltzmann working point), zero background stimulus, and a uniform isothermal background. A total of 5\,468 brain voxels fire over the 20~ms ON window, with a mean of 3.73 spikes per firing voxel and a peak focal-voxel rate of 300~Hz. The firing zone (Figs.~\ref{fig:yaakub_neural_latency_iso}--\ref{fig:yaakub_neural_snap_rast}) is centred on the geometric focus and extends a few millimetres axially and laterally with a graded falloff that mirrors the underlying tension field. The iso-50 firing-rate volume covers 7\,025\,mm$^3$ and the iso-25 volume 8\,523\,mm$^3$, both substantially exceeding the acoustic $-6$\,dB focal volume (161\,mm$^3$, Table~\ref{tab:yaakub_summary}), because the strain field elongates the firing zone along the beam axis through the CTX-500 focal aspect ratio. This shows that the strain-driven mechanosensitive pathway alone (without invoking intramembrane cavitation, calcium coupling, or thermosensors) is sufficient to trigger focal firing at the calibrated Yaakub exposure, and that the focal voxel sits about $2.42\times \Thalf$ on the Piezo1 Boltzmann curve, well inside the upper knee where small strain shifts produce large changes in spike count (\S\ref{sec:results_sensitivity}). As a cell-type-independent corollary, the conductance-weighted passive depolarisation field $\Delta V_{\mathrm{passive}}(\mathbf{x})$ peaks at $\sim 53$\,mV at the focal voxel (i.e., the focal-voxel membrane settles to $V_{\mathrm{ss}} \approx -17$\,mV in the absence of HH spiking). The firing zone for any cell-type model drops out as the iso-surface of $\Delta V_{\mathrm{passive}}$ at that model's spike threshold, decoupling the strain-driven input from the threshold-dependent output.

\subsection{Mechanism-comparison sweep on the real Halle\,/\allowbreak\,dACC field}
\label{sec:results_mechanism_sweep}

To isolate the mechanism-comparison axis we kept the upstream Halle\,/\allowbreak\,dACC strain field fixed (the same field that drives the canonical run, with peak $\varepsilon_{\mathrm{eq}}^{\max} \approx 2.6 \times 10^{-5}$) and uniformly rescaled it to six target peak strains $\varepsilon \in \{2{\times}10^{-6}, 5{\times}10^{-6}, 10^{-5}, 2{\times}10^{-5}, 5{\times}10^{-5}, 10^{-4}\}$ that walk the focal-voxel tension from sub-threshold ($T \approx 0.5$\,mN/m, $0.18\Thalf$) through the Boltzmann knee ($\sim$5\,mN/m, the canonical working point) to saturation ($\sim$25\,mN/m). At each strain level we ran the same five mechanism scenarios on the rescaled field. These were (S1) a single-compartment neuron with Piezo1 and TREK-1, (S2) a three-compartment neuron (dendrite, soma, AIS) with Piezo1 and TREK-1 and the manuscript's dendrite-heavy density, (S3) a three-compartment neuron with intramembrane cavitation added on all compartments, (S4) a three-compartment neuron with Piezo1 inactivation, intracellular Ca\textsuperscript{2+} pool, and somatic SK counter-current added, and (S5) all four pathways simultaneously, with TRPV1 and TRPV4 thermosensors active on the soma at body temperature. Piezo1 conductance density is set to $\bar{g} = 0.05$~mS/cm$^2$ in the sweep (lower than the canonical $\bar{g}=0.10$) so the focal voxel sits near the spike-initiation threshold for the single-compartment baseline.

\begin{figure}[htbp]
  \centering
  \includegraphics[width=\textwidth]{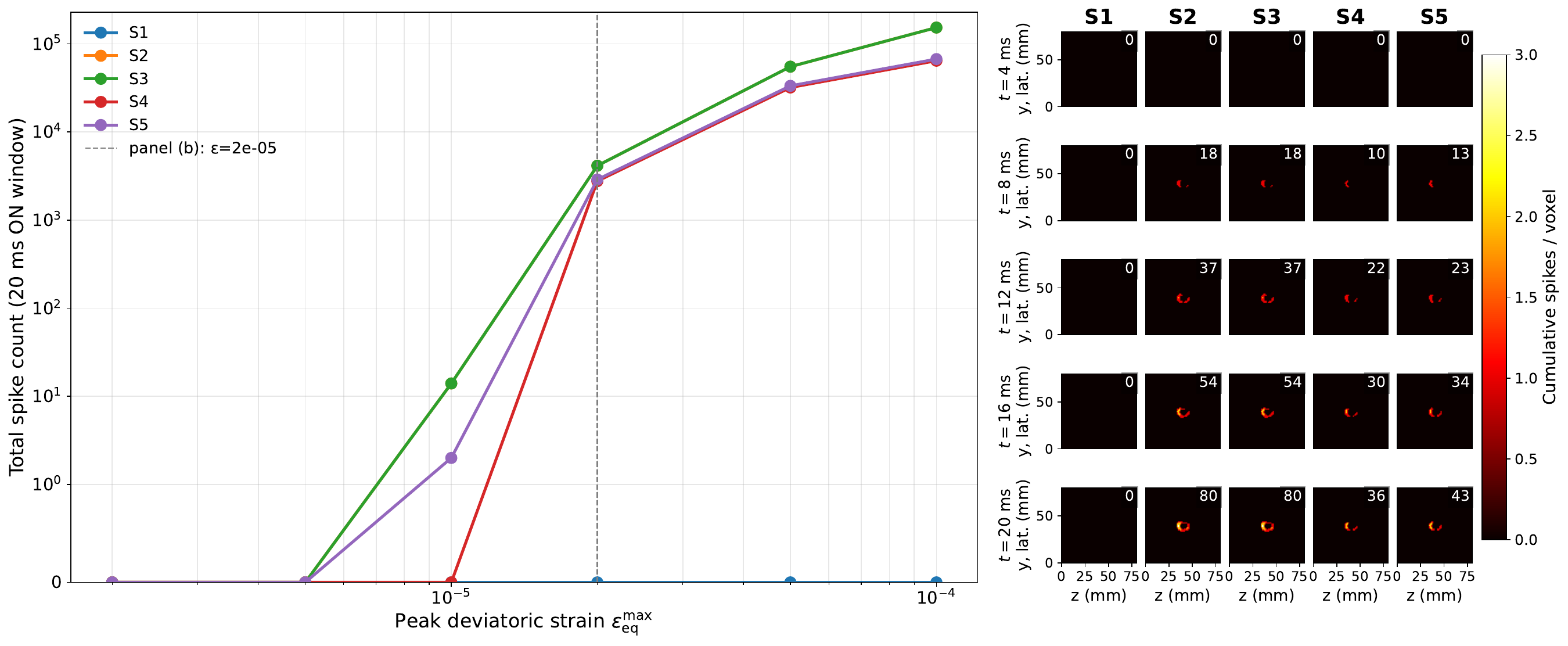}
  \caption{Mechanism-comparison sweep on the real Halle\,/\allowbreak\,dACC strain field. (a)~Total spike count over a 20~ms window as a function of peak deviatoric strain, for five mechanism scenarios. The field is uniformly rescaled to walk the focal-voxel tension from sub-threshold through the Boltzmann knee to saturation. Vertical dashed line, canonical working point $\varepsilon \approx 2.6\times 10^{-5}$. (b)~Cumulative firing maps on the focal $y$--$z$ plane at five within-burst times $t \in \{4, 8, 12, 16, 20\}$~ms (rows) for each of the five mechanism scenarios (columns) at the canonical working point (matching the unscaled Halle\,/\allowbreak\,dACC field, Table~\ref{tab:yaakub_summary}). Per-panel annotation is the cumulative spike count on this slice up to time $t$. At this working point the dendrite-heavy multi-compartment neuron model (S2) fires strongly while the single-compartment baseline (S1) remains subthreshold, so the multi-compartment geometry is necessary for firing rather than merely amplifying it. The Piezo1-inactivation\,/\allowbreak\,Ca\textsuperscript{2+}\,/\allowbreak\,SK scenarios (S4, S5) reduce focal firing by $\sim$30--33\% at the canonical working point and by up to $\sim$58\% at saturation. Intramembrane cavitation (S3 vs S2) and TRP thermosensors (S5 vs S4) contribute marginally at sub-MPa carrier and body temperature, as predicted.}
  \label{fig:tips_mechanisms}
\end{figure}

Figure~\ref{fig:tips_mechanisms}a shows the total spike count over a 20~ms window as a function of peak strain. Three regimes are visible. Below $\varepsilon \approx 5\times 10^{-6}$ ($T \approx 1.2$\,mN/m, sub-$\Thalf$), all scenarios are subthreshold and produce no spikes. Around $\varepsilon \approx 10^{-5}$ ($T \approx 2.5$\,mN/m, just inside the lower Boltzmann knee), the multi-compartment scenarios (S2, S3) begin firing at a handful of focal voxels while the single-compartment baseline (S1) remains subthreshold. The dendrite-heavy Piezo1 distribution and AIS axial coupling are required for any firing at this conductance. The single-compartment baseline (S1) stays subthreshold across the entire sweep, so the multi-compartment geometry, with its lower strain-threshold for spike initiation and more efficient recruitment of off-focal voxels, is necessary rather than merely amplifying. At the canonical working point ($\varepsilon \approx 2.6\times 10^{-5}$, the closest sweep level to the unscaled Halle\,/\allowbreak\,dACC field) S1 produces no spikes, while S2 and S3 each produce 4\,142 spikes (S3 matching S2 to numerical precision as discussed below), and S4/S5 produce 2\,770 / 2\,889 spikes (a $\sim$30--33\% reduction relative to S2 from Piezo1 inactivation and SK counter-current). Above $\varepsilon \approx 5\times 10^{-5}$ the curves separate further. S4 and S5 produce $\sim$40--58\% fewer spikes than S2/S3 at the same drive, with the inhibition growing with strain. This is the predicted Piezo1-inactivation / Ca\textsuperscript{2+} / SK signature. The inactivation gate cuts off the depolarising drive within $\sim$12~ms, and the Ca\textsuperscript{2+} pool, charged during the brief active window, opens SK channels that hyperpolarise the soma for the remainder of the burst. The intramembrane-cavitation contribution at sub-MPa carrier pressures (S3 vs S2) is essentially zero across the entire sweep (S3 reproduces S2 to numerical precision), consistent with the cycle-averaged $\langle \Delta C_m \rangle / C_{m,0} \approx 5\%$ predicted at 100~kPa by the SONIC reduction. The TRP contribution at body temperature (S5 vs S4) is comparably small ($\sim$5\% from a modest TRPV4 tonic current, with TRPV1 essentially closed). Figure~\ref{fig:tips_mechanisms}b shows firing-map slices through the focal plane at the canonical working point ($\varepsilon \approx 2.6\times 10^{-5}$). S1 produces no spikes, S2 / S3 produce a clearly delineated focal lobe, and S4 / S5 reduce that lobe in per-voxel spike count and extent through the inactivation / Ca\textsuperscript{2+} / SK pathway. Spatially, this suppression concentrates at the high-tension centre. The focal voxel ($T = 6.52$~mN/m) sits above the Piezo1 inactivation half-tension ($T_{1/2,i} = 4.5$~mN/m) and self-inactivates, so under inactivation the focal centre falls silent while the surrounding $\sim$3--4.5~mN/m shell continues to fire, converting the filled focal lobe into an annulus with a quiescent centre (Supplementary Fig.~\ref{fig:supp_focus_ring}). Because this focus$\to$ring topology is a shape rather than a magnitude, it is robust to the uncalibrated $K_A\cdot\alpha$ tension scale and supplies a falsifiable signature distinguishing inactivation-dominated from purely activation-driven recruitment.

This sweep compares the five channel-level scenarios (S1--S5) on a shared neuron. The two relay mechanisms, the astrocytic TRPA1\,$\to$\allowbreak\,glutamate\,$\to$\allowbreak\,NMDA pathway (\S\ref{sec:results_astrocyte}) and the mechanosensitive synaptic\,/\allowbreak\,voltage-gated transmission pathway (supplementary~\S\ref{sec:supp_synaptic}), act through their own pre- and post-synaptic readouts rather than as drop-in channel scenarios on the strain-walk, and are therefore scored separately via closed-form per-case excitation scalars rather than included in this comparison.

\subsection{Source-classified sensitivity}
\label{sec:results_sensitivity}

The parameter source classification (section~\ref{sec:methods_provenance}) gives a principled basis for sensitivity analysis. Literature-anchored parameters can be shifted within their reported uncertainty bands, calibrated parameters only constrain the model qualitatively and their absolute values are not themselves meaningful, and assumed parameters are by construction free. We applied $\pm 25\%$ shifts to six representative parameters individually and re-ran the S5 full-pathway scenario on the real Halle / dACC end-of-pulse strain field ($\varepsilon_{\mathrm{eq}}^{\max} = 2.61 \times 10^{-5}$, $\alpha = 1000$, baseline spike count from the regenerated sensitivity table). Results are summarised in Table~\ref{tab:sensitivity}. The same table carries a separate sub-block for the sixth (synaptic / voltage-gated) mechanism, scored on its own readout, the relay focal spike count at the focal voxel ($T = 6.5$~mN/m), since that pathway does not feed the strain-driven multi-compartment spike count. There the post-synaptic NMDA stretch enhancement $\alpha_{\mathrm{NMDA}}$ (Maneshi et al.) and its half-tension are the most sensitive constants ($\pm 25\%$ shifts move the focal spike count by $\sim 10$--$13\%$), whereas the presynaptic-release and Na\textsuperscript{+}-activation-shift parameters each contribute only a few percent, consistent with the NMDA conductance carrying the bulk of the relay's depolarising drive at the focal tension.

\input{data/sensitivity_table}

Three observations follow. First, the single most sensitive parameter is the Piezo1 half-activation tension $T_{1/2}$, a literature-anchored value. A $-25\%$ shift (from 2.7 to 2.0~mN/m) raises the spike count by 195\%, and a $+25\%$ shift (to 3.4~mN/m) drops it by 72\%. This is the expected consequence of operating near the Boltzmann knee where the open probability $P_o(T)$ transitions sharply from 0 to 1. On the real Halle / dACC field the focal voxel sits at $T \approx 2.42\times \Thalf$, well inside the upper knee, so small shifts in the threshold parameter amplify into large changes in firing. This sensitivity is intrinsic to the strain-driven Boltzmann-channel pathway and would persist under any choice of upstream FDTD solver or cell-type model. It is the central caveat for absolute-rate predictions and motivates careful per-cell-type calibration of $T_{1/2}$ rather than a one-size-fits-all literature value. Second, the bilayer area-expansion modulus $K_A$ in the strain-to-tension conversion (recorded as a modelling assumption, with $K_A \cdot \alpha$ being the operative single knob) is comparably influential, with $\pm 25\%$ shifts producing $-59\%$ and $+89\%$ responses, again because the $K_A \cdot \alpha \cdot \varepsilon$ product determines where the focal voxel sits relative to the Piezo1 threshold. The dendrite-to-soma Piezo1 density-scale ratio of $5\times$, a literature-anchored value~\citep{lewis2015mechanical}, contributes a comparable-magnitude effect ($+18\%$ / $-25\%$). Third, three of the parameters (AIS sodium density, SK $[\mathrm{Ca}^{2+}]_{1/2}$, and SK $\bar{g}$) alter the total spike count by 0--3\%. The AIS at $\bar{g}_{\mathrm{Na}} = 2000$~mS/cm$^2$ is already so far above its spike-initiation threshold under the focal-voxel drive that $\pm 25\%$ shifts do not change firing, and the $\sim$20~ms simulation window is short enough that the Ca\textsuperscript{2+} pool barely engages SK at this drive amplitude.

The framework's predictions in this regime are dominated by where the focal-voxel tension sits relative to the Piezo1 Boltzmann knee. Two literature-anchored parameters ($T_{1/2}$ and the dendrite-to-soma Piezo1 density ratio) and the strain-to-tension product $K_A \cdot \alpha$ (a modelling-assumption pair, with only the product experimentally identifiable and $\alpha$ itself a calibrated parameter spanning $\sim$200--2000 across the literature bracket, supplementary~\S\ref{sec:supp_alpha}) jointly set the firing threshold, and predictions are robust against $\pm 25\%$ shifts of every other parameter we tested. We therefore present the headline firing numbers at $\alpha = 1000$ as a single working point on the literature-bracketed sweep rather than as a calibrated absolute, and the $\alpha$-sweep in supplementary~\S\ref{sec:supp_alpha} is the primary uncertainty report for absolute spike counts. The amplification is large because the real Halle\,/\allowbreak\,dACC field has a long-tailed lateral falloff that places many voxels within ${\sim}\Thalf$ of the Boltzmann inflection, where small parameter shifts cascade into large firing-zone changes. Meaningful comparison to experimental ISPPA--firing curves therefore requires calibration of the $K_A \cdot \alpha$ product against a tissue- and cell-type-matched dataset combined with explicit per-cell-type uncertainty bands on $T_{1/2}$, both of which the parameter source classification makes explicit.

\subsection{Astrocytic TRPA1 relay on the real Halle\,/\allowbreak\,dACC field}
\label{sec:results_astrocyte}

Scoring the astrocytic pathway (\S\ref{sec:methods_astrocyte}) on the canonical field gives a modest, slow contribution. At the focal voxel ($T_{\max} = 6.52$~mN/m) the astrocytic Ca\textsuperscript{2+} pool charges over hundreds of ms, driving $\sim 9~\mu$M glutamate and an NMDA-mediated $\sim 11$~mV sub-threshold depolarisation that, at the 10\% Yaakub duty cycle, stays below the spike-recruitment threshold (no voxels recruited on its own), weaker and slower than the direct Piezo1 pathway at this sub-MPa carrier, consistent with a complementary rather than dominant role in the theta-burst regime. Its distinguishing signature is that the slow relay integrates the duty-gated drive over its Ca\textsuperscript{2+} time constant, so the astrocytic-NMDA drive grows with cumulative duty $\times$ on-time rather than tracking the carrier, and the Oh et al.\ TRPA1-knockout and AP5 controls are reproduced. Full results, the relay timecourse, the duty-cumulative drive, and the per-voxel excitation map on the canonical field, are given in supplementary~\S\ref{sec:supp_astrocyte_results} (Fig.~\ref{fig:supp_astrocyte_results}), with the mechanism controls in \S\ref{sec:supp_verif_astrocyte}.

% =============================================================================
\section{Discussion}
\label{sec:discussion}

\subsection{Mechanism discrimination outlook}

The biophysical mechanisms underlying transcranial focused ultrasound neuromodulation remain unsettled, with multiple plausible candidates (mechanosensitive-channel gating, intramembrane cavitation, Ca\textsuperscript{2+}-coupled inhibition, thermosensor activation, the astrocytic TRPA1$\to$glutamate$\to$NMDA relay, and mechanosensitive synaptic / voltage-gated transmission) argued for in isolation across the published literature. The framework presented here is, to our knowledge, the first that exercises all six candidates on a common neuron model, with traceable per-parameter source classification, so that differences in predicted firing between candidates can be attributed unambiguously to the mechanism rather than to incidental modelling choices. In the strain regime explored in section~\ref{sec:results_mechanism_sweep}, the Piezo1-inactivation / Ca\textsuperscript{2+} / SK pathway qualitatively suppresses firing relative to the strain-only multi-compartment baseline at moderate drive, consistent with Legon-style high-pressure inhibition outcomes, and the intramembrane-cavitation pathway adds only a marginal contribution at sub-MPa carrier pressures. We note that the NICE bilayer-sonophore has not been directly observed in a living membrane, and the more recent Piezo1 knockdown and conditional-knockout literature~\citep{qiu2019piezo1,yoo2022focused,zhu2023piezo1} constrains the fraction of the in-vivo response an NICE-only mechanism can plausibly carry. These trends are properties of the model under literature-fit parameters and as such are predictions the framework makes. They will need experimental cross-checking before any specific mechanism candidate is preferred.

A qualitatively different hypothesis holds that the nerve impulse is an electromechanical density \emph{soliton} propagating near the lipid membrane's gel--fluid melting transition rather than an ion-channel action potential~\citep{heimburg2005soliton,heimburg2007thermal}, which the framework scores as a closed-form per-case excitability scalar alongside the astrocytic and synaptic pathways. It is chiefly of value as a falsifiable discriminator, since biological membranes melt below body temperature and focal heating therefore moves the membrane \emph{away} from the transition and \emph{reduces} excitability, the opposite sign to the thermosensor and $Q_{10}$ pathways, making focal temperature the experimental axis that most cleanly separates the soliton hypothesis from the electrical candidates.

\subsection{Limitations}

The three-compartment skeleton is a deliberate simplification. Active dendritic spikes, branch-specific channel densities, full axonal propagation with myelinated saltatory conduction, and dendrite morphology beyond the dendrite/soma/AIS lumping are out of scope. The framework targets per-voxel firing-map sweeps where iso-frequency point-neuron equivalence at the AIS is the relevant output, not full morphological electrophysiology. The NICE coupling uses the SONIC cycle-averaged simplification of Lemaire et~al.~\citep{lemaire2019understanding}. The intramembrane gas-pocket pressure $P_{\mathrm{in}}(Z, R)$ of the full Plaksin BLS model is not included, capping the carrier-pressure validity at $\sim$1~MPa. The bilayer area-expansion modulus $K_A$ in the strain-to-tension conversion (Eq.~\ref{eq:tension}) and the multi-scale coupling factor $\alpha$ are recorded as modelling assumptions in the parameter source classification. The predicted absolute spike counts depend on the product $K_A \cdot \alpha$, and meaningful comparison to experimental ISPPA--firing curves will require calibration of this product against a tissue- and cell-type-matched dataset. Finally, all gating is deterministic. Stochastic-gate noise, which can shift firing thresholds at low channel counts, is not modelled.

\subsection{Future work}

Five extensions are natural next steps. Per-compartment temperature coupling (carrying a separate temperature trace at the dendrite, soma, and AIS rather than a single per-neuron scalar) would let the bioheat output of the upstream solver drive Q\textsubscript{10} and TRP gating differentially across the cell, relevant for studies above $\sim$1~K focal heating. Including the full Plaksin gas-pocket pressure in the leaflet model would extend the cycle-averaged-cavitation pathway's validity above 1~MPa for high-intensity protocols. Replacing the deterministic Boltzmann gating with stochastic-channel Markov models would let the framework address experimental observations that are hard to capture in mean-field form, such as the firing-jitter and threshold-stochasticity reported in single-cell ultrasound recordings~\citep{kubanek2018ultrasound}. Direct calibration of the $K_A \cdot \alpha$ product against tissue- and cell-type-matched mechano-current recordings, paired with tissue-level elastography, would constrain the dominant absolute-rate uncertainty identified by the sensitivity analysis (\S\ref{sec:results_sensitivity}). Finally, validation against in-vivo datasets~\citep{tufail2010transcranial,deffieux2013lowintensity,legon2014transcranial} is deferred by design: credibly linking exposure to firing is a large, ideally multi-site programme that stays ongoing whatever tool is used, and a framework carrying exposure through to per-voxel firing is its necessary first step, not a competitor.

% =============================================================================
\section{Conclusion}
\label{sec:conclusion}

We presented an open-source framework that bridges full-wave acoustic propagation, viscoelastic shear-wave displacement, Pennes bioheat, and a multi-compartment Hodgkin--Huxley neuron carrying six interchangeable mechanism modules, namely direct Piezo1\,/\allowbreak\,K2P mechanosensitive-channel gating, Plaksin--Krasovitski intramembrane cavitation, Piezo1 inactivation with intracellular Ca\textsuperscript{2+}\,/\allowbreak\,SK coupling, TRPV1\,/\allowbreak\,TRPV4 thermosensors, an astrocytic TRPA1\,$\to$\allowbreak\,glutamate\,$\to$\allowbreak\,NMDA relay~\citep{oh2019ultrasonic}, and mechanosensitive synaptic / voltage-gated transmission~\citep{tyler2012mechanobiology,maneshi2017mechanical,kubanek2016ultrasound}. Every numerical parameter is classified by source (literature-anchored, calibrated, assumed, or derived), and a $\pm 25\%$ sensitivity sweep identifies the Piezo1 half-activation tension $\Thalf$ and the $K_A \cdot \alpha$ product as the two dominant uncertainties. The framework reproduces five published validations, namely AIS-first initiation~\citep{stuart1997action} ($\sim$9~$\mu$s lead), the duty-cycle excitation/inhibition direction across four (pressure, stimulus) regimes~\citep{plaksin2014intramembrane}, Piezo1 inactivation~\citep{coste2010piezo} ($\tau_{\mathrm{inact}}$ in [6, 18]\,ms), Legon-style high-pressure SK-mediated inhibition~\citep{legon2014transcranial}, and the TRPV1 Q\textsubscript{10}\,$\geq$\,20 floor~\citep{caterina1997capsaicin}. Applied to the Yaakub canonical demonstration on the Halle micro-CT skull (auto-pick aim at the left dACC, surface drive calibrated to reproduce Yaakub's 0.5\,MPa transcranial focal PNP, MI $= 0.69$, $I_{\mathrm{SPPA}} = 8.1$\,W/cm$^2$, $\Delta T_{\mathrm{focus}} = 233$\,mK with a skull-bone peak of 0.94\,K below the ITRUSST 2\,K bound and CEM43 $= 8.1{\times}10^{-4}$\,min, within the ITRUSST consensus safety envelopes), the framework predicts 5\,468 firing voxels with a peak rate of 300\,Hz and an iso-25 firing volume of 8\,523\,mm$^3$, substantially larger than the acoustic $-6$\,dB focal volume (161\,mm$^3$) and elongated along the beam axis by the CTX-500 focal aspect ratio. A real-field mechanism-comparison sweep across six strain levels ($\varepsilon \in [2\times 10^{-6}, 10^{-4}]$) and five scenarios shows that the dendrite-heavy three-compartment neuron model is necessary for firing at the canonical working point, where the single-compartment baseline remains subthreshold, the Piezo1-inactivation / Ca\textsuperscript{2+} / SK pathway reduces firing by $\sim 30$--$33\%$ at the canonical working point and $\sim 40$--$58\%$ at saturation, and intramembrane cavitation at sub-MPa carrier pressures and TRP at body temperature contribute marginally, as predicted. The astrocytic TRPA1\,$\to$\allowbreak\,glutamate\,$\to$\allowbreak\,NMDA relay adds a slow, non-Piezo contribution whose drive accumulates with cumulative sonication on-time rather than tracking the carrier, a complementary, weaker contributor in the theta-burst regime but the one mechanism whose duty $\times$ on-time dependence offers a distinguishing experimental signature. Mechanosensitive synaptic and voltage-gated transmission, by contrast, acts on the fast per-pulse synaptic and spike-generating machinery, multiplying excitation through tension-raised presynaptic release, stretch-enhanced NMDA current, and a hyperpolarising Na\textsuperscript{+}/Ca\textsuperscript{2+} activation shift, each exposed as a closed-form per-case scalar gain for inexpensive mechanism scoring. End-to-end run time is $\sim 17$~min per transcranial scenario on a single workstation. The framework provides a starting point for future studies that aim to discriminate among ultrasound-neuromodulation mechanism candidates by direct comparison against experimental ultrasound-neuromodulation sequences.

% =============================================================================
% Back matter (iopjournal.cls macros)
% =============================================================================

\funding{This work was supported in part by NIH grants R01-EB037345 and R01-EB036295, and by the Raynor Cerebellum Project.}

\data{The simulation code, validation scripts, and the per-parameter source classification are openly available under the Apache~2.0 license at \url{https://github.com/pinton-lab/acousto-mechano-neuron}.}

\clearpage
\appendix
\renewcommand{\thesection}{S\arabic{section}}
\setcounter{section}{0}
\renewcommand{\thetable}{S\arabic{table}}
\renewcommand{\thefigure}{S\arabic{figure}}
\renewcommand{\theequation}{S\arabic{equation}}
\setcounter{table}{0}\setcounter{figure}{0}\setcounter{equation}{0}
\section*{Supplementary information}
\addcontentsline{toc}{section}{Supplementary information}
\section{Channel library and per-channel parameters}
\label{sec:supp_channel_library}

The default channel library covers five mechanosensitive channels with parameters drawn from single-channel and patch-clamp recordings, listed in Table~\ref{tab:channels}.  The fifth, TRPA1, is the astrocytic sensor of the gliotransmitter relay detailed in \S\ref{sec:supp_astrocyte}, and its mechanogating constants are flagged \emph{Estimate} pending a direct tension-clamp fit.  Six neuron-type-specific channel profiles cover CNS targets relevant to tFUS, namely cortical pyramidal, cortical interneuron, hippocampal pyramidal, thalamic relay, cerebellar Purkinje, and striatal MSN.  Piezo1's causal role in ultrasound-evoked neural responses is supported by knockdown reducing ultrasound-evoked currents in cultured neurons~\citep{qiu2019piezo1}, conditional cortical knockout attenuating ultrasound-evoked motor responses in mice~\citep{zhu2023piezo1}, and pharmacological/genetic dissection identifying Piezo1 as a primary contributor to ultrasound-evoked Ca\textsuperscript{2+} accumulation in cortical neurons~\citep{yoo2022focused}.

\begin{table}[H]
    \caption{Default mechanosensitive channel parameters. $T_{1/2}$ is the Boltzmann half-activation tension. $\Delta A$ is the gate-area parameter in the Boltzmann exponent, namely the change in protein-occupied area between closed and open states rather than the channel's physical footprint. $\bar{g}$ is the population conductance density. $E_{\mathrm{rev}}$ is the reversal potential. $\tau_{\mathrm{act}}$ is the fixed activation time constant.  Piezo1 also carries an inactivation gate with $\tau_{\mathrm{inact}}=12$~ms, $T_{1/2,i}=4.5$~mN/m, and $\Delta A_i=15$~nm$^2$, defined in \S\ref{sec:supp_channel_extensions}.}
    \label{tab:channels}
    \footnotesize
    \begin{tabular}{@{}lccccccl@{}}
        \br
        Channel & Type & $T_{1/2}$ & $\Delta A$ & $\bar{g}$ & $E_{\mathrm{rev}}$ & $\tau_{\mathrm{act}}$ & Reference \\
                &      & (mN/m) & (nm$^2$) & (mS/cm$^2$) & (mV) & (ms) & \\
        \mr
        Piezo1 & Cation & 2.7 & 18.0 & 0.005 & 0     & 2.0 & \cite{coste2010piezo,lewis2015mechanical} \\
        TRAAK  & K\textsuperscript{+}     & 5.0 & 12.0 & 0.010 & $-$90 & 3.0 & \cite{brohawn2014mechanosensitivity} \\
        TREK-1 & K\textsuperscript{+}     & 3.5 & 14.0 & 0.020 & $-$90 & 5.0 & \cite{honore2007neuronal} \\
        TREK-2 & K\textsuperscript{+}     & 4.0 & 13.0 & 0.015 & $-$90 & 4.0 & \cite{bang2000trek2,honore2007neuronal} \\
        TRPA1  & Cation & 6.0 & 15.0 & 0.020 & 0     & 3.0 & \cite{oh2019ultrasonic,karashima2010trpa1} \\
        \br
    \end{tabular}
\end{table}

\begin{figure}[htbp]
    \centering
    \includegraphics[width=\textwidth]{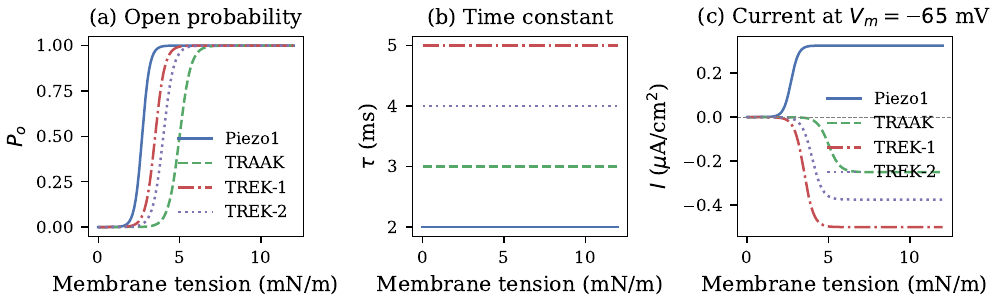}
    \caption{Biophysical properties of the four neuronal mechanosensitive channels Piezo1, TRAAK, TREK-1, and TREK-2.  The astrocytic TRPA1 sensor of Table~\ref{tab:channels} is characterised separately in \S\ref{sec:supp_astrocyte}.  (a)~Steady-state open probability $P_o(T)$ from the Boltzmann form.  (b)~Activation time constant $\tau$, set to the literature-reported single value per channel, with $\tau_{\mathrm{act}} = 2$, 3, 5, 4~ms for Piezo1, TRAAK, TREK-1, and TREK-2 as in Table~\ref{tab:channels}.  These tension-independent constants supersede an earlier bell-shaped $\tau(T)$ fallback retained in code only for channels where no $\tau_{\mathrm{act}}$ is supplied.  (c)~Macroscopic current density at resting potential $V_m = -65$~mV.  Positive values indicate the inward depolarising current carried by Piezo1, negative values the outward hyperpolarising current of the K2P channels TRAAK, TREK-1, and TREK-2.  The opposing polarities create a competition that determines the net effect of mechanical stimulation in any cell expressing both classes.}
    \label{fig:supp_channel_properties}
\end{figure}

% =============================================================================
\section{Channel-level extensions: inactivation gating}
\label{sec:supp_channel_extensions}

A channel that inactivates under sustained tension acquires a second Boltzmann gate $i$,
\begin{equation}
    P_o(T,t) = w(t)\,i(t),\qquad
    \frac{di}{dt} = \frac{i_\infty(T) - i}{\tau_{\mathrm{inact}}},\quad
    i_\infty(T) = \bigl(1 + \exp[\Delta A_i(T - T_{1/2,i})/k_BT]\bigr)^{-1},
    \label{eq:piezo_inact}
\end{equation}
so that the mechano current becomes $I_k = \bar{g}_k\,w_k\,i_k\,(V - E_{\mathrm{rev},k})$.  For Piezo1 we use $\tau_{\mathrm{inact}} = 12$~ms~\citep[Fig.~2]{coste2010piezo}, $T_{1/2,i} = 4.5$~mN/m, $\Delta A_i = 15$~nm$^2$.  Channels without inactivation, namely the K2P family in this library, hold $i$ at unity and reduce to the single-gate Boltzmann form.

The activation gate uses a fixed time constant $\tau_{\mathrm{act}}$ rather than the tension-dependent bell form $\tau_w(T) = \tau_0/[4P_o(1-P_o) + 0.01]$ inherited from earlier mechanosensitive HH formulations.  The bell form is discarded because it approaches $100\,\tau_0 \approx 200$~ms at tensions far from $\Thalf$, contradicting the $\sim$5~ms post-pulse closure reported in single-channel recordings~\citep{coste2010piezo}.  The fixed form used here is
\begin{equation}
    \tau_w(T) = \tau_{\mathrm{act}}.
    \label{eq:tau_act}
\end{equation}
For Piezo1 the literature-fit value is $\tau_{\mathrm{act}} = 2$~ms~\citep{coste2010piezo}.  Per-channel values for the K2P family are listed in Table~\ref{tab:channels}.

% =============================================================================
\section{Three-compartment dendrite-soma-AIS coupling}
\label{sec:supp_multicomp}

Compartments are coupled by axial conductances normalised to the soma area in the Pinsky--Rinzel manner~\citep{pinsky1994intrinsic}, giving
\begin{align}
    C_m\,\dot{V_s} &= -I_{\mathrm{ion},s} - I_{\mathrm{mech},s} + g_{ds}(V_d - V_s) + g_{sa}(V_a - V_s) + I_{\mathrm{stim},s}, \nonumber \\
    C_m\,\dot{V_d} &= -I_{\mathrm{ion},d} - I_{\mathrm{mech},d} + g_{ds}\tfrac{A_s}{A_d}(V_s - V_d) + I_{\mathrm{stim},d}, \nonumber \\
    C_m\,\dot{V_a} &= -I_{\mathrm{ion},a} - I_{\mathrm{mech},a} + g_{sa}\tfrac{A_s}{A_a}(V_s - V_a) + I_{\mathrm{stim},a},
    \label{eq:multicomp}
\end{align}
where the area ratios enforce charge conservation across each junction.  The literature-parameterised cortical-pyramidal preset uses $\bar{g}_{\mathrm{Na}} = 30, 120, 2000$~mS/cm$^2$ for dendrite, soma, AIS respectively~\citep{stuart1995initiation,mainen1996influence,kole2008action}, with $A_d : A_s : A_a = 5 : 1 : 0.05$, $g_{ds} = 1$~mS/cm$^2$ and $g_{sa} = 50$~mS/cm$^2$ of soma area.  AIS Na density is on the lower end of the 2000--5000~mS/cm$^2$ literature range.  We chose 2000 because explicit RK4 integration becomes the stability bottleneck above that density.

% =============================================================================
\section{NICE leaflet ODE}
\label{sec:supp_nice}

Following the SONIC reduction~\citep{lemaire2019understanding} of Plaksin \& Krasovitski~\citep{plaksin2014intramembrane}, we resolve the carrier-frequency leaflet displacement $Z$ with a damped oscillator augmented by two non-linear walls,
\begin{equation}
    m_{\mathrm{eff}}\,\ddot{Z} + 2\zeta m_{\mathrm{eff}}\omega_0\,\dot{Z} + m_{\mathrm{eff}}\omega_0^2\,Z = P_{\mathrm{ac}}(t) + P_{\mathrm{LJ}}(Z) + P_S(Z),
    \label{eq:nice_ode}
\end{equation}
\begin{equation}
\begin{aligned}
    P_{\mathrm{LJ}}(Z) &= \begin{cases}
        K_{\mathrm{LJ}}\bigl[(\delta_0/(\delta_0+Z))^5 - (\delta_0/(\delta_0+Z))^3\bigr] & Z < 0 \\
        0 & Z \geq 0,
    \end{cases} \\[6pt]
    P_S(Z) &= \begin{cases}
        -2\sigma_{\mathrm{eff}}\bigl[1/(\delta_0+Z) - 1/\delta_0 + Z/\delta_0^2\bigr] & Z > 0 \\
        0 & Z \leq 0,
    \end{cases}
\end{aligned}
\end{equation}
with $K_{\mathrm{LJ}} = \alpha_{\mathrm{LJ}} m_{\mathrm{eff}} \omega_0^2 \delta_0$ and $\sigma_{\mathrm{eff}} = \alpha_S m_{\mathrm{eff}} \omega_0^2 \delta_0^2 / 2$ at default values $\alpha_{\mathrm{LJ}} = 0.01$ and $\alpha_S = 1.0$.  Here $P_{\mathrm{LJ}}$ resists leaflet collision at $Z < 0$ and $P_S$ resists over-expansion at $Z > 0$.  The default oscillator parameters follow the Plaksin and SONIC values, with $\delta_0 = 2$\,nm, $m_{\mathrm{eff}} = 0.16$\,kg/m$^2$, $\omega_0 = 2\pi f_c$ at carrier frequency $f_c = 500$\,kHz, and damping ratio $\zeta = 0.1$ giving $Q \approx 5$.  The instantaneous capacitance is $C_m(Z) = C_{m,0}\,\delta_0/(\delta_0+Z)$, and the cycle-averaged perturbation is computed by integrating Eq.~\ref{eq:nice_ode} with RK4 at 256 sub-steps per carrier cycle over 8 cycles, then averaging $C_m(Z(t)) - C_{m,0}$ over the final cycle.  At 1~MPa drive amplitude $|Z| < 3\delta_0$ and $\langle \Delta C_m \rangle / C_{m,0} \approx 0.55$, qualitatively matching~\citep[Fig.~2]{plaksin2014intramembrane}.  The cycle-averaged capacitance enters the per-compartment HH equation as
\begin{equation}
    [C_m + \langle \Delta C_m \rangle(P_{\mathrm{env}})] \dot V + V \cdot \tfrac{d\langle \Delta C_m \rangle}{dt}
    = -I_{\mathrm{Na}} - I_K - I_L - I_{\mathrm{mech}} + I_{\mathrm{axial}} + I_{\mathrm{stim}},
    \label{eq:hh_nice}
\end{equation}
the displacement-current term being non-zero only at envelope ramps.

% =============================================================================
\section{Calcium pool and SK channel}
\label{sec:supp_calcium}

When the Ca\textsuperscript{2+} pathway is enabled, every compartment maintains an intracellular Ca\textsuperscript{2+} concentration that obeys
\begin{equation}
    \frac{d[\mathrm{Ca}^{2+}]_i}{dt} = -k_{\mathrm{flux}}\sum_k f_{\mathrm{Ca},k}\,I_k - \frac{[\mathrm{Ca}^{2+}]_i - [\mathrm{Ca}^{2+}]_{\mathrm{rest}}}{\tau_{\mathrm{Ca}}},
    \label{eq:ca_pool}
\end{equation}
with $f_{\mathrm{Ca},k}$ the per-channel Ca\textsuperscript{2+} permeability fraction, equal to 0.10 for Piezo1, $\tau_{\mathrm{Ca}} = 50$~ms~\citep{mainen1996influence,borggraham1999interpretations}, $[\mathrm{Ca}^{2+}]_{\mathrm{rest}} = 0.05~\mu$M, and $k_{\mathrm{flux}} = 0.5~\mu$M$\cdot$cm$^2$/($\mu$A$\cdot$ms) lumping Ca\textsuperscript{2+} valence, sub-membrane shell volume, and bulk-cytosol buffering.  An SK-type Ca\textsuperscript{2+}-activated K\textsuperscript{+} channel population on the soma adds a Hill-gated counter-current,
\begin{equation}
    I_{\mathrm{KCa}} = \bar{g}_{\mathrm{KCa}}\,
        \frac{[\mathrm{Ca}^{2+}]_i^{n_H}}{[\mathrm{Ca}^{2+}]_i^{n_H} + [\mathrm{Ca}^{2+}]_{1/2}^{n_H}}\,(V - E_K),
    \label{eq:kca}
\end{equation}
with $[\mathrm{Ca}^{2+}]_{1/2} = 0.5~\mu$M~\citep{kohler1996sk}, $n_H = 4$, $\bar{g}_{\mathrm{KCa}} = 0.3$~mS/cm$^2$, $E_K = -100$~mV.

% =============================================================================
\section{TRPV1 / TRPV4 thermosensors}
\label{sec:supp_trp}

When the thermosensitive pathway is enabled, each compartment carries two TRP channels, TRPV1 and TRPV4, whose open probability is a Boltzmann function of the local absolute temperature,
\begin{equation}
    P_{o,j}(T) = \bigl(1 + \exp[(T_{1/2,j} - T)/k_{T,j}]\bigr)^{-1},
    \qquad j \in \{\mathrm{TRPV1}, \mathrm{TRPV4}\}.
    \label{eq:trp_supp_boltzmann}
\end{equation}
The macroscopic current density per channel follows the same $(V - E)$ convention as the rest of the model,
\begin{equation}
    I_{\mathrm{TRP},j} = \bar{g}_j\,P_{o,j}(T)\,(V_m - E_{\mathrm{rev},j}),
    \label{eq:trp_supp_current}
\end{equation}
entering the HH equation as $-I_{\mathrm{TRP},j}$.  The temperature $T$ is the local absolute temperature delivered by the Pennes bioheat solver in the main text, $T = T_a + \Delta T(\mathbf{x})$.

Default per-channel parameters are listed in Table~\ref{tab:trp}.  TRPV1's half-activation $T_{1/2} = 43\,^\circ$C and the $Q_{10} \geq 20$ heat-activation floor are literature-anchored to the Caterina capsaicin-receptor characterisation~\citep{caterina1997capsaicin}.  The Boltzmann steepness $k_{T,\mathrm{TRPV1}} = 1.5\,^\circ$C is calibrated so that the simulated $Q_{10} = P_o(T+10)/P_o(T)$ at body temperature comfortably exceeds the Caterina floor.  At 37\,$^\circ$C the model gives $P_o = 0.018$ and at 47\,$^\circ$C $P_o = 0.94$, yielding $Q_{10} \approx 52$.  TRPV4's half-activation $T_{1/2} = 30\,^\circ$C falls within the 27--35\,$^\circ$C literature range and produces a tonic depolarising drive with $P_o \approx 0.91$ at body temperature.  Only the temperature-gated branch of TRPV4 is modelled here.  The channel's reported mechanosensitive component is intentionally omitted so that the mechanical drive into the neuron remains routed through Piezo1 and TREK-1 alone, keeping the mechanical-pathway parameters jointly identifiable.

\begin{table}[H]
    \caption{Default thermosensitive-channel parameters.  $T_{1/2}$, $k_T$, $\bar{g}$, $E_{\mathrm{rev}}$ are the Boltzmann half-activation temperature, steepness, conductance density, and reversal potential.  Source-classification entries appear in \S\ref{sec:supp_provenance}.}
    \label{tab:trp}
    \footnotesize
    \begin{tabular}{@{}lccccl@{}}
        \br
        Channel & $T_{1/2}$ & $k_T$ & $\bar{g}$ & $E_{\mathrm{rev}}$ & Reference \\
                & ($^\circ$C) & ($^\circ$C) & (mS/cm$^2$) & (mV) & \\
        \mr
        TRPV1 & 43 & 1.5 & 0.05 & $+10$ & \cite{caterina1997capsaicin} \\
        TRPV4 & 30 & 3.0 & 0.02 & $\phantom{+}0$ & \cite{guler2002trpv4} \\
        \br
    \end{tabular}
\end{table}

% =============================================================================
\section{Astrocytic TRPA1 \texorpdfstring{$\to$}{->} glutamate \texorpdfstring{$\to$}{->} NMDA relay}
\label{sec:supp_astrocyte}

The astrocytic pathway of main text \S\ref{sec:methods_astrocyte} is implemented as an optional single-pool relay on the shared neuron, the same complexity tier as the calcium/SK and NICE modules.  TRPA1 reuses the Boltzmann-in-tension gate of the mechanosensitive-channel library from Table~\ref{tab:channels}, with $T_{1/2} = 6.0$~mN/m and $\Delta A = 15$~nm$^2$, and a Ca\textsuperscript{2+}-permeability fraction $f_{\mathrm{Ca}} = 0.18$ corresponding to $P_{\mathrm{Ca}}/P_{\mathrm{Na}} \approx 5$--$8$ from Karashima et al.~\citep{karashima2010trpa1}.  Because no direct tension-clamp characterisation of TRPA1 mechanogating exists, $T_{1/2}$ and $\Delta A$ are classified \emph{Estimate} in \S\ref{sec:supp_provenance}, Table~\ref{tab:prov_trpa1}.  They are placed between the Piezo1 value of 2.7~mN/m and the TRAAK value of 5.0~mN/m as a best guess.

The astrocytic Ca\textsuperscript{2+} pool obeys a single-pool tracker driven by the TRPA1 Ca\textsuperscript{2+} current,
\begin{equation}
    \frac{d[\mathrm{Ca}^{2+}]}{dt} = k_{\mathrm{flux}}\,\bar{g}_{\mathrm{TRPA1}}\,P_o(T)\,f_{\mathrm{Ca}}\,(E_{\mathrm{rev}} - V_{\mathrm{astro}}) - \frac{[\mathrm{Ca}^{2+}] - [\mathrm{Ca}^{2+}]_{\mathrm{rest}}}{\tau_{\mathrm{Ca}}},
\end{equation}
with astrocyte resting potential $V_{\mathrm{astro}} = -80$~mV from Bazargani \& Attwell~\citep{bazargani2016astrocyte}, $[\mathrm{Ca}^{2+}]_{\mathrm{rest}} = 0.05~\mu$M, and a slow decay $\tau_{\mathrm{Ca}} = 300$~ms reflecting the hundreds-of-ms-to-seconds timescale of astrocytic Ca\textsuperscript{2+} signalling.  The conversion factor $k_{\mathrm{flux}}$ is \emph{Calibrated} so that the pool peaks $\sim 1~\mu$M at full ultrasound drive, the physiological gliotransmission regime.  Glutamate release is a Hill function of astrocytic Ca\textsuperscript{2+},
\begin{equation}
    [\mathrm{glu}] = G_{\max}\,\frac{[\mathrm{Ca}^{2+}]^{n}}{[\mathrm{Ca}^{2+}]^{n} + [\mathrm{Ca}^{2+}]_{1/2}^{\,n}},
\end{equation}
with $[\mathrm{Ca}^{2+}]_{1/2} = 0.3~\mu$M and $n = 3$ for Ca\textsuperscript{2+}-dependent gliotransmission from Parpura \& Haydon~\citep{parpura2000gliotransmission}, and $G_{\max} = 10~\mu$M.  Glutamate drives a neuronal NMDA conductance combining saturating binding at EC50 $K_{\mathrm{glu}} = 2~\mu$M from Patneau \& Mayer with the Jahr--Stevens~\citep{jahr1990voltage} voltage-dependent Mg\textsuperscript{2+} block,
\begin{equation}
    g_{\mathrm{NMDA}}(V, [\mathrm{glu}]) = \bar{g}_{\mathrm{NMDA}}\,\frac{[\mathrm{glu}]}{[\mathrm{glu}] + K_{\mathrm{glu}}}\,\frac{1}{1 + [\mathrm{Mg}^{2+}]\,e^{-V/16.13}/3.57},
\end{equation}
at $[\mathrm{Mg}^{2+}] = 1$~mM and $E_{\mathrm{rev}} = 0$~mV.  Setting $\bar{g}_{\mathrm{NMDA}} = 0$ recovers the AP5 control, where the astrocytic Ca\textsuperscript{2+} and glutamate transients are unchanged but no neuronal excitation results.  Setting $\bar{g}_{\mathrm{TRPA1}} = 0$ recovers the TRPA1-knockout or antagonist control, abolishing the relay entirely.  Both are enforced by unit tests.

For inexpensive per-case mechanism scoring, for example adding the pathway as a column in a downstream parameter sweep without re-running the forward model, the relay reduces to a closed-form scalar drive built from its causal chain.  The intracellular Ca\textsuperscript{2+} accumulates with duty cycle $D$ and cumulative on-time $t_{\mathrm{on}}$, which sets the glutamate concentration $[\mathrm{glu}]$ through the Hill function above.  The astrocytic-NMDA excitation drive is then the glutamate-occupancy fraction times the NMDA Mg\textsuperscript{2+}-unblock factor,
\begin{equation}
\begin{aligned}
    [\mathrm{Ca}^{2+}] &= [\mathrm{Ca}^{2+}]_{\mathrm{rest}} + k_{\mathrm{flux}}\,I_{\mathrm{Ca}}(T_{\max})\,D\,\tau_{\mathrm{Ca}}\bigl(1 - e^{-t_{\mathrm{on}}/\tau_{\mathrm{Ca}}}\bigr), \\[6pt]
    \mathrm{drive}(T_{\max},\, D,\, t_{\mathrm{on}}) &= \frac{[\mathrm{glu}]}{[\mathrm{glu}] + K_{\mathrm{glu}}}\cdot\frac{1}{1 + [\mathrm{Mg}^{2+}]\,e^{-V_{\mathrm{ref}}/16.13}/3.57},
\end{aligned}
\end{equation}
which is monotone in tension and, through the $(1 - e^{-t_{\mathrm{on}}/\tau_{\mathrm{Ca}}})$ accumulation term, grows with duty cycle and cumulative on-time.  This duty-cumulative dependence distinguishes the slow pathway from the per-pulse channel mechanisms, as shown on the canonical field below.  Per-parameter source classification for TRPA1, the astrocytic relay, and the NMDA receptor appears in \S\ref{sec:supp_provenance}.

\subsection{Astrocytic relay on the canonical Halle\,/\allowbreak\,dACC field}
\label{sec:supp_astrocyte_results}

\begin{figure}[htbp]
    \centering
    \includegraphics[width=\textwidth]{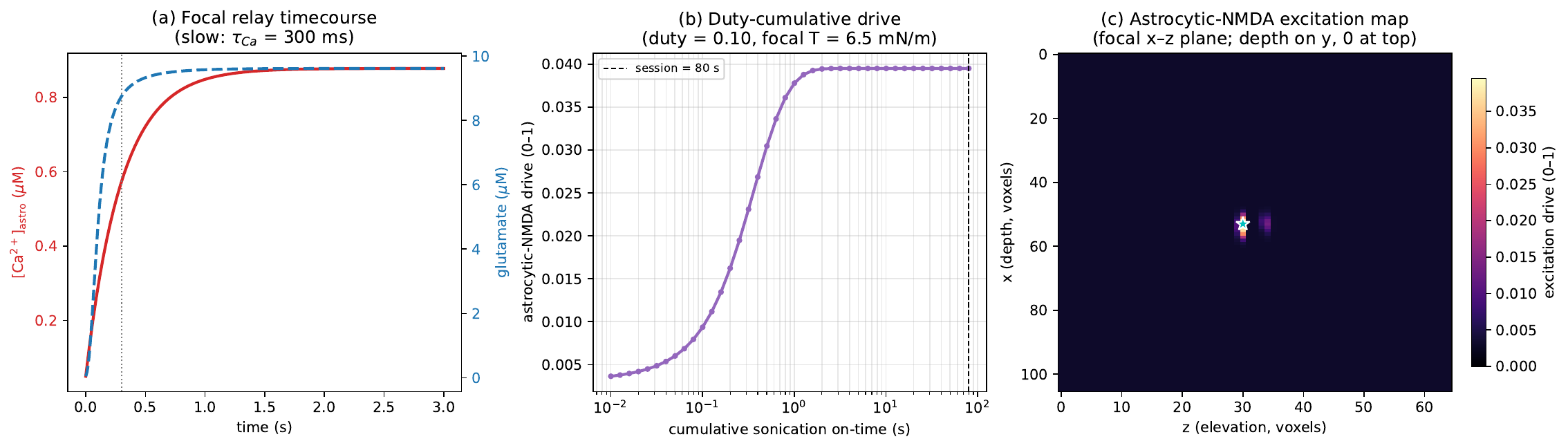}
    \caption{Astrocytic TRPA1 $\to$ Ca\textsuperscript{2+} $\to$ glutamate $\to$ NMDA relay on the canonical Halle\,/\allowbreak\,dACC field.  (a)~Focal-voxel relay timecourse, showing the astrocytic Ca\textsuperscript{2+} (red) and glutamate (blue dashed) rising over hundreds of ms with $\tau_{\mathrm{Ca}} = 300$~ms, the defining slowness of the pathway.  (b)~Astrocytic-NMDA excitation drive versus cumulative sonication on-time.  The slow Ca\textsuperscript{2+} accumulation makes the drive grow with duty $\times$ on-time and saturate only after several seconds (dashed line marks the protocol session duration), unlike the per-pulse channel mechanisms.  (c)~Per-voxel astrocytic-NMDA excitation map on the focal $x$--$z$ plane, with depth $x$ on the vertical axis and 0 at the top (cyan star marks the focal voxel).}
    \label{fig:supp_astrocyte_results}
\end{figure}

Scoring the astrocytic pathway on the same per-voxel peak-tension field that drives the canonical run of main text \S\ref{sec:results_astrocyte} exposes a qualitatively different response profile from the per-pulse channel mechanisms, as shown in Fig.~\ref{fig:supp_astrocyte_results}.  At the focal voxel, where $T_{\max} = 6.52$~mN/m sits just above the TRPA1 mechanogating estimate $T_{1/2} = 6.0$~mN/m, the astrocytic Ca\textsuperscript{2+} pool charges over hundreds of ms to $\sim 0.9~\mu$M and drives glutamate release to $\sim 9~\mu$M (Fig.~\ref{fig:supp_astrocyte_results}a), and a short focal forward simulation gives an NMDA-mediated sub-threshold depolarisation of $\sim 11$~mV.  Because the relay integrates the duty-gated drive over its slow Ca\textsuperscript{2+} time constant, the astrocytic-NMDA drive rises with cumulative sonication on-time and only saturates after several seconds, as in Fig.~\ref{fig:supp_astrocyte_results}b.  This duty $\times$ on-time dependence is absent from the 20~ms-window channel mechanisms and is the distinguishing experimental signature of this pathway.  At the canonical exposure the astrocytic drive stays below the 0.05 recruitment cut-off everywhere, with a peak drive of $\approx 0.04$ and no voxels recruited on the relay's own, as shown in Fig.~\ref{fig:supp_astrocyte_results}c.  It is weaker and slower than the direct Piezo1 pathway at this sub-MPa carrier, consistent with the astrocytic account being a complementary rather than dominant contributor in the theta-burst regime.  Two calibration checks reproduce the Oh et al.\ controls of \S\ref{sec:supp_verif_astrocyte}.  Zeroing the TRPA1 conductance, the knockout or antagonist condition, abolishes the relay output, and zeroing the NMDA conductance, the AP5 condition, abolishes neuronal excitation while the astrocytic Ca\textsuperscript{2+} and glutamate transients persist.

% =============================================================================
\section{Mechanosensitive synaptic and voltage-gated transmission}
\label{sec:supp_synaptic}

The sixth mechanism of main text \S\ref{sec:methods_synaptic} extends mechanosensitivity from the single post-synaptic membrane to synaptic transmission and the spike-generating conductances, bundling three literature-grounded effects on the shared neuron.  Each effect reduces to the unmodified model at zero tension and exposes a closed-form per-case scalar gain for inexpensive mechanism scoring.

\paragraph{Presynaptic release (Tyler).}  Presynaptic membrane tension raises the vesicle release probability~\citep{tyler2008remote,tyler2012mechanobiology} along a saturating sigmoid normalised so $P_r(0) = P_{r,0}$ exactly and $P_r(\infty) = P_{r,\max}$,
\begin{equation}
    P_r(T) = P_{r,0} + (P_{r,\max} - P_{r,0})\,\frac{\sigma\!\left(\tfrac{T - T_{1/2}^{\mathrm{pre}}}{k_{\mathrm{pre}}}\right) - \sigma\!\left(\tfrac{-T_{1/2}^{\mathrm{pre}}}{k_{\mathrm{pre}}}\right)}{1 - \sigma\!\left(\tfrac{-T_{1/2}^{\mathrm{pre}}}{k_{\mathrm{pre}}}\right)},
\end{equation}
with $\sigma$ the logistic function and defaults $P_{r,0} = 0.10$, $P_{r,\max} = 0.60$, $T_{1/2}^{\mathrm{pre}} = 4.0$~mN/m, and $k_{\mathrm{pre}} = 1.5$~mN/m.  Release feeds a first-order cleft glutamate pool $\dot{[\mathrm{glu}]} = (G_s/\tau_{\mathrm{glu}})\,P_r(T)\,\mathbb{1}_{\mathrm{US}} - [\mathrm{glu}]/\tau_{\mathrm{glu}}$, so the steady cleft glutamate under sustained sonication is $G_s\,P_r(T)$, with $G_s = 30~\mu$M at $P_r = 1$ and $\tau_{\mathrm{glu}} = 5$~ms.  Here $G_s$ and the AMPA conductance are \emph{Calibrated} so that the resting baseline-$P_r$ state is sub-threshold and ultrasound-driven release brings the post neuron to firing.

\paragraph{Post-synaptic NMDA stretch (Maneshi).}  Membrane stretch raises NMDA open probability at fixed glutamate and voltage~\citep{maneshi2017mechanical}.  The NMDA conductance of main-text Eq.~\ref{eq:nmda} is multiplied by $f_{\mathrm{stretch}}(T)$ from main-text Eq.~\ref{eq:nmda_stretch}, normalised so $f_{\mathrm{stretch}}(0) = 1$ exactly and $f_{\mathrm{stretch}}(\infty) = 1 + \alpha_{\mathrm{NMDA}}$.  With $\alpha_{\mathrm{NMDA}} = 1$ the NMDA current roughly doubles at saturating stretch, as reported by Maneshi et al., reached with $T_{1/2}^{\mathrm{s}} = 5.0$~mN/m and $k_{\mathrm{s}} = 1.5$~mN/m.  Setting $\alpha_{\mathrm{NMDA}} = 0$ recovers the baseline NMDA receptor.  The post-synaptic glutamate also drives a low-affinity AMPA conductance with EC50 $K_{\mathrm{AMPA}} \approx 300~\mu$M.

\paragraph{Voltage-gated Na\textsuperscript{+}/Ca\textsuperscript{2+} shift (Kubanek).}  Kubanek et al.~\citep{kubanek2016ultrasound} patch-clamped Na\textsuperscript{+}/Ca\textsuperscript{2+}/K\textsuperscript{+} currents under ultrasound and found a shift of the activation voltage dependence that raises the inward current.  We model a tension-proportional hyperpolarising shift of the activation half-voltage, main-text Eq.~\ref{eq:nav_shift}, applied to the activation gate of the Hodgkin--Huxley Na\textsuperscript{+} current, so the net effect is a current \emph{increase}.  The inactivation gate is left unshifted to avoid the partial cancellation that would obscure it.  The sensitivity $k_{\mathrm{mechano}}$ is \emph{Calibrated} because its magnitude is not directly tabulated by Kubanek et al.  The canonical $k_{\mathrm{mechano}} = 0.3$~mV per mN/m yields a $\sim$mV-scale shift and a tens-of-percent current change at a few mN/m, and $k_{\mathrm{mechano}} = 0$ recovers the unmodified neuron exactly.  The corresponding scalar gain is the ratio of the activation open fraction $m_\infty$ at a near-threshold reference voltage with and without the shift.

\paragraph{Scalars and provenance.}  The three per-case scalars, namely presynaptic $P_r(T)/P_r(0)$, post-synaptic $f_{\mathrm{stretch}}(T)$, and the Na\textsuperscript{+} activation-shift current ratio, are monotone non-decreasing in tension and equal to $1$ at zero tension, so their product is a single multiplicative excitation gain usable as a downstream sweep column.  Per-parameter source classification appears in \S\ref{sec:supp_provenance}, in Tables~\ref{tab:prov_mechanonav} and~\ref{tab:prov_mechanosynapse}, with the NMDA stretch parameters $\alpha_{\mathrm{NMDA}}$, $T_{1/2}^{\mathrm{s}}$, and $k_{\mathrm{s}}$ in Table~\ref{tab:prov_nmdareceptor}.

% =============================================================================
\section{Verification}
\label{sec:supp_verification}

This section reports the per-anchor numerical experiments behind the validation summary in the main text, Section~3.1.  Every experiment is implemented as a unit test in the open-source repository, so the figures below are reproducible by re-running the test suite.

\subsection{Single-compartment Hodgkin--Huxley baseline}
\label{sec:supp_verif_hh}

This subsection is the regression check at the original Hodgkin--Huxley squid parameterisation~\citep{hodgkin1952quantitative}, used as a baseline before any temperature-scaling or cortical-pyramidal substitution.  The Pospischil cortical regular-spiking f--I curve at body temperature, the parameterisation actually used for the manuscript's transcranial runs, is validated separately in \S\ref{sec:supp_verif_pospischil_fi}, and the Q\textsubscript{10}-scaling bridge between the two is in \S\ref{sec:supp_verif_q10}.

\begin{figure}[htbp]
    \centering
    \includegraphics[width=0.6\textwidth]{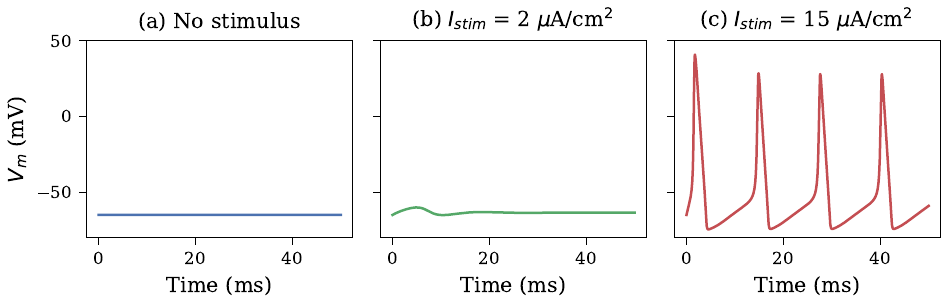}
    \caption{Single-compartment Hodgkin--Huxley validation at $T = 6.3\,^\circ$C.  (a)~No stimulus, giving a stable resting potential.  (b)~Subthreshold stimulus at $I_{\mathrm{stim}} = 2$~$\mu$A/cm$^2$, giving passive depolarisation with no spikes.  (c)~Suprathreshold stimulus at $I_{\mathrm{stim}} = 15$~$\mu$A/cm$^2$, giving repetitive firing at 78~Hz.}
    \label{fig:supp_hh_baseline}
\end{figure}

Figure~\ref{fig:supp_hh_baseline} confirms the resting-state stability and threshold behaviour of the single-compartment Hodgkin--Huxley neuron at the original 6.3\,$^\circ$C parameterisation.  With no stimulus, the membrane potential remains at rest with $|\Delta V| < 0.01$~mV over 50~ms.  A subthreshold current of $I_{\mathrm{stim}} = 2$~$\mu$A/cm$^2$ produces a passive depolarisation without spike generation.  A suprathreshold current of $I_{\mathrm{stim}} = 15$~$\mu$A/cm$^2$ elicits four action potentials in 50~ms at 78~Hz, with peak voltage $\approx 41$~mV and first-spike latency 1.5~ms.

\begin{figure}[htbp]
    \centering
    \includegraphics[width=0.51\textwidth]{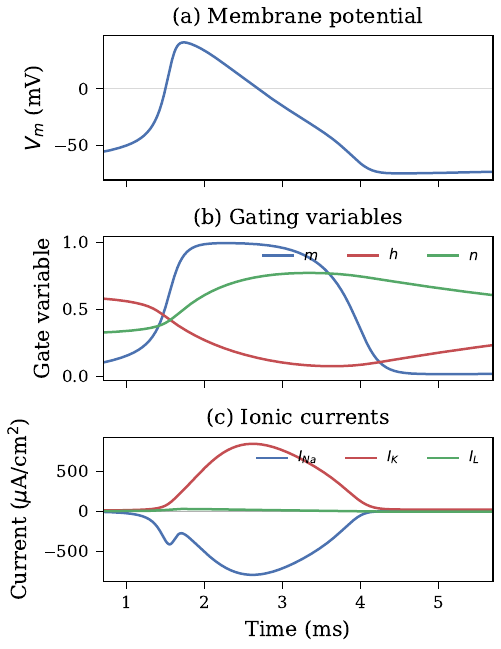}
    \caption{Action potential dynamics at $T = 6.3\,^\circ$C, $I_{\mathrm{stim}} = 15$~$\mu$A/cm$^2$, zoomed onto a single action potential.  (a)~Membrane potential trace.  (b)~HH gating variables $m, h, n$.  (c)~Ionic currents in the standard biophysics convention where outward is positive.  Na is inward and negative during depolarisation, K is outward and positive during repolarisation, and together they balance $-C_m\,dV/dt + I_{\mathrm{stim}}$.}
    \label{fig:supp_gate_dynamics}
\end{figure}

Figure~\ref{fig:supp_gate_dynamics} resolves the gating-variable trajectories during a single action potential.  The canonical sequence is recovered, namely rapid Na$^+$ activation as $m$ rises from $\sim$0.05 to near unity, Na$^+$ inactivation as $h$ drops to $\sim$0.07, and delayed K$^+$ activation as $n$ rises to $\sim$0.77, producing the characteristic overshoot/undershoot of the action-potential waveform.

\subsection{Q\textsubscript{10} temperature scaling}
\label{sec:supp_verif_q10}

\begin{figure}[htbp]
    \centering
    \includegraphics[width=\textwidth]{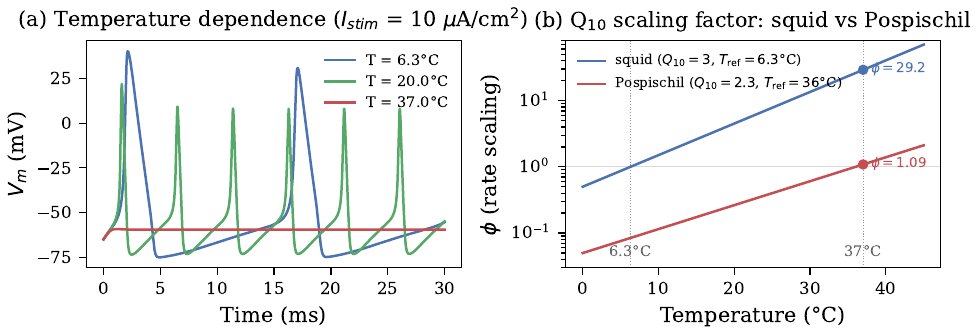}
    \caption{Q\textsubscript{10} temperature scaling.  (a)~Voltage responses to constant current injection at three temperatures using the squid HH model.  At 37\,$^\circ$C the accelerated kinetics with $\phi = 29.2$ cause depolarisation block.  (b)~Q\textsubscript{10} scaling factor $\phi$ versus temperature for the squid and Pospischil cortical models.}
    \label{fig:supp_q10}
\end{figure}

Figure~\ref{fig:supp_q10} contrasts the squid HH parameterisation, which enters depolarisation block at body temperature, with the Pospischil cortical regular-spiking model used throughout this paper.  The cortical model uses $Q_{10} = 2.3$ with $T_{\mathrm{ref}} = 36\,^\circ$C, giving $\phi = 1.087$ at 37\,$^\circ$C and preserving normal firing.  The squid model gives $\phi = 29.2$ at 37\,$^\circ$C and ceases to fire.  Use of the cortical model is what makes bioheat-coupled body-temperature simulations meaningful, because temperature perturbations of $\Delta T \sim 1$--$10$~K then produce modest, physiologically interpretable changes in $\phi$ rather than a complete loss of excitability.  The $Q_{10}$ values referred to here are HH rate-constant scaling factors~\citep{hodgkin1952quantitative}.  The separate TRPV1 thermal-activation $Q_{10} \geq 20$ floor of Caterina~\citep{caterina1997capsaicin} used in \S\ref{sec:supp_trp} is a channel-gate-ratio quantity over a 10\,$^\circ$C window, not a rate-constant scaling.

\subsection{Mechanosensitive channel coupling}
\label{sec:supp_verif_mechano}

\begin{figure}[htbp]
    \centering
    \includegraphics[width=\textwidth]{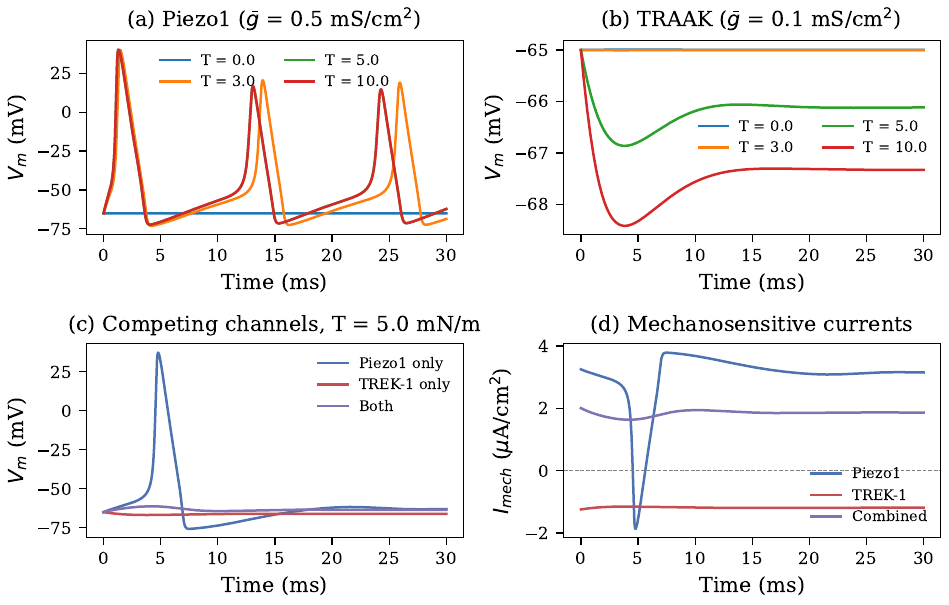}
    \caption{Mechanosensitive channel coupling under sustained tension, using the squid HH parameterisation at $T = 6.3\,^\circ$C.  See \S\ref{sec:supp_verif_q10} for the Q$_{10}$ scaling that motivates this choice.  (a)~Piezo1, a cation channel with $E_{\mathrm{rev}} = 0$, at four tension levels with $\bar{g} = 0.5$~mS/cm$^2$.  Saturating activation produces repetitive firing.  (b)~TRAAK, a K\textsuperscript{+} channel with $E_{\mathrm{rev}} = -90$, at four tension levels with $\bar{g} = 0.1$~mS/cm$^2$.  The membrane is driven below rest.  (c--d)~Co-expression of Piezo1 and TREK-1 at equal conductance densities of $\bar{g} = 0.05$~mS/cm$^2$ each.  The opposing polarities partially cancel.}
    \label{fig:supp_mechano}
\end{figure}

Figure~\ref{fig:supp_mechano} shows the distinct effects of the depolarising Piezo1 and the hyperpolarising TRAAK channel activation on neural dynamics.  At Piezo1 conductance $\bar{g} = 0.5$~mS/cm$^2$ and tension just above $\Thalf$ at 3~mN/m, the channel population drives a depolarising current of $\sim$26~$\mu$A/cm$^2$ at rest, well above the firing threshold, producing three action potentials.  At higher tensions the response saturates with $P_o > 0.99$.  TRAAK at $\bar{g} = 0.1$~mS/cm$^2$ and $T = 5$~mN/m drives the membrane 1.1~mV below rest.  At 10~mN/m the hyperpolarisation grows to 2.3~mV.  Co-expression of both classes at comparable conductance densities reproduces the partial cancellation expected from the opposing reversal potentials.

When Piezo1 and TREK-1 are co-expressed at equal conductance densities of $\bar{g} = 0.05$~mS/cm$^2$ each, the net mechanosensitive current at $T = 5$~mN/m is $\sim 1.9\,\mu$A/cm$^2$, which is depolarising, producing a steady-state membrane potential of $-63.6$~mV.  Despite both channels being near full activation, the net effect is modest because the Piezo1 driving force $E_{\mathrm{rev}} - V_m = +65$~mV and the TREK-1 driving force of $-25$~mV partially cancel.  At the default library conductances of Table~\ref{tab:channels}, TREK-1 at $\bar{g} = 0.020$~mS/cm$^2$ actually dominates over Piezo1 at $\bar{g} = 0.005$~mS/cm$^2$.  The maximum Piezo1 depolarising current is $0.325\,\mu$A/cm$^2$ while the maximum TREK-1 hyperpolarising current is $0.500\,\mu$A/cm$^2$.  This predicts that, at default densities, cortical pyramidal neurons carrying Piezo1 with the TREK-1 profile would be net inhibited by membrane tension, consistent with experimental observations of mixed excitatory and inhibitory responses to focused ultrasound~\citep{kubanek2018ultrasound}.

\subsection{Strain-to-tension conversion}
\label{sec:supp_verif_strain}

\begin{figure}[htbp]
    \centering
    \includegraphics[width=\textwidth]{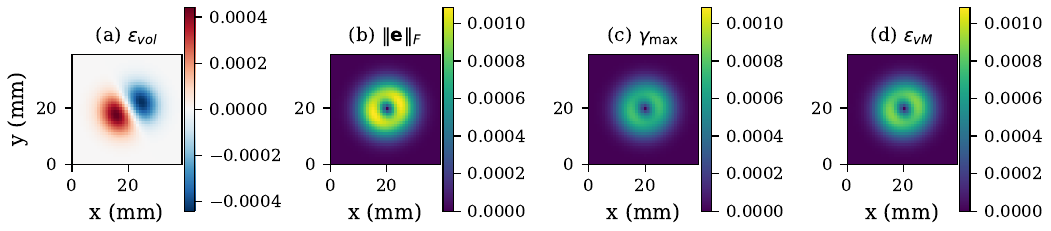}
    \caption{Strain invariants in a central $x$--$y$ slice from a Gaussian displacement field.  The volumetric strain captures dilation/compression.  The deviatoric norm, maximum shear, and von~Mises equivalent highlight distortional deformation, which is the input to the membrane-tension model.}
    \label{fig:supp_strain_invariants}
\end{figure}

\begin{figure}[htbp]
    \centering
    \includegraphics[width=\textwidth]{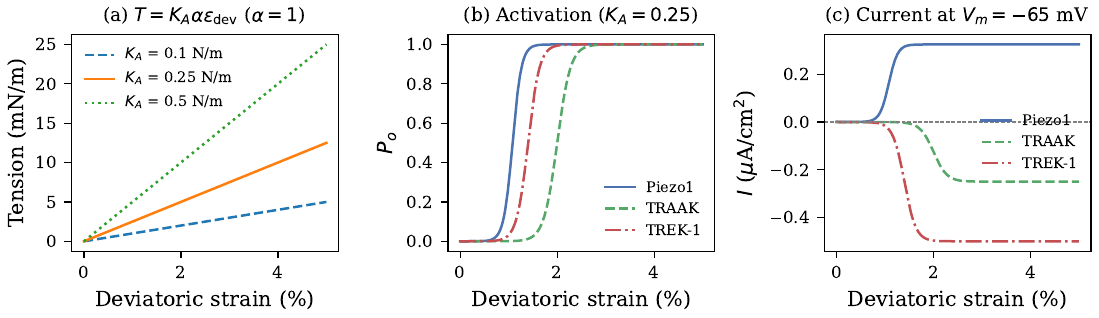}
    \caption{Strain-to-tension and channel response.  (a)~Membrane tension as a function of deviatoric strain for three values of the area-expansion modulus $K_A$, with the multi-scale coupling factor $\alpha = 1$ used didactically so the bare $K_A$ sensitivity is visible, since $\alpha$ enters as a multiplicative prefactor and is the same for all three curves.  (b)~Channel open probability versus deviatoric strain at $K_A = 0.25$~N/m.  (c)~Macroscopic current density at $V_m = -65$~mV for the same conditions.}
    \label{fig:supp_strain_to_tension}
\end{figure}

Figure~\ref{fig:supp_strain_invariants} illustrates the standard scalar strain invariants computed from a synthetic Gaussian-displacement field, showing the distinct spatial signatures of volumetric vs.\ deviatoric strain.  Figure~\ref{fig:supp_strain_to_tension} reports the strain-to-tension conversion of main text Eq.~\ref{eq:tension} for three values of the bilayer area-expansion modulus $K_A$, with the corresponding open probability and current at resting potential.  The equivalent strain that puts a typical Piezo1 at $\Thalf$ is in the $10^{-3}$ range under $K_A \approx 0.25$~N/m and unit coupling.  Transcranial-shear-FDTD strain fields reach this threshold over a millimetre-scale focal region.

\subsubsection{Origin and calibration of the multi-scale coupling factor $\alpha$}
\label{sec:supp_alpha}

The strain-to-tension conversion of main-text Eq.~\ref{eq:tension},
\begin{equation}
T = K_A \cdot \alpha \cdot \varepsilon_{\mathrm{eq}},
\label{eq:supp_tension_alpha}
\end{equation}
introduces a dimensionless gain $\alpha$ that bridges the FDTD voxel-scale tissue strain $\varepsilon_{\mathrm{eq}}$ and the nanometre-scale bilayer areal strain experienced by a mechanosensitive channel.  $\alpha$ is a calibrated parameter, classified as such in \S\ref{sec:supp_provenance}, not a first-principles quantity, and its value depends on cell type, tissue medium, and the experimental preparation against which it is fitted.  We adopt a literature-bracketed sweep rather than a single committed value.  At the iso-potential and hippocampal-preparation end of the bracket, Rawicz, Sukharev, and Salahshoor report $\alpha \approx 200$~\citep{rawicz2000effect,sukharev2012mechanosensitive,salahshoor2020transcranial} from planar bilayer area-expansion measurements and continuum mechanics, conditions that do not include cytoskeletal stress concentration, dendrite morphology, or viscoelastic load.  At the cortical-pyramidal end of the bracket, an additional factor of $\sim 5$ is plausible from the dendrite-heavy Piezo1 distribution alone, with Lewis \& Grandl~\citep{lewis2015mechanical} reporting a $\sim 5\times$ apical-dendrite-to-soma Piezo1 cluster density ratio.  Adding viscoelastic stress concentration at the membrane scale places the cortical-pyramidal upper bracket at $\alpha \approx 1000$.  Direct experimental calibration of $\alpha$ on neurons in tissue under ultrasound has not been reported.  The headline numbers in the canonical demonstration are presented at the $\alpha = 1000$ working point as a point on the bracket, not as a calibrated absolute.

Two consequences are operational.  First, $K_A$ and $\alpha$ enter only as the product $K_A \cdot \alpha$, so only this product is experimentally identifiable.  The sensitivity analysis in Table~\ref{tab:sensitivity} shows a $\pm 25\%$ shift in $K_A$ moves the firing count by $-59\%$ / $+89\%$, the dominant uncertainty in absolute spike-count predictions.  Second, sweeping $\alpha$ across the [200, 2000] literature bracket on the canonical Halle\,/\allowbreak\,dACC field walks the focal-voxel tension from sub-threshold to saturation.  At $\alpha \lesssim 460$ the focal-voxel tension is below $\Thalf$ and only the focal voxel fires, or none does.  At the canonical $\alpha = 1000$ working point the focal tension sits at $\sim 2.42\,\Thalf$ on the upper Boltzmann knee where mechanism scenarios diverge most strongly, and at $\alpha \gtrsim 2000$ firing saturates and the spatial gradient washes out.  This bracket is the primary uncertainty report for absolute spike-count predictions and is more informative than any single committed $\alpha$ value.  A direct experimental falsifier is paired shear-wave elastography of the tissue $\mu$ and single-cell mechano-current recording in the same preparation under transcranial ultrasound, back-fitting $K_A \cdot \alpha$ from the observed firing threshold.  High-density extracellular recordings under TUS protocols matching the simulated exposure would provide an independent calibration route.

The post-FDTD linear rescaling of the pressure field used to land the in-brain peak at Yaakub's reported 0.5\,MPa transcranial PNP in \S\ref{sec:results_tips} is a separate calibration step from the $\alpha$ bracket.  Acoustic propagation is linear in the source amplitude, so rescaling the propagation map by a single factor $s = p_{\mathrm{target}} / p_{\mathrm{observed,in\text{-}brain}}$ is exact and avoids re-running the FDTD.  The rescale factor and the observed-vs-target peak are recorded in the run's provenance metadata.  The downstream stages of ARF, shear-FDTD, bioheat, and neural pick up the rescaled field automatically.  The rescaled headline firing numbers are reported at $\alpha = 1000$.  The sweep in this section's accompanying figure shows how those numbers slide as $\alpha$ is varied within the literature bracket.

\subsection{Piezo1 Boltzmann fit against patch-clamp literature}
\label{sec:supp_verif_piezo1_lit}

\begin{figure}[htbp]
    \centering
    \includegraphics[width=\textwidth]{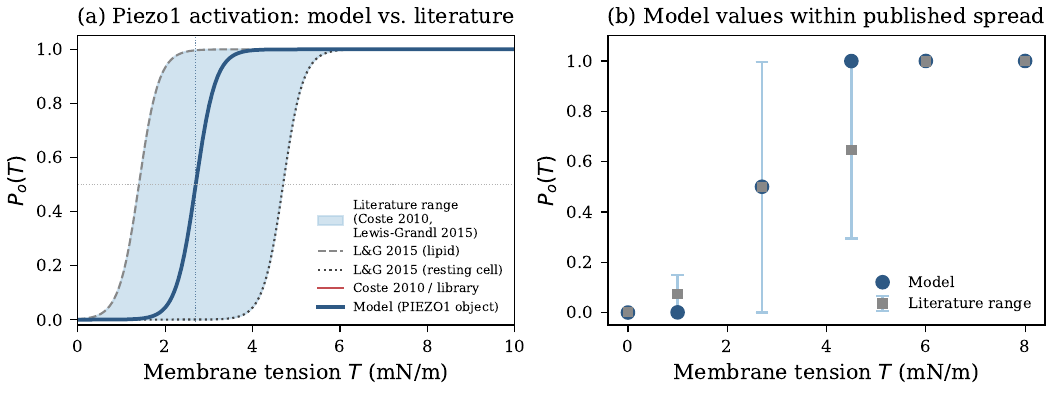}
    \caption{Piezo1 open-probability curve from the library compared with the patch-clamp literature uncertainty band~\citep{lewis2015mechanical,coste2010piezo}.  (a)~Continuous $P_o(T)$ overlay, where the shaded band spans the broader cytoskeleton-disrupted range $T_{1/2} \in [1.4, 4.7]$~mN/m at $\Delta A = 18$~nm$^2$, with the Coste and library fit at $T_{1/2} = 2.7$~mN/m drawn as a solid red reference.  (b)~Discrete check at six canonical tensions, with a marker for the model value and an error bar for the literature spread at that tension.  The library Boltzmann fit sits within the literature band across the full 0--10~mN/m range.}
    \label{fig:supp_piezo1_lit}
\end{figure}

Figure~\ref{fig:supp_piezo1_lit} overlays the library's $P_o(T)$ curve against the literature uncertainty band assembled from cell-attached patch-clamp records.  Lewis \& Grandl~\citep{lewis2015mechanical} report Boltzmann fits with $T_{1/2}$ in the 2.4--3.0~mN/m range for cytoskeleton-intact records, broadening to 1.4--4.7~mN/m across cytoskeleton-disrupted preparations, with $\Delta A$ in the 15--22~nm$^2$ range.  Coste et al.~\citep{coste2010piezo} report a comparable steepness in HEK293 patches.  The library curve sits well within the literature band across the full 0--10~mN/m range, with $|P_o^{\mathrm{model}}(T) - 0.5(P_o^{\mathrm{lo}}(T) + P_o^{\mathrm{hi}}(T))| < 0.25$ at the half-activation point and $< 0.05$ in the saturated regime at $T \geq 6$~mN/m, confirming that the chosen Boltzmann gating parameters are quantitatively consistent with published Piezo1 records.

\subsection{Pospischil regular-spiking f--I curve}
\label{sec:supp_verif_pospischil_fi}

\begin{figure}[htbp]
    \centering
    \includegraphics[width=0.95\textwidth]{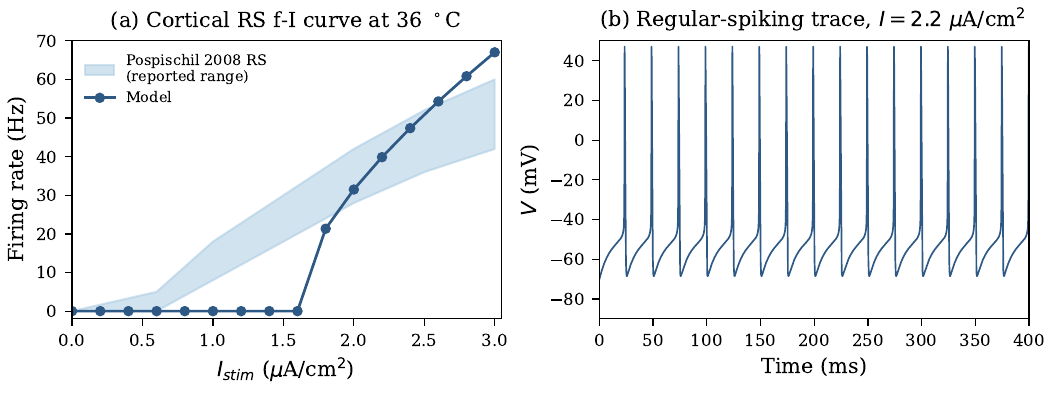}
    \caption{Cortical regular-spiking validation at the Pospischil reference temperature of $36\,^\circ$C with $\phi = 1.0$.  (a)~Steady-state firing rate vs.\ injected somatic current.  The shaded band is the regular-spiking reported range from~\citep[Fig.~3A]{pospischil2008minimal}.  The model's rheobase under a $\geq 2$-spike criterion is somewhat above the upper edge of the band, at $\approx 1.7$ against $0.6$--$1.0\,\mu$A/cm$^2$, and the firing rate at $I_{\mathrm{stim}} = 3\,\mu$A/cm$^2$ is approximately 67~Hz, near the upper edge of the 30--60~Hz reported range.  (b)~Sample suprathreshold voltage trace at $I = 2.2\,\mu$A/cm$^2$, showing the characteristic monotone non-bursting RS firing pattern.}
    \label{fig:supp_pospischil_fi}
\end{figure}

Figure~\ref{fig:supp_pospischil_fi} sweeps the somatic stimulus current density $I_{\mathrm{stim}} \in [0, 3]\,\mu$A/cm$^2$ at the Pospischil reference temperature of $36\,^\circ$C with $\phi = 1.0$, which enables direct comparison with the f--I curve in~\citep[Fig.~3A]{pospischil2008minimal}.  At mammalian body temperature $37\,^\circ$C the scaling factor rises to $\phi = 1.087$, modestly accelerating firing rates without altering the curve's qualitative shape.  The sweep runs over a 1500~ms simulation, discarding the first 200~ms as a transient and computing the steady-state firing rate from the inverse mean inter-spike interval.  When repetitive firing is scored by the $\geq 2$-spike inter-spike-interval criterion over the 1.3~s analysis window, the model's effective rheobase is approximately $1.7\,\mu$A/cm$^2$, somewhat above the $0.6$--$1.0\,\mu$A/cm$^2$ rheobase reported in~\citep[Fig.~3A]{pospischil2008minimal}.  The difference is attributable in part to a stricter spike-pair scoring threshold than the single-spike definition used there.  The saturated firing rate at $I_{\mathrm{stim}} = 3\,\mu$A/cm$^2$ is approximately 67~Hz, near the upper edge of the 30--60~Hz reported regular-spiking range.  The curve is monotone with no bursting and the voltage trace in panel~(b) shows the characteristic non-adapting regular-spiking pattern, as expected from the RS parameterisation.

\subsection{Three-compartment dendrite-soma-AIS validation: AIS-first initiation}
\label{sec:supp_verif_multicomp}

\begin{figure}[htbp]
    \centering
    \includegraphics[width=0.95\textwidth]{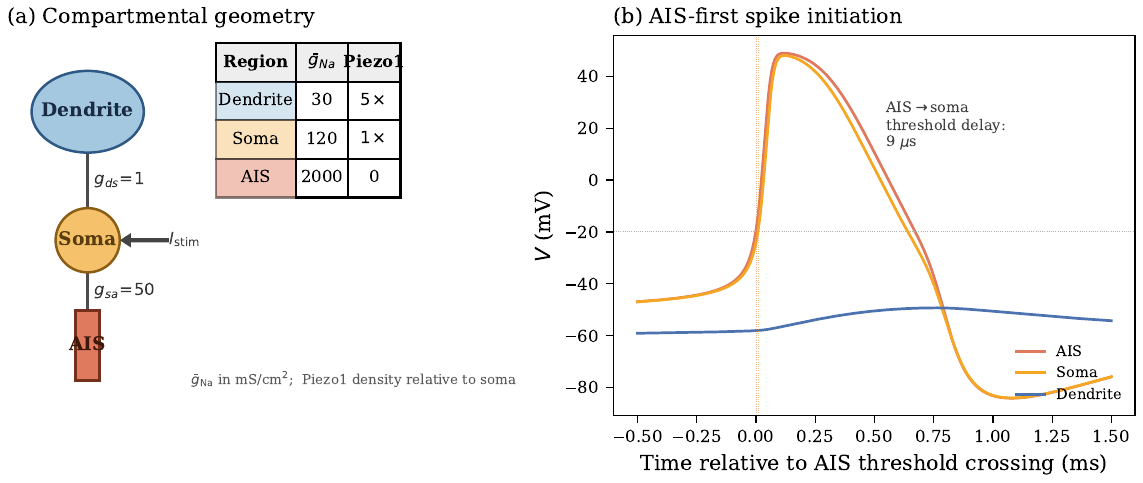}
    \caption{Three-compartment dendrite-soma-AIS validation.  (a)~Geometry and per-compartment parameters of the literature-parameterised cortical-pyramidal preset.  (b)~Voltage traces around the first spike under $I_{\mathrm{stim,soma}} = 12$~$\mu$A/cm$^2$, with time origin at the AIS threshold crossing of $-20$~mV.  The high-density AIS reaches threshold $\sim 9\,\mu$s before the soma despite the soma being the stimulation site.  The dendrite then receives a passive backpropagated depolarisation of $\sim 10$~mV thanks to the $A_s/A_d = 1/5$ area ratio.}
    \label{fig:supp_multicomp}
\end{figure}

Figure~\ref{fig:supp_multicomp} shows the three-compartment geometry and the voltage traces around the first spike under somatic current injection, validating the AIS-first prediction~\citep{stuart1997action}.  With the literature-parameterised cortical-pyramidal preset, where the dendrite is at $\bar{g}_{\mathrm{Na}}=30$~mS/cm$^2$ with large surface area and $5\times$ Piezo1 density, the soma at $\bar{g}_{\mathrm{Na}}=120$~mS/cm$^2$ with $1\times$ Piezo1, the AIS at $\bar{g}_{\mathrm{Na}}=2000$~mS/cm$^2$ with no Piezo1, and axial coupling $g_{ds}=1$~mS/cm$^2$ and $g_{sa}=50$~mS/cm$^2$ of soma area, the high-density AIS reaches the $-20$~mV threshold $\sim 9\,\mu$s before the soma despite the soma being the stimulation site.  The dendrite then receives a passive backpropagated depolarisation of only $\sim 10$~mV, attenuated by the area-ratio factor $A_s/A_d = 1/5$ at the dendrite-soma junction.  Setting all three compartments identical with zero coupling recovers the single-compartment trace bit-for-bit, providing a regression check.

\subsection{Intramembrane-cavitation pathway: leaflet dynamics, capacitance saturation, and duty-cycle direction-of-effect}
\label{sec:supp_verif_nice}

\begin{figure}[htbp]
    \centering
    \includegraphics[width=\textwidth]{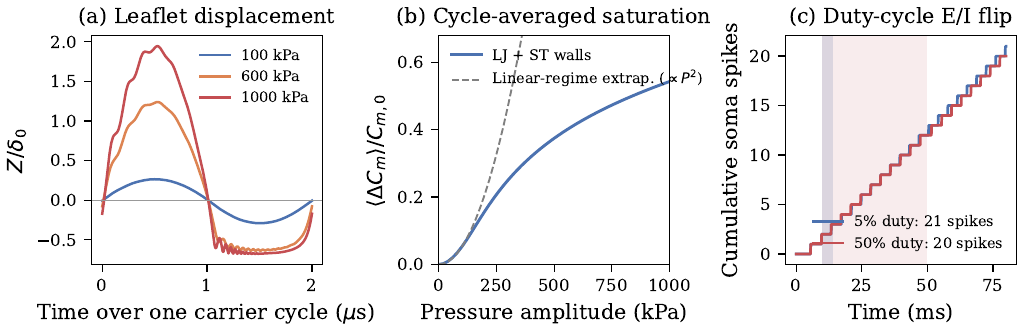}
    \caption{Intramembrane-cavitation validation.  (a)~Leaflet displacement $Z(t)$ over a single carrier cycle at 100, 600, and 1000~kPa drive amplitude, taken as the final cycle of an 8-cycle integration with $\delta_0 = 2$~nm.  At 100~kPa the response is nearly sinusoidal in the linear regime.  At 600--1000~kPa the trace is asymmetric, with the LJ molecular wall capping the compressive excursion and the surface-tension wall capping the expansion.  (b)~Cycle-averaged $\langle \Delta C_m \rangle / C_{m,0}$ versus pressure amplitude (solid blue) compared with the $P^2$ extrapolation calibrated at 100~kPa (dashed grey).  The two agree below $\sim$300~kPa and bend apart sharply above.  (c)~Cumulative somatic spike count under matched 100~kPa pressure amplitude with 5\% duty cycle as a single 4~ms ON pulse versus 50\% duty cycle as a single 40~ms ON pulse and matched 16~$\mu$A/cm$^2$ suprathreshold somatic stimulus.  Pulse-on windows shaded.  At 80~ms the 5\%-duty trial has fired one more spike than the 50\%-duty trial, matching~\citep[Fig.~6]{plaksin2014intramembrane}.}
    \label{fig:supp_nice_dynamics}
\end{figure}

Figure~\ref{fig:supp_nice_dynamics} validates the cycle-averaged intramembrane-cavitation pathway against three independent predictions from~\citep{plaksin2014intramembrane}.  Panel (a) shows the leaflet displacement $Z(t)$ over a single carrier cycle at three drive amplitudes spanning the linear and saturating regimes, where at 100~kPa the response is nearly sinusoidal.  At 600--1000~kPa the trace becomes asymmetric, with the LJ molecular wall capping the compressive excursion below $|Z| \approx 0.7\,\delta_0$ and the surface-tension wall capping the expansion at $|Z| \lesssim 2\,\delta_0$.  Panel (b) shows the cycle-averaged $\langle \Delta C_m \rangle / C_{m,0}$ versus pressure amplitude, illustrating the sub-quadratic saturation that the non-linear walls produce above $\sim$300~kPa, agreeing with the $P^2$ extrapolation calibrated at 100~kPa below the saturation onset and bending sharply away above it, asymptoting to $\sim 0.55$ at 1~MPa as predicted in~\citep[Fig.~2]{plaksin2014intramembrane}.  Panel (c) shows the cumulative somatic spike count under matched 100~kPa pressure amplitude with 5\% versus 50\% duty cycle, where the 5\%-duty trial has fired more spikes than the 50\%-duty trial by 80~ms, matching the direction predicted in~\citep[Fig.~6]{plaksin2014intramembrane}.

\subsection{Astrocytic TRPA1 relay: knockout, AP5, and slow-timescale controls}
\label{sec:supp_verif_astrocyte}

\begin{figure}[htbp]
    \centering
    \includegraphics[width=\textwidth]{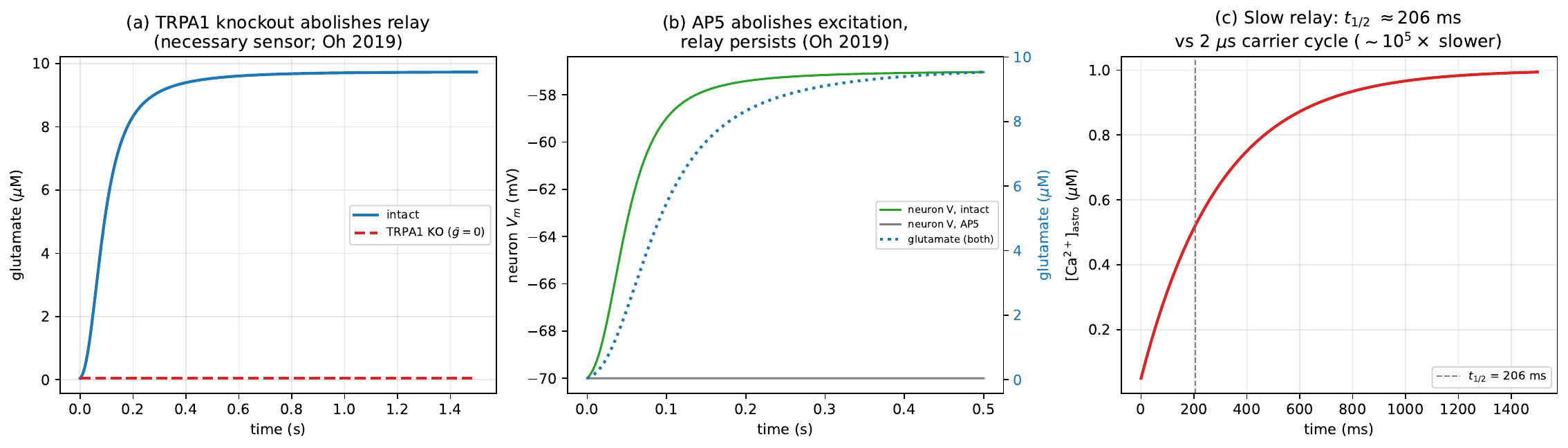}
    \caption{Astrocytic TRPA1-relay mechanism controls from Oh et al.\ 2019, at a 10~mN/m focal-tension drive.  (a)~TRPA1 knockout at $\bar{g}_{\mathrm{TRPA1}} = 0$ (red dashed) abolishes glutamate release relative to the intact relay (blue), confirming TRPA1 as the necessary sensor.  (b)~AP5 at $\bar{g}_{\mathrm{NMDA}} = 0$ (grey) abolishes the neuronal depolarisation that the intact pathway produces (green), while the astrocytic glutamate transient (blue dotted) is identical in both conditions, since the block acts downstream of release.  (c)~The astrocytic Ca\textsuperscript{2+} half-rise time of $t_{1/2} \approx 206$~ms is $\sim 10^5\times$ slower than the carrier cycle, the slow-relay signature.  Each panel corresponds to a mechanism-control assertion in the open-source test suite.}
    \label{fig:supp_astrocyte_valid}
\end{figure}

The astrocytic relay of \S\ref{sec:supp_astrocyte} is validated against the three qualitative controls reported by Oh et al.~\citep{oh2019ultrasonic}, each implemented as a unit test in the open-source repository.  Unlike the literature curve-matches above, these are mechanism-control anchors that pin the causal structure of the pathway rather than a single quantitative datum.  Figure~\ref{fig:supp_astrocyte_valid} shows all three on a focal-tension drive of 10~mN/m.  In panel (a), abolishing the TRPA1 conductance $\bar{g}_{\mathrm{TRPA1}} \to 0$, the knockout or HC-030031 or A-967079 antagonist condition, collapses the relay output, with glutamate falling from a $\sim 9.7~\mu$M intact peak to $< 0.05~\mu$M and confirming TRPA1 as the necessary sensor.  In panel (b), the AP5 condition $\bar{g}_{\mathrm{NMDA}} \to 0$ abolishes the neuronal depolarisation from $\sim 11$~mV to $0$~mV while the astrocytic Ca\textsuperscript{2+} and glutamate transients are unchanged, since the block acts downstream of gliotransmitter release.  In panel (c), the astrocytic Ca\textsuperscript{2+} rises with a half-time $t_{1/2} \approx 206$~ms, roughly five orders of magnitude slower than the 2~$\mu$s carrier cycle of a 500~kHz sonication, confirming that the relay integrates over the sonication rather than tracking the carrier.  Two further tests assert that the cheap scalar drive of \S\ref{sec:supp_astrocyte} is monotone in peak tension and grows with duty cycle and cumulative on-time.

\subsection{Mechanosensitive synaptic / voltage-gated transmission: monotone, baseline-preserving gains}
\label{sec:supp_verif_synaptic}

\begin{figure}[htbp]
    \centering
    \includegraphics[width=\textwidth]{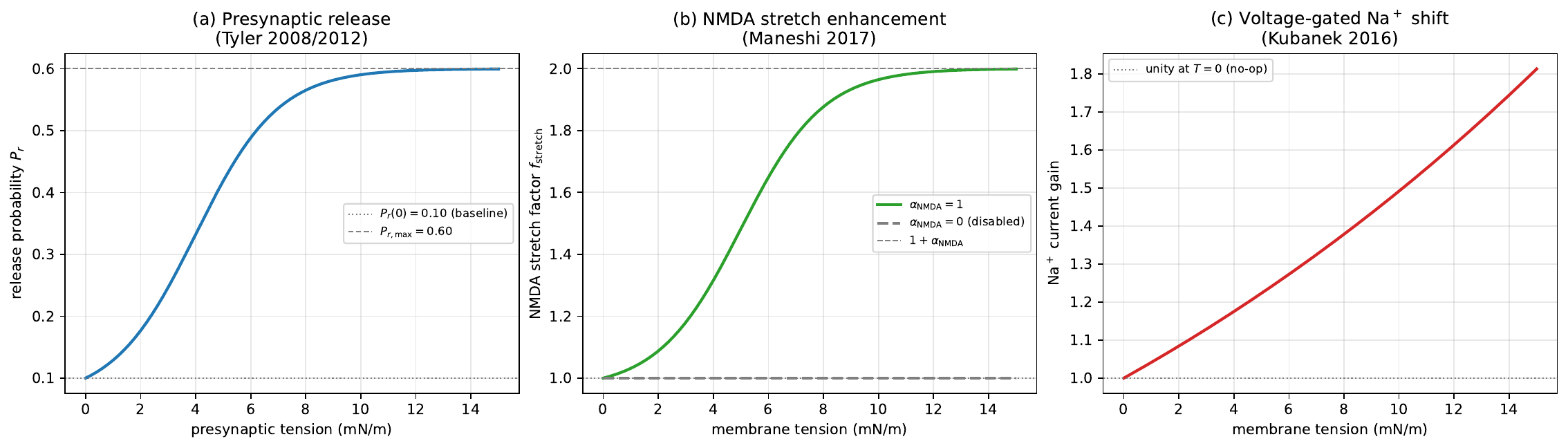}
    \caption{Mechanosensitive synaptic / voltage-gated transmission controls.  (a)~Presynaptic release probability $P_r(T)$ from Tyler, with baseline $0.10$ at zero tension and saturating to $P_{r,\max} = 0.60$.  (b)~Post-synaptic NMDA stretch factor from Maneshi, unity at zero tension and saturating to $1 + \alpha_{\mathrm{NMDA}} = 2$.  The $\alpha_{\mathrm{NMDA}} = 0$ case (grey dashed) stays flat at $1$, recovering the baseline receptor.  (c)~Voltage-gated Na\textsuperscript{+} current gain from Kubanek, unity at zero tension and monotone increasing.  All three gains equal $1$ or baseline at zero tension, the no-ultrasound no-op control, and each panel corresponds to assertions in the open-source test suite.}
    \label{fig:supp_synapse_valid}
\end{figure}

The synaptic / voltage-gated pathway of \S\ref{sec:supp_synaptic} bundles three literature-grounded gain factors, each validated against its source and implemented as a unit test in the open-source repository.  As with the astrocytic relay, these are mechanism-control anchors.  The shared requirement is that every factor reduces to the unmodified neuron at zero tension, the no-ultrasound no-op, and rises monotonically thereafter, so the pathway can never spuriously alter the baseline model.  Figure~\ref{fig:supp_synapse_valid} shows all three.  In panel (a), presynaptic release $P_r(T)$ rises from the resting $P_{r,0} = 0.10$ and saturates at $P_{r,\max} = 0.60$, following Tyler~\citep{tyler2008remote,tyler2012mechanobiology}.  In panel (b), the post-synaptic NMDA stretch factor $f_{\mathrm{stretch}}(0) = 1$ exactly and saturates to $1 + \alpha_{\mathrm{NMDA}} = 2$, a current doubling at saturating stretch reported by Maneshi~\citep{maneshi2017mechanical}.  Setting $\alpha_{\mathrm{NMDA}} = 0$ recovers the baseline receptor identically.  In panel (c), the voltage-gated Na\textsuperscript{+} current gain is unity at zero tension and increases monotonically with the Kubanek activation shift~\citep{kubanek2016ultrasound}, reaching $\sim 1.8\times$ at 15~mN/m at the canonical $k_{\mathrm{mechano}}$.  The relay-level controls, namely silent at zero tension, firing at high tension, and excitation monotone in tension, are likewise asserted in the test suite.

\begin{figure}[htbp]
    \centering
    \includegraphics[width=\textwidth]{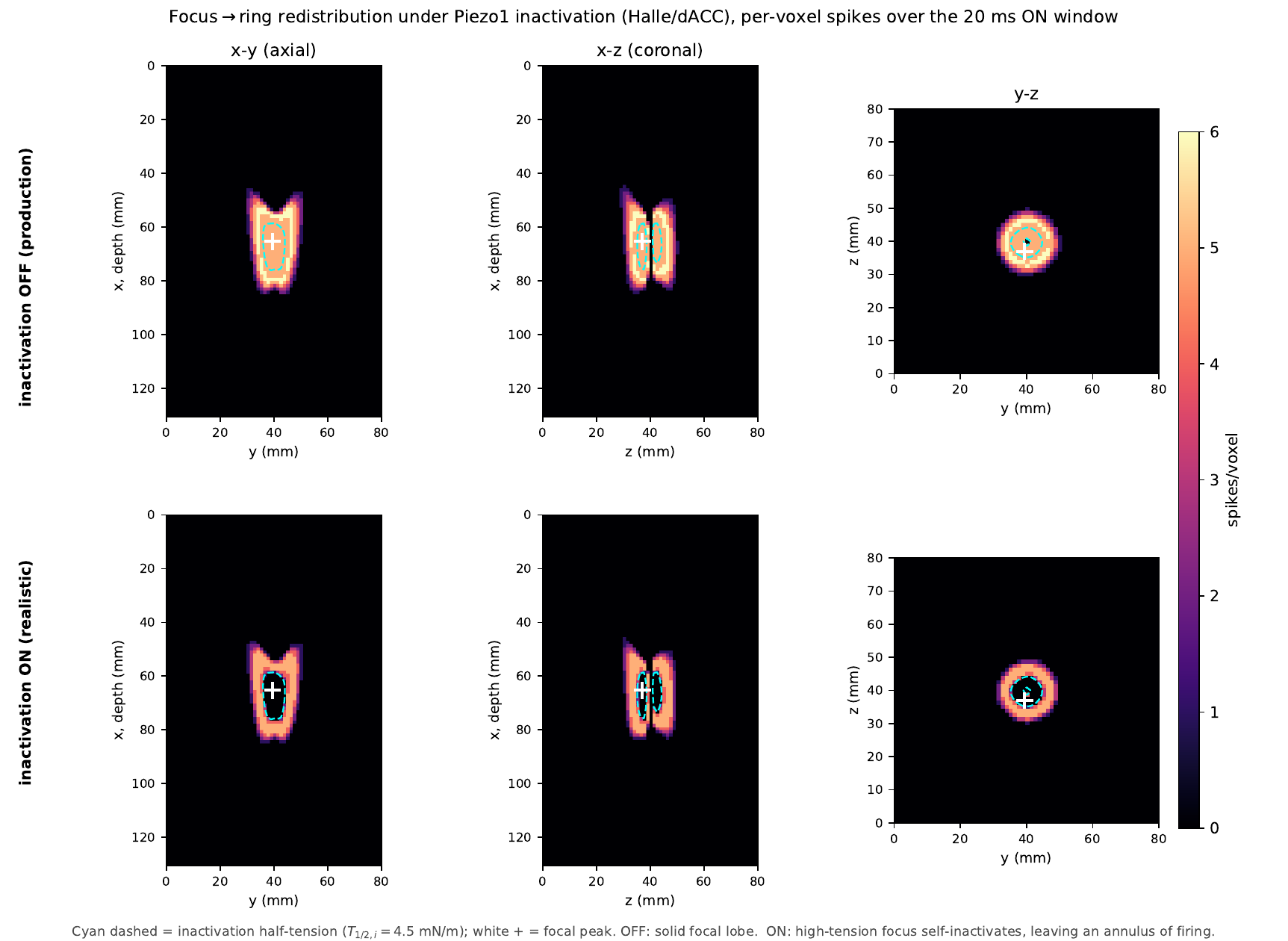}
    \caption{Spatial focus$\to$ring redistribution under Piezo1 inactivation on the Halle\,/\allowbreak\,dACC canonical field, showing per-voxel spikes over the 20\,ms ON window.  \textbf{Top}, inactivation OFF for the S2 production scenario, shows a solid focal lobe, brightest at the focal peak (white $+$).  \textbf{Bottom}, inactivation ON, shows the high-tension centre inside the cyan dashed $T_{1/2,i} = 4.5$~mN/m contour self-inactivating and falling silent, leaving an annulus of firing in the surrounding $\sim$3--4.5~mN/m shell, clearest in the $y$--$z$ view.  The columns are three orthogonal planes through the focal voxel.  The total firing-voxel count falls by $\sim$20\% here from inactivation alone, so the effect is primarily a \emph{redistribution} of the firing topology rather than an overall suppression.  Because it is a shape, not a magnitude, this focus$\to$ring signature is robust to the uncalibrated $K_A\cdot\alpha$ tension scale and offers a falsifiable discriminator between inactivation-dominated and purely activation-driven recruitment.}
    \label{fig:supp_focus_ring}
\end{figure}

% =============================================================================
\section{Full parameter source classification}
\label{sec:supp_provenance}

This section enumerates the source-classification entry for every numerical parameter used by the framework.  Each entry is classified as one of five kinds.  A Literature entry is fit to a published measurement and carries a citation.  A Calibrated entry is chosen so the model reproduces a qualitative literature observation, with the calibration target named.  An Assumption is a deliberate modelling choice not pinned by data.  An Estimate is a best guess from sparse or indirect data, weaker than an assumption, an example being the TRPA1 and synaptic mechanogating constants for which no direct tension-clamp fit exists.  A Derived entry is computed from other parameters.  An automated check enforces that every Literature entry carries a non-empty reference and every Calibrated entry carries non-empty calibration-target notes.  Entries are grouped by source category within each table.

Grouped by pathway, literature-anchored entries dominate the channel kinetics, covering the Piezo1 parameters $\Thalf$, $\Delta A$, $\tau_{\mathrm{act}}$, and $\tau_{\mathrm{inact}}$, the TRPV1 parameters $T_{1/2}$ and $E_{\mathrm{rev}}$, the SK parameters $[\mathrm{Ca}^{2+}]_{1/2}$ and $n_H$, and the dendrite-to-soma Piezo1 density ratio of $5\times$ from Lewis \& Grandl~\citep{lewis2015mechanical}.  Calibrated entries cover the NICE leaflet inertia, the TRP Boltzmann steepness, and the AIS sodium density, which is chosen at the lower end of the cited 2000--5000~mS/cm$^2$ literature range so that AIS-first initiation is preserved within the explicit-RK4 stability bound.  Assumed entries are dominated by the bilayer area-expansion modulus $K_A$, the multi-scale coupling factor $\alpha$, the Pinsky--Rinzel area ratios, and the AIS Piezo1 density, which is set to zero.

\input{data/provenance_tables}

% =============================================================================
\section{Numerical validation details}
\label{sec:supp_numerics}

\subsection{RK4 convergence}

\begin{figure}[htbp]
    \centering
    \includegraphics[width=0.42\textwidth]{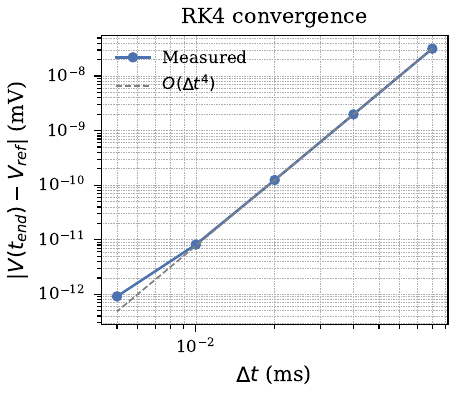}
    \caption{RK4 convergence test for the single-compartment integrator.  Error in terminal voltage decreases as $O(\Delta t^4)$, shown by the dashed reference line, when halving $\Delta t$.  Deviation from the ideal slope at the smallest steps reflects floating-point round-off in $V_{\mathrm{ref}}$ at $\sim 10^{-13}$~mV.}
    \label{fig:supp_rk4_convergence}
\end{figure}

The fourth-order Runge--Kutta integrator is verified by halving the time step and comparing terminal-voltage error to a fine-grid reference.  Figure~\ref{fig:supp_rk4_convergence} shows the empirical convergence of the single-compartment integrator, where error decreases approximately as $O(\Delta t^4)$ over the dominant range of step sizes, with the asymptotic ratio degraded slightly by gate-clamping non-linearities.  At the working time step $\Delta t = 0.01$~ms the residual error in terminal voltage is $\sim 0.02$~mV, negligible against the $\sim$100~mV action-potential amplitude.

% =============================================================================

\bibliographystyle{unsrt}
\bibliography{references}

\end{document}

%% file: data/sensitivity_table.tex
% Auto-generated by scripts/run_sensitivity.py — do not edit by hand.
\begin{table}
    \caption{Source-classified sensitivity at the S5 full-pathway baseline on the real Halle / dACC end-of-pulse strain field ($\varepsilon_{\mathrm{eq}}^{\max} = 2.61{\times}10^{-5}$, $\alpha = 1000$, 14\,275 baseline spikes). Each row shifts a single parameter by $\pm 25\%$ from its default and reports total spike count and fractional change, sorted by absolute deviation. The Piezo1 half-activation tension $T_{1/2}$ and the strain-tension coupling $K_A$ jointly dominate sensitivity by setting where the focal-voxel tension lies relative to the Boltzmann knee.}
    \label{tab:sensitivity}
    \footnotesize
    \centering
    \begin{tabular}{@{}lccccc@{}}
        \toprule
        Parameter & Source & $-25\%$ count & $+25\%$ count & $\Delta_{\!-}\,(\%)$ & $\Delta_{\!+}\,(\%)$ \\
        \midrule
        Piezo1 $T_{1/2}$ (mN/m) & Literature & 42\,049 & 3\,963 & +195 & -72 \\
        $K_A$ (N/m) & Assumption & 5\,861 & 26\,931 & -59 & +89 \\
        Dendrite Piezo1 scale & Literature & 10\,678 & 16\,802 & -25 & +18 \\
        SK $\bar{g}$ (mS/cm$^2$) & Assumption & 14\,651 & 13\,930 & +3 & -2 \\
        SK $[\mathrm{Ca}^{2+}]_{1/2}$ ($\mu$M) & Literature & 13\,977 & 14\,583 & -2 & +2 \\
        AIS Na density (mS/cm$^2$) & Literature & 14\,274 & 14\,282 & -0 & +0 \\
        % M6-BEGIN (synaptic / voltage-gated; scripts/run_sensitivity.py)
        \midrule
        \multicolumn{6}{@{}p{\linewidth}@{}}{\textit{Synaptic / voltage-gated transmission (M6), relative to 31 relay focal spikes over a 300~ms window at the focal voxel ($T = 6.5$~mN/m):}} \\
        \midrule
        NMDA stretch $\alpha_{\mathrm{NMDA}}$ & Calibrated & 27 & 34 & -13 & +10 \\
        NMDA $T_{1/2}^{\mathrm{s}}$ (mN/m) & Estimate & 33 & 27 & +6 & -13 \\
        Presynaptic $P_{r,\max}$ & Estimate & 29 & 32 & -6 & +3 \\
        Glutamate scale $G_s$ ($\mu$M) & Calibrated & 29 & 32 & -6 & +3 \\
        Presynaptic $T_{1/2}^{\mathrm{pre}}$ (mN/m) & Estimate & 31 & 30 & +0 & -3 \\
        Na$^+$ shift $k_{\mathrm{mechano}}$ (mV per mN/m) & Calibrated & 30 & 32 & -3 & +3 \\
        AMPA $\bar{g}$ (mS/cm$^2$) & Calibrated & 30 & 31 & -3 & +0 \\
        % M6-END
        % SOLITON-BEGIN (alternative mechanism; scripts/run_sensitivity.py)
        \midrule
        \multicolumn{6}{@{}p{\linewidth}@{}}{\textit{Heimburg-Jackson soliton (alternative thermodynamic-membrane mechanism), $-25\%$/$+25\%$ columns are the dimensionless excitability $\xi = \kappa_{\mathrm{rel}}(T)\,|P/\rho c^2|$ relative to $\xi_0 = $ $3.18{\times}10^{-5}$ at the focal voxel ($P = 0.50$~MPa, $T = 37^\circ$C). The $+25\%$ $T_m$ shift reaches $T_m = 37.5^\circ$C $\geq$ body temperature, where the heating-reduces-excitability sign no longer holds; its large $\Delta_{\!+}$ reflects $T_m$ crossing the operating point rather than a physical regime:}} \\
        \midrule
        Transition temp $T_m$ ($^\circ$C) & Literature & $1.92{\times}10^{-5}$ & $2.08{\times}10^{-4}$ & -40 & +555 \\
        Peak half-width $\sigma_T$ ($^\circ$C) & Estimate & $2.07{\times}10^{-5}$ & $5.27{\times}10^{-5}$ & -35 & +66 \\
        Excess-$c_p$ ratio $c_p^{\max}$ & Estimate & $3.70{\times}10^{-5}$ & $2.84{\times}10^{-5}$ & +17 & -10 \\
        % SOLITON-END
        \bottomrule
    \end{tabular}
\end{table}

%% file: data/provenance_tables.tex
% Auto-generated by scripts/render_provenance.py -- do not edit by hand.
% Place \input{data/provenance_tables} inside supplementary.tex.
% (supplementary.tex preamble already loads \usepackage{float}.)

\begin{table}[H]
    \caption{Source classification: Piezo1.}
    \label{tab:prov_piezo1}
    \footnotesize
    \centering
    \begin{tabular}{@{}llp{6.5cm}@{}}
        \toprule
        Parameter & Source & Reference / notes \\
        \midrule
        \texttt{Ca\_permeability\_fraction} & Literature & Coste et al. 2010 (P\_Ca / P\_Na \textasciitilde{}5) --- \textasciitilde{}10\% of cation current is Ca\textsuperscript{2}\textsuperscript{+} in physiological gradients \\
        \texttt{E\_rev} & Literature & Coste et al. 2010 (non-selective cation, \textasciitilde{}0 mV) \\
        \texttt{T\_half} & Literature & Coste et al. 2010, Science 330:55-60; Lewis \& Grandl 2015 --- Mid-point of activation across multiple cell types \\
        \texttt{delta\_A} & Literature & Coste et al. 2010 (Boltzmann fit to single-channel records) --- Effective gating area change from $\Delta$A$\cdot$T fit \\
        \texttt{g\_max} & Literature & Coste et al. 2010 (single-channel slope) --- Single-channel conductance \textasciitilde{}25 pS in physiological gradients \\
        \texttt{tau\_act} & Literature & Coste et al. 2010 Fig. 2 (activation rise $\tau$ \textasciitilde{}1-2 ms) --- Phase-5 fixed-tau replacement for the bell-shaped $\tau$\_w(T) \\
        \texttt{tau\_inact} & Literature & Coste et al. 2010 Fig. 2 (inactivation $\tau$ 10-15 ms) \\
        \texttt{g\_density} & Assumption & --- --- Sparse expression assumption; user typically overrides per cell type \\
        \texttt{inact\_T\_half} & Assumption & --- --- Set above activation T\_half; not directly fit in Coste 2010 \\
        \texttt{inact\_delta\_A} & Assumption & --- --- Comparable steepness to activation; placeholder pending dedicated fit \\
        \texttt{tau\_0} & Assumption & --- --- Single time constant subsuming activation kinetics; superseded by tau\_act when set \\
        \bottomrule
    \end{tabular}
\end{table}

\begin{table}[H]
    \caption{Source classification: TRAAK.}
    \label{tab:prov_traak}
    \footnotesize
    \centering
    \begin{tabular}{@{}llp{6.5cm}@{}}
        \toprule
        Parameter & Source & Reference / notes \\
        \midrule
        \texttt{E\_rev} & Literature & K\textsuperscript{+} Nernst at physiological [K] --- -90 mV typical \\
        \texttt{T\_half} & Literature & Brohawn et al. 2014, PNAS 111:3614-3619 \\
        \texttt{delta\_A} & Literature & Brohawn et al. 2014 \\
        \texttt{g\_max} & Literature & Brohawn et al. 2014 \\
        \texttt{tau\_act} & Literature & Brohawn et al. 2014 (single-channel mechanical activation rise) --- Phase-5 fixed-tau replacement for the bell-shaped $\tau$\_w(T) \\
        \texttt{g\_density} & Assumption & Estimated from neural expression levels; user-tunable \\
        \texttt{tau\_0} & Assumption & Legacy bell-shape baseline; superseded by tau\_act on the library preset \\
        \texttt{Ca\_permeability\_fraction} & Derived & K2P channels are K\textsuperscript{+}-selective; no Ca\textsuperscript{2}\textsuperscript{+} flux \\
        \texttt{inact\_T\_half} & Derived & Not applicable; tau\_inact is None \\
        \texttt{inact\_delta\_A} & Derived & Not applicable; tau\_inact is None \\
        \texttt{tau\_inact} & Derived & K2P channels don't inactivate; field defaults to None (gate held at 1) \\
        \bottomrule
    \end{tabular}
\end{table}

\begin{table}[H]
    \caption{Source classification: TREK-1.}
    \label{tab:prov_trek_1}
    \footnotesize
    \centering
    \begin{tabular}{@{}llp{6.5cm}@{}}
        \toprule
        Parameter & Source & Reference / notes \\
        \midrule
        \texttt{E\_rev} & Literature & K\textsuperscript{+} Nernst \\
        \texttt{T\_half} & Literature & Honor\'{e} 2007, Nat Rev Neurosci 8:251-261 \\
        \texttt{delta\_A} & Literature & Honor\'{e} 2007 \\
        \texttt{g\_max} & Literature & Honor\'{e} 2007 \\
        \texttt{tau\_act} & Literature & Honor\'{e} 2007 (mid-range of 3-10 ms activation $\tau$ reported in review) --- Phase-5 fixed-tau replacement for the bell-shaped $\tau$\_w(T) \\
        \texttt{g\_density} & Assumption & Cortical pyramidal neuron expression range \\
        \texttt{tau\_0} & Assumption & Legacy bell-shape baseline; superseded by tau\_act on the library preset \\
        \texttt{Ca\_permeability\_fraction} & Derived & K2P channels are K\textsuperscript{+}-selective; no Ca\textsuperscript{2}\textsuperscript{+} flux \\
        \texttt{inact\_T\_half} & Derived & Not applicable; tau\_inact is None \\
        \texttt{inact\_delta\_A} & Derived & Not applicable; tau\_inact is None \\
        \texttt{tau\_inact} & Derived & K2P channels don't inactivate; field defaults to None \\
        \bottomrule
    \end{tabular}
\end{table}

\begin{table}[H]
    \caption{Source classification: TREK-2.}
    \label{tab:prov_trek_2}
    \footnotesize
    \centering
    \begin{tabular}{@{}llp{6.5cm}@{}}
        \toprule
        Parameter & Source & Reference / notes \\
        \midrule
        \texttt{E\_rev} & Literature & K\textsuperscript{+} Nernst \\
        \texttt{T\_half} & Literature & Kang \& Kim 2024 review \\
        \texttt{delta\_A} & Literature & Kang \& Kim 2024 \\
        \texttt{g\_max} & Literature & Kang \& Kim 2024 \\
        \texttt{tau\_act} & Literature & Kang \& Kim 2024 (TREK-2 kinetics; similar to TREK-1) --- Phase-5 fixed-tau replacement for the bell-shaped $\tau$\_w(T) \\
        \texttt{g\_density} & Assumption & Cortical pyramidal neuron expression range \\
        \texttt{tau\_0} & Assumption & Legacy bell-shape baseline; superseded by tau\_act on the library preset \\
        \texttt{Ca\_permeability\_fraction} & Derived & K2P channels are K\textsuperscript{+}-selective; no Ca\textsuperscript{2}\textsuperscript{+} flux \\
        \texttt{inact\_T\_half} & Derived & Not applicable; tau\_inact is None \\
        \texttt{inact\_delta\_A} & Derived & Not applicable; tau\_inact is None \\
        \texttt{tau\_inact} & Derived & K2P channels don't inactivate; field defaults to None \\
        \bottomrule
    \end{tabular}
\end{table}

\begin{table}[H]
    \caption{Source classification: TRPV1.}
    \label{tab:prov_trpv1}
    \footnotesize
    \centering
    \begin{tabular}{@{}llp{6.5cm}@{}}
        \toprule
        Parameter & Source & Reference / notes \\
        \midrule
        \texttt{E\_rev} & Literature & Slightly Ca\textsuperscript{2}\textsuperscript{+}-preferring non-selective cation, \textasciitilde{}10 mV \\
        \texttt{T\_half} & Literature & Caterina et al. 1997, 2000 (heat activation threshold) \\
        \texttt{k} & Calibrated & Boltzmann steepness chosen so Po(47)/Po(37) > 20 matching the literature Q10 $\geq$ 20 floor (Caterina 1997). \\
        \texttt{g\_density} & Assumption & Conservative mid-range density when expressed \\
        \bottomrule
    \end{tabular}
\end{table}

\begin{table}[H]
    \caption{Source classification: TRPV4.}
    \label{tab:prov_trpv4}
    \footnotesize
    \centering
    \begin{tabular}{@{}llp{6.5cm}@{}}
        \toprule
        Parameter & Source & Reference / notes \\
        \midrule
        \texttt{E\_rev} & Literature & Non-selective cation, \textasciitilde{}0 mV \\
        \texttt{T\_half} & Literature & Watanabe et al. 2002 (warm-activation threshold) --- Within the 27-35$^\circ$C literature range \\
        \texttt{k} & Calibrated & Q10 $\approx$ 10 around threshold, less steep than TRPV1 \\
        \texttt{g\_density} & Assumption & Conservative mid-range density when expressed \\
        \bottomrule
    \end{tabular}
\end{table}

\begin{table}[H]
    \caption{Source classification: TRPA1.}
    \label{tab:prov_trpa1}
    \footnotesize
    \centering
    \begin{tabular}{@{}llp{6.5cm}@{}}
        \toprule
        Parameter & Source & Reference / notes \\
        \midrule
        \texttt{Ca\_permeability\_fraction} & Literature & Karashima et al. 2010; Wang et al. 2008 (P\_Ca/P\_Na $\approx$ 5--8) --- \textasciitilde{}15--20\% of TRPA1 cation current is Ca\textsuperscript{2}\textsuperscript{+} at physiological gradients. \\
        \texttt{E\_rev} & Literature & Karashima et al. 2010 (non-selective cation, E\_rev $\approx$ 0 mV) \\
        \texttt{g\_max} & Literature & Karashima et al. 2010; Wang et al. 2008 (TRPA1 single-channel \textasciitilde{}tens of pS) \\
        \texttt{g\_density} & Assumption & Astrocytic TRPA1 surface density; user overrides per preparation. \\
        \texttt{tau\_0} & Assumption & Legacy bell-shape baseline; superseded by tau\_act. \\
        \texttt{T\_half} & Estimate & Oh et al. 2019, Curr Biol 29:3386 (US activates astrocytic TRPA1) --- Half-activation tension for TRPA1 mechanogating is not directly quantified; set between Piezo1 (2.7) and TRAAK (5.0) mN/m as a best guess pending a dedicated tension-clamp fit. \\
        \texttt{delta\_A} & Estimate & Gating-area sensitivity guessed comparable to other MS channels (\textasciitilde{}15 nm\textsuperscript{2}). \\
        \texttt{tau\_act} & Estimate & Channel gating is fast (ms); the slow relay signature lives in the downstream Ca/glutamate dynamics, not in TRPA1 gating. \\
        \bottomrule
    \end{tabular}
\end{table}

\begin{table}[H]
    \caption{Source classification: strain\_to\_tension.}
    \label{tab:prov_strain_to_tension}
    \footnotesize
    \centering
    \begin{tabular}{@{}llp{6.5cm}@{}}
        \toprule
        Parameter & Source & Reference / notes \\
        \midrule
        \texttt{K\_A} & Assumption & Lipid-bilayer area-expansion modulus, range 0.1-0.5 N/m --- Default 0.25 N/m is the centre of the published range (Rawicz 2000, Sukharev 2012); the actual coupling from tissue-scale strain to per-channel membrane tension subsumes geometric concentration, cytoskeletal mechanics, and membrane curvature effects. User must calibrate for the specific tissue/cell-type/exposure under study. \\
        \texttt{coupling\_factor} & Assumption & Dimensionless multi-scale amplification $\alpha$ relating far-field tissue strain to local membrane area change. Code default 1.0; manuscript-canonical $\alpha$=1000 (cortical-pyramidal-calibrated, 5$\times$ the iso-potential/hippocampal literature $\alpha$$\approx$200). Calibrate per tissue/cell-type/exposure. \\
        \bottomrule
    \end{tabular}
\end{table}

\begin{table}[H]
    \caption{Source classification: NICEModel.}
    \label{tab:prov_nicemodel}
    \footnotesize
    \centering
    \begin{tabular}{@{}llp{6.5cm}@{}}
        \toprule
        Parameter & Source & Reference / notes \\
        \midrule
        \texttt{C\_m0} & Literature & Standard neuronal specific capacitance, 1 $\mu$F/cm\textsuperscript{2} \\
        \texttt{delta\_0} & Literature & Lipid bilayer half-thickness (Plaksin 2014) \\
        \texttt{f\_carrier} & Literature & Typical TUS carrier (Legon, Tyler) --- 500 kHz default; user supplies their protocol value \\
        \texttt{f\_resonant} & Literature & Plaksin et al. 2014 (leaflet resonance \textasciitilde{}5 MHz) \\
        \texttt{lj\_strength} & Calibrated & Dimensionless prefactor for the LJ-like wall on Z<0; calibrated so the wall contribution stays <5\% of linear restoring at |Z| < 0.1$\cdot$$\delta$\_0 \\
        \texttt{m\_eff} & Calibrated & SONIC operating range, Lemaire et al. 2019 --- Calibrated so |Z|(100 kPa) $\approx$ 0.6 nm $\to$ $\langle$$\Delta$C\_m$\rangle$/C\_m0 $\approx$ 5\% \\
        \texttt{st\_strength} & Calibrated & Dimensionless prefactor for the surface-tension wall on Z>0; default 1.0 doubles asymptotic restoring on the expansion side, extending the quantitative validity range to \textasciitilde{}1 MPa (Plaksin \& Krasovitski 2014 Fig. 2). \\
        \texttt{damping\_ratio} & Assumption & $\zeta$ = 0.1 (Q $\approx$ 5); Plaksin-consistent moderate damping \\
        \bottomrule
    \end{tabular}
\end{table}

\begin{table}[H]
    \caption{Source classification: SolitonModel.}
    \label{tab:prov_solitonmodel}
    \footnotesize
    \centering
    \begin{tabular}{@{}llp{6.5cm}@{}}
        \toprule
        Parameter & Source & Reference / notes \\
        \midrule
        \texttt{T\_m} & Literature & Heimburg 2007, Thermal Biophysics of Membranes; Heimburg \& Jackson 2005, PNAS 102:9790 --- Membrane melting transition. Biological membranes melt \textasciitilde{}10-15$^\circ$C below body temperature; default 30$^\circ$C is representative (lung-surfactant T\_m $\approx$ 30$^\circ$C, the system Heimburg \& Jackson fit). Must be < 37$^\circ$C for the heating-reduces-excitability sign. \\
        \texttt{c} & Literature & Soft-tissue speed of sound; matches repo FDTD C0 = 1540 m/s --- $\rho$c\textsuperscript{2} is the adiabatic bulk modulus for the strain conversion. \\
        \texttt{rho} & Literature & Soft-tissue/water density; matches repo FDTD RHO0 = 1000 kg/m\textsuperscript{3} --- Used in the bulk-modulus strain $\varepsilon$ = P/($\rho$c\textsuperscript{2}). \\
        \texttt{peak\_shape} & Assumption & Gaussian (default) vs Lorentzian shape for the normalised c\_p peak. \\
        \texttt{c\_p\_max} & Estimate & Heimburg 2007 ($\kappa$\_A\^{}T $\propto$ $\Delta$c\_p proportionality) --- Dimensionless peak-to-baseline excess-compressibility ratio encoding Heimburg's c\_p $\leftrightarrow$ compressibility divergence near T\_m. Tuned amplitude (default 10$\times$). \\
        \texttt{sigma\_T} & Estimate & Heimburg 2007 (broadened multicomponent-membrane transition) --- Half-width of the c\_p / compressibility peak, a few $^\circ$C. Biological transitions are broad; specific value tuned. \\
        \bottomrule
    \end{tabular}
\end{table}

\begin{table}[H]
    \caption{Source classification: CalciumDynamics.}
    \label{tab:prov_calciumdynamics}
    \footnotesize
    \centering
    \begin{tabular}{@{}llp{6.5cm}@{}}
        \toprule
        Parameter & Source & Reference / notes \\
        \midrule
        \texttt{Ca\_rest} & Literature & Cortical pyramidal resting [Ca\textsuperscript{2}\textsuperscript{+}]\_i, \textasciitilde{}50 nM \\
        \texttt{tau\_decay} & Literature & Mainen \& Sejnowski 1996; Borg-Graham 1999 --- Single-pool decay $\tau$ in cortical models, range 30-80 ms \\
        \texttt{flux\_factor} & Calibrated & Lumped sub-membrane shell volume $\times$ buffering fraction $\times$ 1/(z$\cdot$F). Default 0.5 lies within published current-to-concentration coupling factors used in cortical Ca\textsuperscript{2}\textsuperscript{+}-pool models. \\
        \bottomrule
    \end{tabular}
\end{table}

\begin{table}[H]
    \caption{Source classification: KCaChannel.}
    \label{tab:prov_kcachannel}
    \footnotesize
    \centering
    \begin{tabular}{@{}llp{6.5cm}@{}}
        \toprule
        Parameter & Source & Reference / notes \\
        \midrule
        \texttt{Ca\_half} & Literature & K\"{o}hler et al. 1996 (SK channel) \\
        \texttt{E\_K} & Literature & K\textsuperscript{+} Nernst, matches Pospischil RS \\
        \texttt{n\_hill} & Literature & Hirschberg et al. 1998 \\
        \texttt{g\_density} & Assumption & Cortical pyramidal SK density range, 0.1--1.0 mS/cm\textsuperscript{2} (Sah 1996) \\
        \bottomrule
    \end{tabular}
\end{table}

\begin{table}[H]
    \caption{Source classification: AstrocyteRelay.}
    \label{tab:prov_astrocyterelay}
    \footnotesize
    \centering
    \begin{tabular}{@{}llp{6.5cm}@{}}
        \toprule
        Parameter & Source & Reference / notes \\
        \midrule
        \texttt{Ca\_rest} & Literature & Bazargani \& Attwell 2016 (resting astrocytic [Ca\textsuperscript{2}\textsuperscript{+}] \textasciitilde{}50--100 nM) \\
        \texttt{V\_astro} & Literature & Bazargani \& Attwell 2016, Nat Neurosci (astrocyte V\_rest $\approx$ -80 mV) --- Sets the Ca\textsuperscript{2}\textsuperscript{+} driving force through TRPA1 (non-spiking cell). \\
        \texttt{tau\_decay} & Literature & Bazargani \& Attwell 2016; Oh et al. 2019 (astrocytic Ca\textsuperscript{2}\textsuperscript{+} transients decay over 100s of ms) --- Slow relay: hundreds of ms, not per-carrier-cycle. \\
        \texttt{flux\_factor} & Calibrated & Chosen so the astrocytic Ca\textsuperscript{2}\textsuperscript{+} pool peaks \textasciitilde{}1 $\mu$M at full US drive (physiological gliotransmission regime), given TRPA1 g\_density, Ca permeability, and the -80 mV driving force. \\
        \texttt{glu\_Ca\_half} & Estimate & Parpura \& Haydon 2000 (Ca-dependent gliotransmission threshold) --- Half-release [Ca\textsuperscript{2}\textsuperscript{+}] a few hundred nM; tuned for gliotransmission onset. \\
        \texttt{glu\_max} & Estimate & Peak cleft glutamate scale [$\mu$M] from gliotransmission; order-of-magnitude. \\
        \texttt{glu\_n\_hill} & Estimate & Cooperative Ca\textsuperscript{2}\textsuperscript{+}-triggered release; Hill coefficient \textasciitilde{}2--3 (vesicular sensor). \\
        \bottomrule
    \end{tabular}
\end{table}

\begin{table}[H]
    \caption{Source classification: NMDAReceptor.}
    \label{tab:prov_nmdareceptor}
    \footnotesize
    \centering
    \begin{tabular}{@{}llp{6.5cm}@{}}
        \toprule
        Parameter & Source & Reference / notes \\
        \midrule
        \texttt{E\_rev} & Literature & NMDA non-selective cation, E\_rev $\approx$ 0 mV \\
        \texttt{K\_glu} & Literature & Patneau \& Mayer 1990 (NMDA glutamate EC50 \textasciitilde{}1--3 $\mu$M) \\
        \texttt{Mg} & Literature & Jahr \& Stevens 1990 (physiological extracellular [Mg\textsuperscript{2}\textsuperscript{+}] $\approx$ 1 mM) \\
        \texttt{mg\_block\_v\_scale} & Literature & Jahr \& Stevens 1990 (voltage-dependent Mg\textsuperscript{2}\textsuperscript{+} block: 16.13 mV, 3.57 mM) --- g(V) $\propto$ 1/(1 + [Mg]$\cdot$exp(-V/16.13)/3.57). \\
        \texttt{alpha\_NMDA} & Calibrated & Maneshi et al. 2017, Sci Rep (membrane stretch activates NMDARs); Singh et al. 2012 --- Max fractional stretch enhancement of NMDA current. Default 1.0 $\to$ current roughly doubles at saturating stretch, matching Maneshi 2017 (stretch $\approx$ doubled NMDA current). f\_stretch($\infty$) = 1 + alpha\_NMDA. \\
        \texttt{g\_max} & Assumption & Post-synaptic NMDA conductance density; user overrides per cell type. \\
        \texttt{T\_half\_stretch} & Estimate & Maneshi et al. 2017 --- Half-enhancement membrane tension [mN/m]. Mapped to the MS-channel tension range pending a dedicated stretch-clamp NMDA fit. \\
        \texttt{k\_stretch} & Estimate & Boltzmann slope [mN/m] of the stretch sigmoid; comparable steepness to MS channels. \\
        \bottomrule
    \end{tabular}
\end{table}

\begin{table}[H]
    \caption{Source classification: MechanoNav.}
    \label{tab:prov_mechanonav}
    \footnotesize
    \centering
    \begin{tabular}{@{}llp{6.5cm}@{}}
        \toprule
        Parameter & Source & Reference / notes \\
        \midrule
        \texttt{k\_a} & Literature & Hodgkin-Huxley-type Nav activation Boltzmann slope --- Activation slope factor [mV]. \\
        \texttt{v\_half0} & Literature & Representative mammalian Nav activation midpoint (Hodgkin-Huxley-type) --- Baseline activation half-voltage [mV] for the scalar gain. \\
        \texttt{k\_mechano} & Calibrated & Kubanek et al. 2016, Sci Rep (US shifts Nav/Cav voltage dependence) --- Hyperpolarising activation-shift sensitivity [mV per (mN/m)]. K\_MECHANO\_NAV = 0.3 is calibrated so a few-mN/m tension yields a \textasciitilde{}mV-scale activation shift and a tens-of-percent Na\textsuperscript{+} current change, the order Kubanek report (not a directly tabulated value). The neuron field k\_mechano\_nav defaults to 0 (pathway off) to preserve the baseline neuron; set it to K\_MECHANO\_NAV to enable. \\
        \texttt{v\_eval} & Assumption & Near-threshold reference voltage [mV] at which nav\_current\_gain reads the open fraction. \\
        \bottomrule
    \end{tabular}
\end{table}

\begin{table}[H]
    \caption{Source classification: MechanoSynapse.}
    \label{tab:prov_mechanosynapse}
    \footnotesize
    \centering
    \begin{tabular}{@{}llp{6.5cm}@{}}
        \toprule
        Parameter & Source & Reference / notes \\
        \midrule
        \texttt{E\_ampa} & Literature & AMPA non-selective cation reversal, \textasciitilde{}0 mV \\
        \texttt{K\_ampa} & Literature & AMPA is low-affinity (glutamate EC50 \textasciitilde{} hundreds of $\mu$M) --- AMPA glutamate half-saturation [$\mu$M]. \\
        \texttt{Pr\_baseline} & Literature & Tyler 2008; central-synapse release probability range 0.1--0.3 --- Resting vesicle release probability at zero tension. \\
        \texttt{tau\_glu} & Literature & Cleft glutamate clearance / uptake time constant (few ms) --- First-order glutamate pool decay [ms]. \\
        \texttt{g\_ampa} & Calibrated & Post-synaptic AMPA conductance density [mS/cm\textsuperscript{2}]; calibrated with glu\_scale. \\
        \texttt{glu\_scale} & Calibrated & Sustained cleft glutamate [$\mu$M] at full release; calibrated with g\_ampa so the resting (baseline-Pr) state is sub-threshold and US-driven release brings the post neuron to firing. \\
        \texttt{Pr\_max} & Estimate & Tyler 2012 (mechanical perturbation raises release probability) --- Saturating release probability at high presynaptic tension. \\
        \texttt{T\_half\_pre} & Estimate & Tyler 2012 --- Half-modulation presynaptic tension [mN/m]; in the MS-channel range. \\
        \texttt{k\_pre} & Estimate & Boltzmann slope [mN/m] of the release-probability sigmoid. \\
        \bottomrule
    \end{tabular}
\end{table}

\begin{table}[H]
    \caption{Source classification: cortical\_pyramidal\_3comp.}
    \label{tab:prov_cortical_pyramidal_3comp}
    \footnotesize
    \centering
    \begin{tabular}{@{}llp{6.5cm}@{}}
        \toprule
        Parameter & Source & Reference / notes \\
        \midrule
        \texttt{dendrite\_piezo\_scale} & Literature & Lewis \& Grandl 2015; Dalghi 2019 --- Qualitative dendrite-heavy Piezo1 expression \\
        \texttt{g\_Na\_ais} & Literature & Kole 2008 (lower end of 2000-5000 mS/cm\textsuperscript{2} range) --- Default chosen for RK4 stability at dt = 1 $\mu$s; user can raise to 5000 \\
        \texttt{g\_Na\_dend} & Literature & Stuart \& Sakmann 1995 \\
        \texttt{g\_Na\_soma} & Literature & Mainen \& Sejnowski 1996 \\
        \texttt{area\_ratios} & Assumption & Pinsky-Rinzel-style lumped ratios A\_d:A\_s:A\_a = 5:1:0.05 \\
        \texttt{g\_dend\_soma} & Assumption & Lumped axial conductance, mS/cm\textsuperscript{2} of soma \\
        \texttt{g\_soma\_ais} & Assumption & Tighter coupling than dend-soma, reflecting AIS proximity \\
        \bottomrule
    \end{tabular}
\end{table}